\documentclass[11pt,a4paper]{article}
\pdfoutput=1
\usepackage{jheppub}
\usepackage[T1]{fontenc}
\usepackage{fix-cm}
\usepackage{lmodern}
\usepackage{amsmath,amssymb,mathtools,mathrsfs,empheq}
\usepackage{physics}
\usepackage{bbm}
\usepackage{bm}
\usepackage{hhline}

\newcommand\nn{\nonumber}
\newcommand\tx{\tilde x}
\newcommand\tu{\tilde u}
\newcommand\mn{\mathfrak n}
\newcommand\tq{\tilde q}
\newcommand\tB{\tilde B}
\newcommand\tn{\tilde n}
\newcommand\tv{\tilde v}

\newcommand\mm{\mathfrak m}
\newcommand\tmm{\tilde{\mathfrak m}}
\newcommand\tpsi{\widetilde{\psi}}
\newcommand\mkt{\mathfrak{t}}
\newcommand\tPsi{\widetilde{\Psi}}
\newcommand\tDelta{\tilde{\Delta}}
\newcommand\tmn{\tilde{\mathfrak{n}}}

\newcommand\mF{\mathcal{F}}

\makeatletter
\newcommand*{\rom}[1]{\expandafter\@slowromancap\romannumeral #1@}
\makeatother

\def\ri{{\rm i}}

\usepackage{tikz}

\usepackage{xcolor}

\newcommand{\mg}{\mathfrak{g}}

\newcommand\fft[2]{\frac{#1}{#2}}

\def\SO{{\rm SO}}

\def\SU{{\rm SU}}
\def\U{{\rm U}}

\usepackage{tensor}

\title{Large $N$ Partition Functions of 3d Holographic SCFTs}
\author[a]{Nikolay Bobev,}
\author[a]{Junho Hong,}
\author[b]{and Valentin Reys}

\affiliation[a]{Instituut voor Theoretische Fysica, KU Leuven, \\
	Celestijnenlaan 200D, B-3001 Leuven, Belgium}

\affiliation[b]{Université Paris-Saclay, CNRS, CEA, \\
	Institut de physique théorique, 91191, Gif-sur-Yvette, France}

\emailAdd{nikolay.bobev@kuleuven.be}
\emailAdd{junho.hong@kuleuven.be}
\emailAdd{valentin.reys@ipht.fr}

\abstract{We study the $S^1\times\Sigma_\mg$ topologically twisted index and the squashed sphere partition function of various 3d $\mathcal N\geq2$ holographic superconformal field theories arising from M2-branes. Employing numerical techniques in combination with well-motivated conjectures we provide compact closed-form expressions valid to all orders in the perturbative $1/N$ expansion for these observables. We also discuss the holographic implications of our results for the topologically twisted index for the dual M-theory Euclidean path integral around asymptotically AdS$_4$ solutions of 11d supergravity. In Lorentzian signature this leads to a prediction for the corrections to the Bekenstein-Hawking entropy of a class of static asymptotically AdS$_4$ BPS black holes.}

\begin{document}
	
\maketitle

%%%%%%%%%%%%%%%%%%%%%%%%
\section{Introduction}
\label{sec:intro}
%%%%%%%%%%%%%%%%%%%%%%%%

Supersymmetric localization is an invaluable tool to explore precision holography and has been applied with great success to many AdS/CFT setups. The key technical advantage is that supersymmetric localization often reduces the path integral of supersymmetric QFTs to finite dimensional matrix models. In favorable circumstances, and for specific background sources in the path integral, the resulting matrix models can be evaluated explicitly. The utility of this approach to the path integral of supersymmetric QFTs is clear for holographic SCFTs since it allows for the large $N$ and strong coupling calculation of various physics observables. These results can in turn be used in holography to test the AdS/CFT correspondence and to provide concrete calculational tools to go beyond the leading supergravity approximation. Numerous examples of this approach to top-down holography have been explored recently, see \cite{Pestun:2016zxk,Zaffaroni:2019dhb} for reviews. Our goal in this work is to continue this quest in the context of AdS$_4$/CFT$_3$ and derive new results for the partition functions of holographic 3d $\mathcal{N}=2$ SCFTs realized on the worldvolume of M2-branes.

To set the context of our discussion we begin with a summary of some supersymmetric localization results for the 3d $\mathcal N=6$ U$(N)_k\times$U$(N)_{-k}$ Chern-Simons-matter theory constructed by Aharony-Bergman-Jafferis-Maldacena (ABJM) \cite{Aharony:2008ug}, which is the prime example of a 3d holographic SCFT. The supersymmetric localization matrix model for the ABJM theory on $S^3$ \cite{Kapustin:2009kz} can be evaluated to all orders in the perturbative $1/N$ expansion and the result is compactly written in terms of an Airy function which  receives additional exponentially suppressed non-perturbative corrections. This result has been deduced by employing the relation of this matrix model to topological strings on local $\mathbb{P}^1\times\mathbb{P}^1$ \cite{Fuji:2011km}, and also by using an ideal Fermi-gas formulation \cite{Marino:2011eh}.  According to the AdS/CFT correspondence, the large $N$ $S^3$ ABJM partition function should be dual to the M-theory path integral around the Euclidean AdS$_4\times S^7/\mathbb Z_k$ solution of 11d supergravity.  This duality has been confirmed at the leading $N^\fft32$ order in \cite{Drukker:2010nc}, see also \cite{Herzog:2010hf}, but also at the first $N^\fft12$ sub-leading order \cite{Bobev:2020egg,Bobev:2021oku} and at the universal $\log N$ order \cite{Bhattacharyya:2012ye}. Beyond the $\log N$ correction, the $S^3$ ABJM partition function written in terms of an Airy function provides a remarkable prediction for the Euclidean M-theory path integral to all orders in the perturbative $1/N$ expansion, which seems highly non-trivial to confirm  directly in M-theory.\footnote{See \cite{Dabholkar:2014wpa,Hristov:2022lcw} for attempts, based on supersymmetric localization of 4d supergravity on Euclidean AdS$_4$, at a holographic derivation of the full Airy function capturing the all-order $1/N$ expansion of the ABJM $S^3$ partition function.}

These impressive results motivated several generalizations to include deformations of the ABJM theory on $S^3$ as well as the study of the $S^3$ partition function of other holographic SCFTs with similar methods. For instance, the $S^3$ partition function of the ABJM theory deformed by two real masses \cite{Nosaka:2015iiw} and the partition function on the squashed sphere, $S^3_b$, with the specific value of the squashing parameter $b=\sqrt{3}$ \cite{Hatsuda:2016uqa} can also be expressed in terms of an Airy function that captures the all-order $1/N$ expansion of these observables in the large $N$ limit. Similar results for the $S^3$ partition function in terms of an Airy function have been derived also for large $N$ holographic SCFTs arising from D-branes and orientifold planes in type IIA string theory \cite{Mezei:2013gqa}. Recently, it was conjectured that the all-order $S^3_b$ partition function for the large $N$ ABJM theory deformed by three real masses and general squashing parameter is also given in terms of an Airy function, see \cite{Bobev:2022jte,Bobev:2022eus,Hristov:2022lcw}.

Given these results one may wonder if similar compact expressions can be obtained for the partition functions of the ABJM theory on other compact Euclidean manifolds to all orders in the $1/N$ expansion. Indeed, it was recently shown in \cite{Bobev:2022jte,Bobev:2022eus,Bobev:2022wem}, see also \cite{Liu:2017vll} for earlier work, that this is possible for the topologically twisted index (TTI) and the superconformal index (SCI) of the theory. The TTI is the partition function of the theory on $S^1\times\Sigma_{\mg}$ with a partial topological twist on the Riemann surface $\Sigma_{\mg}$ of genus $\mg$, see \cite{Benini:2015noa,Benini:2015eyy,Benini:2016hjo,Closset:2016arn}. The large $N$ expansion of this index for the ABJM theory admits a simple compact resummation in terms of elementary functions that captures the full answer up to exponentially suppressed non-perturbative contributions, see \cite{Bobev:2022jte,Bobev:2022eus}.\footnote{See the discussion around \eqref{eq:calB1calB2} below as well as Section 6 of \cite{Bobev:2022eus} for a more precise definition of what we mean when we discuss all order TTI results in the large $N$ limit.} Similar results can be derived for the all-order large $N$ SCI of the ABJM theory in the so-called Cardy-like limit of small angular momentum fugacity \cite{Bobev:2022wem}. These results can be successfully compared with calculations in the dual supergravity theory. The leading $N^\fft32$ term in the ABJM TTI accounts for the Bekenstein-Hawking entropy of a dual magnetically charged static BPS AdS$_4$ black hole upon an appropriate change of ensembles and application of $\mathcal{I}$-extremization \cite{Benini:2015eyy}. The first subleading $N^\fft12$ term in the TTI can be matched with the correction from the Wald entropy formula to the Bekenstein-Hawking entropy of the black hole \cite{Bobev:2020egg,Bobev:2021oku}, while the universal $\log N$ contribution can be compared with a 1-loop supergravity calculation \cite{Liu:2017vbl}. The all-order ABJM TTI obtained in \cite{Bobev:2022jte,Bobev:2022eus} is consistent with these supergravity results and furthermore should encode the all-order perturbative quantum correction to the dual magnetically charged AdS$_4$ black hole entropy. Put differently, these large $N$ results for the ABJM TTI provide a precise prediction for the M-theory path integral around asymptotically Euclidean AdS$_4$ ``black saddles'' with an $S^7/\mathbb Z_k$ internal manifold to all orders in the perturbative $1/N$ expansion \cite{Bobev:2020pjk,Bobev:2022jte,Bobev:2022eus}. This is analogous to the role played by the $S^3$ ABJM partition function for the M-theory path integral around Euclidean AdS$_4\times S^7/\mathbb Z_k$. Similar results are available for the ABJM SCI dual to rotating asymptotically AdS black hole solutions \cite{Bobev:2022wem}.

It is natural to ask whether the results summarized above for the deformed $S^3$ partition function and the TTI can be extended to other 3d holographic SCFTs. In this work we show that this is indeed possible. Using the numerical techniques of \cite{Bobev:2022jte,Bobev:2022eus} we perform extensive numerical studies of the TTI for four classes of 3d SCFTs with known gravitational duals in M-theory. The examples we consider are the 3d $\mathcal{N}=4$ ADHM theory, as well as the Chern-Simons-matter theories dual to AdS$_4\times N^{0,1,0}$, AdS$_4\times V^{5,2}$, and AdS$_4\times Q^{1,1,1}$, and appropriate orbifolds thereof. In all these examples the very precise numerical data allows us to conjecture closed form analytic expressions for the all-order large $N$ TTI which are valid up to exponentially suppressed terms in the large $N$ limit. Our results amount to a significant improvement of the calculations in \cite{Hosseini:2016tor,Hosseini:2016ume,PandoZayas:2020iqr} where the leading $N^{\fft32}$ term and the universal $\log N$ term in the TTI for these theories were studied. Moreover, building on the recent work \cite{Chester:2021gdw,Minahan:2021pfv,Bobev:2022jte,Bobev:2022eus}, we conjecture an Airy function type formula for the large $N$ squashed $S^3$ partition function of the 3d $\mathcal{N}=4$ ADHM model deformed by a single real mass parameter. Finally, we consider the mABJM theory which is a 3d $\mathcal{N}=2$ SCFT obtained from the ABJM model by a superpotential mass deformation, see \cite{Jafferis:2011zi,Bobev:2018uxk,Bobev:2018wbt}, and has a holographic description in terms of the CPW AdS$_4$ vacuum of M-theory \cite{Corrado:2001nv}. Using the results for the large $N$ ABJM TTI and squashed and mass-deformed $S^3$ partition function of the ABJM theory in \cite{Bobev:2022jte,Bobev:2022eus} we deduce the corresponding expressions for the mABJM model. 

Our field theory results have important implications for holography which we discuss in some detail. In particular, we obtain predictions for the corrections to the Bekenstein-Hawking entropy of various static BPS AdS$_4$ black hole solutions in M-theory, or alternatively predictions for the path integral of M-theory on Euclidean asymptotically AdS$_4$ gravitational backgrounds. We also use our results in conjunction with the higher-derivative supergravity analysis in \cite{Bobev:2020egg,Bobev:2021oku} to calculate the coefficients of four-derivative terms in 4d $\mathcal{N}=2$ minimal gauged supergravity obtained as a consistent truncation on the Sasaki-Einstein manifolds we study. As a by-product of our analysis we present predictions for the subleading $N^{\fft12}$ in the partition function of the four holographic SCFTs placed on the so-called spindle geometries recently discussed in the literature.

The rest of this paper is organized as follows. In Section~\ref{sec:TTI} we introduce the four different 3d $\mathcal N\geq2$ holographic SCFTs of interest and derive closed-form expressions for their TTI to all orders in the perturbative $1/N$ expansion based on the numerical techniques used in \cite{Bobev:2022jte,Bobev:2022eus} to study the ABJM TTI. In Section~\ref{sec:S3} we consider the large $N$ $S^3_b$ partition function of 3d holographic SCFTs with real mass deformations and discuss the challenges for calculating it for theories with $\mathcal N=2$ supersymmetry. The $S^3$ partition function and the TTI for the 3d $\mathcal N=2$ mABJM theory are discussed in Section~\ref{sec:mABJM}. We continue in Section~\ref{sec:holo} with a discussion on the holographic implications of our results for the Euclidean M-theory path integral and the all-order quantum corrections to the entropy of supersymmetric AdS$_4$ black holes. We summarize our main results and discuss some open questions in Section~\ref{sec:discussion}. Various technical details on the evaluation of the TTI and the dual AdS$_4$ description for four of the holographic SCFTs of interest in this work are provided in the Appendices.

%\NPB{Make a general statement about the imaginary part of the TTI somewhere.}

%%%%%%%%%%%%%%%%%
\section{Topologically twisted index}
\label{sec:TTI}
%%%%%%%%%%%%%%%%%

In this section we investigate the topologically twisted index (TTI) of various 3d $\mathcal N\geq2$ holographic SCFTs on $S^1\times\Sigma_{\mg}$, with $\Sigma_{\mg}$ a Riemann surface of genus $\mg$. Using supersymmetric localization, the path integral representation of the TTI can be recast as a matrix model. For generic $\mathcal N=2$ SCFTs this was first studied in~\cite{Hosseini:2016tor}, based on the earlier work of~\cite{Benini:2015noa,Benini:2015eyy}, and explicit examples have then been studied in~\cite{Hosseini:2016ume,PandoZayas:2020iqr}. We will first present the matrix model for the TTI of general $\mathcal N=2$ SCFTs and then move on to specific examples in the following subsections.

Consider an $\mathcal N=2$ Chern-Simons-matter quiver gauge theory with $p$ nodes, each of which represents a gauge group U$(N)_{k_a}$ with Chern-Simons (CS) levels $k_a~(a=1,\cdots,p)$. The theory may also include $\mathcal N=2$ chiral multiplets $\Psi_s~(s=1,\cdots,I)$ in the $R_s$ representation of the gauge group $\otimes_{a=1}^p$U$(N)_{k_a}$. In this work, we focus on the case where $R_s$ is either a pair of bi-fundamental and anti-bi-fundamental, an adjoint, or a pair of fundamental and anti-fundamental representations. To be specific, the $\mathcal N=2$ SCFT of interest may include
\begin{itemize}
	\item Pairs of chiral multiplets $\Psi_{(a,b)}$ and $\Psi_{(b,a)}$ transforming in the $(\boldsymbol{N},\overline{\boldsymbol{N}})$ and $(\overline{\boldsymbol{N}},\boldsymbol{N})$ representations of U$(N)_{k_a}\times$U$(N)_{k_b}$, with associated magnetic fluxes $\mn_{\Psi_{(a,b)}}$ and $\mn_{\Psi_{(b,a)}}$ and fugacities $y_{\Psi_{(a,b)}}$ and $y_{\Psi_{(b,a)}}$, respectively,
	
	\item Chiral multiplets $\Psi_{(a,a)}$ transforming in the adjoint representation of U$(N)_{k_a}$, with associated magnetic flux $\mn_{\Psi_{(a,a)}}$ and fugacity $y_{\Psi_{(a,a)}}$,
	
	\item Pairs of chiral multiplets $\Psi_{a}$ and $\tPsi_{a}$ transforming in the $\boldsymbol{N}$ and $\overline{\boldsymbol N}$ representations of U$(N)_{k_a}$, with associated magnetic fluxes $\mn_{\Psi_a}$ and $\mn_{\widetilde\Psi_a}$ and fugacities $y_{\Psi_a}$ and $y_{\widetilde\Psi_a}$, respectively.

\end{itemize}
Here the magnetic fluxes on $\Sigma_{\mg}$ and the fugacities of the chiral multiplets are associated with the Cartan subgroup of the global symmetry group of the SCFT. The latter consists of the U(1)$_R$ superconformal R-symmetry together with various U(1) flavor symmetries. For simplicity, we will refer to them as flavor magnetic fluxes and flavor fugacities from here on. \\

The $S^1\times S^2$ TTI of the SCFT with the above matter content reads~\cite{Hosseini:2016tor,Hosseini:2016ume,PandoZayas:2020iqr}
\begin{align}
\label{general:TTI}
		Z_{S^1\times S^2}&=\fft{1}{(N!)^p}\sum_{\mm_1,\ldots,\mm_p\in\mathbb Z^N}\int_{\mathcal C}\prod_{a=1}^p\Bigg[\prod_{i=1}^N\fft{dx_{a,i}}{2\pi \mathrm{i}x_{a,i}}(x_{a,i})^{k_a\mm_{a,i}+\mkt_a}(\xi_a)^{\mm_{a,i}}\prod_{i\neq j}^N\bigg(1-\fft{x_{a,i}}{x_{a,j}}\bigg)\Bigg]\nonumber \\
		&\quad\times\prod_{i,j=1}^N\Bigg[\prod_{\text{bi-fund pairs }\Psi_{(a,b)}\,\&\,\Psi_{(b,a)}}\bigg(\fft{\sqrt{\fft{x_{a,i}}{x_{b,j}}y_{\Psi_{(a,b)}}}}{1-\fft{x_{a,i}}{x_{b,j}}y_{\Psi_{(a,b)}}}\bigg)^{\mm_{a,i}-\mm_{b,j}-\mn_{\Psi_{(a,b)}}+1}\\
		&\kern15em\times\bigg(\fft{\sqrt{\fft{x_{b,j}}{x_{a,i}}y_{\Psi_{(b,a)}}}}{1-\fft{x_{b,j}}{x_{a,i}}y_{\Psi_{(b,a)}}}\bigg)^{-\mm_{a,i}+\mm_{b,j}-\mn_{\Psi_{(b,a)}}+1}\Bigg]\nonumber \\
		&\quad\times\prod_{i,j=1}^N\Bigg[\prod_{\text{adjoint }\Psi_{(a,a)}}\bigg(\fft{\sqrt{\fft{x_{a,i}}{x_{a,j}}y_{\Psi_{(a,a)}}}}{1-\fft{x_{a,i}}{x_{a,j}}y_{\Psi_{(a,a)}}}\bigg)^{\mm_{a,i}-\mm_{a,j}-\mn_{\Psi_{(a,a)}}+1}\Bigg] \nonumber \\
		&\quad\times\prod_{i=1}^N\Bigg[\prod_{\text{fund }\Psi_a}\bigg(\fft{\sqrt{x_{a,i}y_{\Psi_a}}}{1-x_{a,i}y_{\Psi_a}}\bigg)^{\mm_{a,i}-\mn_{\Psi_a}+1}\prod_{\text{anti-fund }\tPsi_a}\bigg(\fft{\sqrt{\fft{1}{x_{a,i}}y_{\widetilde\Psi_a}}}{1-\fft{1}{x_{a,i}}y_{\widetilde\Psi_a}}\bigg)^{-\mm_{a,i}-\mn_{\widetilde\Psi_a}+1}\Bigg]\,, \nonumber
\end{align}
where we have also included contributions from the U(1)$_a$ topological symmetry written in terms of the associated fluxes $\mkt_a$ and fugacities $\xi_a$. Before considering specific SCFTs, let us collect some general remarks on the above matrix model for the TTI:
\begin{itemize}
	\item The contour $\mathcal C$ in (\ref{general:TTI}) captures the contribution from so-called Jeffrey-Kirwan residues, which can be written in terms of the Bethe Ansatz (BA) formula that we will briefly review for various examples in the following subsections. We refer to~\cite{Benini:2015noa,Benini:2015eyy} for a more detailed discussion related to the choice of this contour. 
	
	\item The generalization of the $S^1\times S^2$ TTI (\ref{general:TTI}) to the $S^1\times\Sigma_{\mg}$ TTI with a Riemann surface of genus $\mg$ can be obtained following~\cite{Benini:2016hjo}. We will provide the resulting explicit relation between $Z_{S^1\times S^2}$ and $Z_{S^1\times\Sigma_{\mg}}$ for various examples in the following subsections.
	
	\item We introduce flavor chemical potentials $\Delta$ in terms of the flavor fugacities as $y=e^{\mathrm{i}\pi\Delta}$. We then define the ``superconformal $\Delta$-configuration'' as the set of flavor chemical potentials that extremizes the large $N$ limit of the Bethe potential, which will be introduced explicitly below for various SCFTs. Then, since the Bethe potential as a function of chemical potentials becomes proportional to the $S^3$ free energy as a function of trial $R$-charges in the large $N$ limit (provided chemical potentials are identified with trial $R$-charges~\cite{Azzurli:2017kxo}),
%\footnote{In \cite{Azzurli:2017kxo} chemical potentials are identified with trial $R$-charges multiplied by $\pi$ but here they are mapped identically since we pull out the factor of $\pi$ from the definition of chemical potentials in advance as $y=e^{\mathrm{i}\pi\Delta}$.}
and since the latter is extremized at the exact superconformal $R$-charges~\cite{Jafferis:2010un,Jafferis:2011zi}, the superconformal $\Delta$-configuration so defined precisely matches the exact superconformal $R$-charges. It is worth mentioning, however, that the flavor chemical potentials for the $S^1\times\Sigma_\mg$ TTI and the trial $R$-charges for the $S^3$ partition function are only formally related to each other by the map between the Bethe potential and the free energy~\cite{Azzurli:2017kxo}.
	
	\item We define the ``universal $\mn$-configuration'', also called the universal twist, as the set of flavor magnetic fluxes that ensures a non-zero magnetic flux only for the U(1)$_R$ superconformal R-symmetry and vanishing magnetic fluxes for all other U(1) flavor symmetries. This universal twist is achieved simply by setting the $\mn$-configuration to be proportional to the superconformal $\Delta$-configuration just defined~\cite{Azzurli:2017kxo,Bobev:2017uzs}.
\end{itemize}
%

%%%%%
\subsection{$\mathcal N=4$ ADHM theory dual to AdS$_4\times S^7/\mathbb Z_{N_f}$}
\label{sec:TTI:ADHM}
%%%%%

In this subsection we consider the TTI of the $\mathcal N=4$ SCFT with gauge group U($N$), one adjoint hypermultiplet, $N_f$ fundamental hypermultiplets, and no Chern-Simons term for the gauge field. Decomposing the $\mathcal{N} = 4$ multiplets into $\mathcal N=2$ ones as in~\cite{Kapustin:2010xq}, the theory consists of three adjoint chiral multiplets $\Psi_I~(I=1,2,3)$ and $N_f$ pairs of fundamental and anti-fundamental chiral multiplets $\psi_q$ and $\tpsi_q~(q=1,\ldots,N_f)$ in an $\mathcal{N} = 2$ language. The superpotential is given as
\begin{equation}
	W=\Tr\Bigg[\sum_{q=1}^{N_f}\tpsi_q\Psi_3\psi_q+\Psi_3[\Psi_1,\Psi_2]\Bigg] \, .\label{ADHM:W}
\end{equation}
This SCFT can be realized as the low-energy effective worldvolume theory of $N$ M2-branes probing a $\mathbb C^2\times(\mathbb C^2/\mathbb Z_{N_f})$ singularity and enjoys $\SU(2)\times \SU(N_f)$ flavor symmetry in addition to the $\SO(4)$ R-symmetry~\cite{Benini:2009qs,Bashkirov:2010kz,Mezei:2013gqa,Grassi:2014vwa}. For $N_f=1$, it is therefore equivalent to the ABJM theory with CS level $k=1$ realized as the low-energy effective worldvolume theory of $N$ M2-branes probing $\mathbb C^4/\mathbb Z_k$~\cite{Aharony:2008ug}. The $\mathcal{N}=4$ SCFT is often called the ADHM theory after the authors of~\cite{Atiyah:1978ri}, or simply the $N_f$ model as in~\cite{Grassi:2014vwa,Minahan:2021pfv}.

To evaluate the ADHM TTI, we first briefly introduce its BA formulation in Section~\ref{sec:TTI:ADHM:BA}. Then in Section~\ref{sec:TTI:ADHM:analytic} we review the analytic calculation of the ADHM TTI in the large $N$ limit based on the BA formula. Finally in Sections~\ref{sec:TTI:ADHM:num-result} and~\ref{sec:TTI:ADHM:num-derive} we study the BA formulation numerically and deduce an analytic expression for the ADHM TTI to all orders in the perturbative $1/N$ expansion.

%%%%%
\subsubsection{Bethe Ansatz formulation}\label{sec:TTI:ADHM:BA}
%%%%%

Using the general formula (\ref{general:TTI}), the $S^1 \times S^2$ TTI of the ADHM theory reduces to the following matrix model:
\begin{equation}
\begin{split}
	Z^\text{ADHM}_{S^1\times S^2}(N,N_f,\Delta,\mn)&=\fft{1}{N!}\sum_{\mm\in\mathbb Z^N}\oint_{\mathcal C}\prod_{i=1}^N\fft{dx_i}{2\pi \mathrm{i}x_i}\,x_i^{\mathfrak t}\,\xi^{\mm_i}\prod_{i\neq j}^N\left(1-\fft{x_i}{x_j}\right)\\
	&\quad\times\prod_{I=1}^3\prod_{i,j=1}^N\left(\fft{\sqrt{\fft{x_i}{x_j}y_I}}{1-\fft{x_i}{x_j}y_I}\right)^{\mm_i-\fft12\mn_I+\fft12}\left(\fft{\sqrt{\fft{x_j}{x_i}y_I}}{1-\fft{x_j}{x_i}y_I}\right)^{-\mm_i-\fft12\mn_I+\fft12}\\
	&\quad\times\prod_{i=1}^N\left(\fft{\sqrt{x_iy_q}}{1-x_iy_q}\right)^{N_f(\mm_i-\mn_q+1)}\left(\fft{\sqrt{\fft{1}{x_i}y_{\tq}}}{1-\fft{1}{x_i}y_{\tq}}\right)^{N_f(-\mm_i-\mn_{\tq}+1)} \, ,\label{ADHM:TTI:1}
\end{split}
\end{equation}
where we have assigned the same fugacities $y_q$ and $y_{\tilde q}$ and magnetic fluxes $\mn_q$ and $\mn_{\tilde q}$ for the $N_f$ pairs of fundamental and anti-fundamental multiplets for simplicity. From the marginality of the ADHM superpotential (\ref{ADHM:W}) under global symmetries, the chemical potentials and magnetic fluxes are constrained as 
\begin{equation}
	\Delta_q+\Delta_{\tq}+\Delta_3=\Delta_1+\Delta_2+\Delta_3=2 \, ,\qquad \mn_q+\mn_{\tq}+\mn_3=\mn_1+\mn_2+\mn_3=2 \, .\label{ADHM:constraints}
\end{equation}
Here we have simplified the notation compared to (\ref{general:TTI}) by denoting\footnote{The fugacity $\Delta_m$ and its corresponding background flux $\mkt$ are defined below \eqref{general:TTI}. The different prescription for $\Delta_m$ for an even/odd $N$ allows for the proper integer choice (\ref{ADHM:integer}) below, see footnote 12 of \cite{Hosseini:2016tor} for a related discussion.}
\begin{equation}
\begin{alignedat}{2}
	y_{\Psi_{(a,b)}}&\to y_I=e^{\mathrm{i}\pi\Delta_I}\, ,&\quad y_{\Psi_a}&\to y_q=e^{\mathrm{i}\pi\Delta_q}\, ,\\
	y_{\tPsi_a}&\to y_{\tq}=e^{\mathrm{i}\pi\Delta_{\tq}}\, ,&\quad \xi_a&\to\xi=e^{\mathrm{i}\pi(\Delta_m-N-1+2\lfloor\fft{N+1}{2}\rfloor)}\, .
\end{alignedat}
\end{equation}
The superconformal $\Delta$-configuration and the universal $\mn$-configuration correspond to
\begin{equation}
\begin{alignedat}{3}
	\Delta_1&=\Delta_2=\fft12\,,&\qquad\Delta_3&=1\,,&\qquad\Delta_m&=0\,,\\
	\mn_1&=\mn_2=\fft12\,,&\qquad \mn_3&=1\,,&\qquad \mkt&=0\,,\label{ADHM:constraints:sc}
\end{alignedat}
\end{equation}
which will be confirmed later by extremizing the large $N$ Bethe potential with respect to the chemical potentials under the constraint (\ref{ADHM:constraints}). For convenience, we use the notation $\Delta$ and $\mn$ to collectively represent all the flavor chemical potentials and magnetic fluxes as
\begin{equation}
	\Delta=(\Delta_I,\Delta_q,\Delta_{\tq},\Delta_m) \, , \qquad \mn=(\mn_I,\mn_q,\mn_{\tq},\mkt)\, .\label{ADHM:Delta:mn}
\end{equation}

\medskip

To obtain the BA formula for the $S^1\times S^2$ ADHM TTI, we first rewrite the matrix model (\ref{ADHM:TTI:1}) as
\begin{equation}
	\begin{split}
		Z^\text{ADHM}_{S^1\times S^2}(N,N_f,\Delta,\mn)&=\fft{1}{N!}\oint_{\mathcal C}\prod_{i=1}^N\fft{dx_i}{2\pi \mathrm{i}x_i}\,x_i^{\mathfrak t}\,\prod_{i\neq j}^N\left(1-\fft{x_i}{x_j}\right)\prod_{i=1}^N\fft{(e^{\mathrm{i}B_i})^M}{e^{\mathrm{i}B_i}-1}\\
		&\quad\times\prod_{I=1}^3\prod_{i,j=1}^N\left(\fft{\sqrt{\fft{x_i}{x_j}y_I}}{1-\fft{x_i}{x_j}y_I}\right)^{-\fft12\mn_I+\fft12}\left(\fft{\sqrt{\fft{x_j}{x_i}y_I}}{1-\fft{x_j}{x_i}y_I}\right)^{-\fft12\mn_I+\fft12}\\
		&\quad\times\prod_{i=1}^N\left(\fft{\sqrt{x_iy_q}}{1-x_iy_q}\right)^{N_f(-\mn_q+1)}\left(\fft{\sqrt{\fft{1}{x_i}y_{\tq}}}{1-\fft{1}{x_i}y_{\tq}}\right)^{N_f(-\mn_{\tq}+1)},\label{ADHM:TTI:2}
	\end{split}
\end{equation}
in terms of a large integer cut-off $M$ $(\mm_i\leq M-1)$ and using the BA operators
\begin{equation}
	e^{\mathrm{i}B_i}=\xi\left(\fft{\sqrt{x_iy_q}}{1-x_iy_q}\right)^{N_f}\left(\fft{\sqrt{\fft{1}{x_i}y_{\tq}}}{1-\fft{1}{x_i}y_{\tq}}\right)^{-N_f}\prod_{I=1}^3\prod_{j=1}^N\left(\fft{x_i-x_jy_I}{x_j-x_iy_I}\right).\label{ADHM:B}
\end{equation}
Using Cauchy's theorem to evaluate the contour integrals in (\ref{ADHM:TTI:2}) and using the constraints (\ref{ADHM:constraints}), we obtain the following BA formulation of the $S^1\times S^2$ ADHM TTI:
\begin{align}
\label{ADHM:TTI:BA}
		Z^\text{ADHM}_{S^1\times S^2}(N,N_f,\Delta,\mn)&=y_q^{\fft{NN_f(1-\mn_q)}{2}}y_{\tq}^{\fft{NN_f(1-\mn_{\tq})}{2}}\prod_{I=1}^3y_I^{-\fft{N^2}{2}\mn_I} \nonumber \\
		&\quad\times\sum_{\{x_i\}\in\text{BAE}}\fft{1}{\det\mathbb B}\fft{\prod_{i=1}^Nx_i^{N+\mathfrak t}\prod_{i\neq j}(1-x_i/x_j)}{\prod_{I=1}^3\prod_{i,j=1}^N(x_j-x_iy_I)^{-\fft12\mn_I+\fft12}(x_i-x_jy_I)^{-\fft12\mn_I+\fft12}} \nonumber \\
		&\kern6em\times\prod_{i=1}^N\fft{x_i^{N_f(1-\fft{\mn_q+\mn_{\tq}}{2})}}{(1-x_iy_q)^{N_f(1-\mn_q)}(x_i-y_{\tq})^{N_f(1-\mn_{\tq})}} \, ,
\end{align}
where the sum is taken over solutions to the Bethe Ansatz Equations (BAE) $e^{\mathrm{i}B_i}=1$. Introducing $x_i=e^{\mathrm{i}u_i}$, the latter can be written as
\begin{align}
\label{ADHM:BAE}
	2\pi n_i&=\pi\left(\Delta_m-N-1+2\left\lfloor\fft{N+1}{2}\right\rfloor\right)+\mathrm{i}N_f\left(\text{Li}_1(e^{\mathrm{i}(-u_i+\pi\Delta_{\tq})})-\text{Li}_1(e^{\mathrm{i}(-u_i-\pi\Delta_q)})\right) \nonumber \\
	&\quad+\fft{N_f\pi}{2}(2-\Delta_q-\Delta_{\tq})+\mathrm{i}\sum_{I=1}^3\sum_{j=1}^N\left(\text{Li}_1(e^{\mathrm{i}(u_j-u_i+\pi\Delta_I)})-\text{Li}_1(e^{\mathrm{i}(u_j-u_i-\pi\Delta_I)})\right) \nonumber \\
	&\quad+N\pi\qquad\text{with}\quad n_i\in\mathbb Z \, .
\end{align}
One can check that the BAE (\ref{ADHM:BAE}) are obtained by differentiating the following Bethe potential with respect to $u_i$,
\begin{equation}
	\begin{split}
		\mathcal V&=\sum_{i=1}^N\left(2n_i+N+1-2\left\lfloor\fft{N+1}{2}\right\rfloor-\Delta_m\right)\pi u_i\\
		&\quad+\fft12\sum_{I=1}^3\sum_{i,j=1}^N\bigg[\text{Li}_2(e^{\mathrm{i}(u_j-u_i+\pi\Delta_I)})-\text{Li}_2(e^{\mathrm{i}(u_j-u_i-\pi\Delta_I)})\bigg]\\
		&\quad+N_f\sum_{i=1}^N\bigg[\text{Li}_2(e^{\mathrm{i}(-u_i+\Delta_{\tq})})-\text{Li}_2(e^{\mathrm{i}(-u_i-\Delta_q)})\bigg]-\fft{N_f\pi}{2}\sum_{i=1}^N(2-\Delta_q-\Delta_{\tq})u_i \, .
	\end{split}\label{ADHM:Beth}
\end{equation}
To do so, one uses the inversion formula of the polylogarithm
\begin{equation}
	\text{Li}_n(e^{2\pi \mathrm{i}x})+(-1)^n\text{Li}_n(e^{-2\pi \mathrm{i}x})=-\fft{(2\pi \mathrm{i})^n}{n!}B_n(x)\quad\text{where}~\begin{cases}
		0\leq\Re[x]<1 & \Im[x]\geq0\\
		0<\Re[x]\leq1 & \Im[x]<0
	\end{cases} \, , \label{polylog:inverse}
\end{equation}
given in terms of the Bernoulli polynomials $B_n(x)$ and under the assumptions
\begin{equation}
	0<\Re[u_j-u_i+\pi\Delta_I]<2\pi\,,\qquad -2\pi<\Re[u_j-u_i-\pi\Delta_I]<0\,.
\end{equation}
In the BA formula (\ref{ADHM:TTI:BA}) we have also introduced the Jacobian matrix $\mathbb B$ as 
\begin{equation}
	\begin{split}
		\mathbb B=\fft{\partial(e^{\mathrm{i}B_1},\cdots,e^{\mathrm{i}B_N})}{\partial(\log x_1,\cdots,\log x_N)}\, ,\qquad \mathbb B\big|_\text{BAE}=\fft{\partial(B_1,\cdots,B_N)}{\partial(u_1,\cdots, u_N)}\, .
	\end{split}\label{ADHM:Jacobian}
\end{equation}
The components of this matrix are given explicitly as
\begin{align}
\label{eq:BB}
		\mathbb B_{i,j}\big|_\text{BAE}&=\delta_{ij}\left[x_i\sum_{I=1}^3\sum_{k=1}^N\left(\fft{1}{x_i-x_ky_I}+\fft{y_I}{x_k-x_iy_I}\right)+N_fx_i\left(\fft{1}{x_i-y_{\tq}}-\fft{1}{x_i-y_q^{-1}}\right)\right] \nonumber \\
		&\quad+x_j\sum_{I=1}^3\left(\fft{-y_I}{x_i-x_jy_I}-\fft{1}{x_j-x_iy_I}\right) \, .
\end{align}

\medskip

To evaluate the ADHM TTI using the BA formula, one must first find all solutions to the BAE (\ref{ADHM:BAE}) and then substitute the BAE solution back into (\ref{ADHM:TTI:BA}). Before evaluating the ADHM TTI in this way, we summarize some of the key properties of the BA formulation:
\begin{itemize}
	\item We focus on a particular solution, $\mathcal{B}_1$, to the BAE (\ref{ADHM:BAE}), which will be reviewed in Section~\ref{sec:TTI:ADHM:analytic}, and on its contribution to the ADHM TTI through the BA formula (\ref{ADHM:TTI:BA}). The solution $\mathcal{B}_1$ provides the dominant $N^{3/2}$ contribution to $\log Z_{S^1\times S^2}$ in the large $N$ limit and our results below should be viewed as extending this solution beyond the leading $N^{3/2}$ order. The full TTI is in general a sum over all BAE solutions, $\mathcal{B}_i$, i.e. $Z_{S^1\times S^2} = e^{\mathcal{B}_1}+e^{\mathcal{B}_2}+\ldots$. Since $\mathcal{B}_1$ is the dominant solution to leading order at large $N$ we can write 
\begin{equation}\label{eq:calB1calB2}
\log Z_{S^1\times S^2} = \mathcal{B}_1 + \log[1+e^{\mathcal{B}_2-\mathcal{B}_1}+\ldots]\,,
\end{equation}
where we have $\mathcal{B}_2-\mathcal{B}_1 = \alpha_1 N^{3/2}+\alpha_2 N^{1/2} +\alpha_3 \log[N]+\alpha_4+\ldots$ with $\alpha_i <0$. While we do not have rigorous mathematical proof that this is the large $N$ structure of the TTI, the known holographic results for the $N^{3/2}$ and $N^{1/2}$ leading and subleading contributions to $\log Z_{S^1\times S^2}$ (summarized in Section~\ref{sec:holo:SUGRAtoSCFT} below) strongly suggest that this structure is correct. We therefore conclude that the contribution to $\log Z_{S^1\times S^2}$ coming from any other BAE solution apart from $\mathcal{B}_1$ is exponentially suppressed in the large $N$ limit. From now on we proceed our analysis with this conclusion in mind and study the all order result for the large $N$ limit of $\mathcal{B}_1$. We will employ the same logic for all other TTI calculations discussed in this paper.

	\item The general $S^1\times\Sigma_{\mg}$ TTI can be obtained from the $S^1\times S^2$ TTI as~\cite{Benini:2016hjo}
	\begin{equation}
		\log Z^\text{ADHM}_{S^1\times\Sigma_{\mg}}(N,N_f,\Delta,\mn)=(1-\mg)\log Z^\text{ADHM}_{S^1\times S^2}\Bigl(N,N_f,\Delta,\fft{\mn}{1-\mg}\Bigr) \, ,\label{ADHM:S2:Riemann}
	\end{equation}
	provided we focus on the contribution from a particular BAE solution to the ADHM TTI through (\ref{ADHM:TTI:BA}). We also note that the $\mg = 1$ case must be treated separately due to the presence of additional fermionic zero-modes on the torus~\cite{Benini:2016hjo}. For this reason, we focus on $\mg \neq 1$ in this paper. 
	
	\item From the structure of the BA formula (\ref{ADHM:TTI:BA}), one can conclude that the logarithm of the ADHM TTI is linear in the flavor magnetic fluxes $\mn$ provided we focus on the contribution from a particular BAE solution. 
	
	\item Let us say $\{x_i^\star\}$ is a solution to the BAE (\ref{ADHM:BAE}) for a given $\Delta=(\Delta_I,\Delta_q,\Delta_{\tq},\Delta_m)$ configuration. Then it is straightforward to show that $\{x_i^\star e^{i\pi\delta}\}$ is a solution to the BAE for a deformed $\Delta=(\Delta_I,\Delta_q-\delta,\Delta_{\tq}+\delta,\Delta_m)$ configuration. Substituting this new solution $\{x_i^\star e^{i\pi\delta}\}$ into the BA formula (\ref{ADHM:TTI:BA}), one can show that the ADHM TTI is independent of $\delta$ up to the overall phase factor $e^{i N\mkt\pi\delta}$. This implies that the modulus of the ADHM TTI is independent of the chemical potentials $(\Delta_q,\Delta_{\tq})$ provided they satisfy the constraint (\ref{ADHM:constraints}) for a given $\Delta_3$; hence we can simply set $\Delta_q=\Delta_{\tq}=1-\Delta_3/2$ without affecting the modulus of the ADHM TTI. Then it also becomes clear from (\ref{ADHM:TTI:BA}) that the magnetic fluxes $(\mn_q,\mn_{\tq})$ can be chosen as $\mn_q=\mn_{\tq}=1-\mn_3/2$ to satisfy the constraint (\ref{ADHM:constraints}) for a given $\mn_3$ without affecting the modulus of the ADHM TTI. 
\end{itemize}
%

%%%%%
\subsubsection{Analytic approach for the large $N$ limit}\label{sec:TTI:ADHM:analytic}
%%%%%

In this subsection we briefly review the analytic evaluation of the ADHM TTI in the large $N$ limit using the BA formulation (\ref{ADHM:TTI:BA}), following~\cite{Hosseini:2016tor,Hosseini:2016ume}.

\medskip

The first step is to find the solution of the BAE (\ref{ADHM:BAE}) in the large $N$ limit. For that purpose, we introduce the large $N$ ansatz
\begin{equation}
	u_i=\mathrm{i}N^\fft12t_i+v_i \, .\label{ADHM:ansatz}
\end{equation}
This discrete ansatz is then made continuous by introducing an eigenvalue distribution $u:[t_{\ll},t_{\gg}]\subseteq\mathbb R\to\mathbb C$ and its real part $v:[t_{\ll},t_{\gg}]\to\mathbb R$ satisfying
\begin{equation}
	u(t(i))=\mathrm{i} N^\fft12t(i)+v(t(i))=u_i\,,\label{ADHM:ansatz:conti}
\end{equation}
where we have also introduced a continuous function $t:[1,N]\to[t_{\ll},t_{\gg}]$ that maps discrete indices $1,2,\cdots,N$ to the real interval $[t_{\ll},t_{\gg}]$. We then introduce the eigenvalue density $\rho:[t_{\ll},t_{\gg}]\to\mathbb R$ as
\begin{equation}
	\rho(t)=\fft{1}{N-1}\sum_{i=1}^N\delta(t-t_i)\quad\Leftrightarrow\quad di=(N-1)\rho(t)dt\,,\label{ADHM:rho}
\end{equation}
which satisfies the following normalization condition by construction:
\begin{equation}
	\int_{t_\ll}^{t_\gg}dt\,\rho(t)=1\,.\label{ADHM:rho:normal}
\end{equation}

In the continuum limit, the large $N$ BAE solution can be obtained by determining the eigenvalue distribution $u(t)$ and the eigenvalue density $\rho(t)$ along the compact support $[t_\ll,t_\gg]$. These functions can be determined by extremizing the large $N$ limit of the Bethe potential (\ref{ADHM:Beth}) written in terms of the continuous variables, namely\footnote{This is slightly different from equation (3.4) of~\cite{Hosseini:2016ume}, which has the opposite sign for $\Delta_m$. In view of equation (2.8) in that paper, we have fixed this typo in \eqref{ADHM:Beth:largeN}.}
\begin{equation}
\begin{split}
	\mathcal V&=-\mathrm{i}N^\fft32\pi\Delta_m\int_{t_\ll}^{t_\gg} dt\,t\rho(t)+\mathrm{i}N^\fft32\fft{4\pi^3}{3}\sum_{I=1}^3B_3\Bigl(\fft{\Delta_I}{2}\Bigr)\int_{t_\ll}^{t_\gg}dt\,\rho(t)^2\\
	&\quad+\mathrm{i}N^\fft32\fft{N_f\pi}{2}(2-\Delta_{q}-\Delta_{\tq})\int_{t_\ll}^{t_\gg} dt\,\rho(t)|t|\,.
\end{split}\label{ADHM:Beth:largeN}
\end{equation}
We refer the reader to~\cite{Hosseini:2016tor,Hosseini:2016ume} for the derivation of (\ref{ADHM:Beth:largeN}). The resulting large $N$ continuous BAE solution reads 
\begin{equation}
\begin{split}
	\rho(t)&=\fft{2\pi\mu-N_f(2-\Delta_q-\Delta_{\tq})|t|+2\Delta_mt}{2\pi^2\Delta_1\Delta_2\Delta_3}\,,\\
	t_\gg&=\fft{2\pi\mu}{N_f(2-\Delta_q-\Delta_{\tq})-2\Delta_m}\,,\\ t_\ll&=-\fft{2\pi\mu}{N_f(2-\Delta_q-\Delta_{\tq})+2\Delta_m}\,,\label{ADHM:BAE:sol:largeN}
\end{split}
\end{equation}
where $\mu$ is determined by the normalization (\ref{ADHM:rho:normal}) to be
\begin{equation}
	\mu=\sqrt{2N_f\Delta_1\Delta_2\fft{\Delta_3}{2-\Delta_q-\Delta_{\tq}}\bigg(\fft{2-\Delta_q-\Delta_{\tq}}{2}-\fft{\Delta_m}{N_f}\bigg)\bigg(\fft{2-\Delta_q-\Delta_{\tq}}{2}+\fft{\Delta_m}{N_f}\bigg)}\,.\label{ADHM:mu}
\end{equation}
To obtain the solution (\ref{ADHM:BAE:sol:largeN}), the integer $n_i$ in the BAE (\ref{ADHM:BAE}) has been chosen as
\begin{equation}
	n_i=\left\lfloor\fft{N+1}{2}\right\rfloor-i\,,\label{ADHM:integer}
\end{equation}
and we have also assumed
\begin{equation}
	\begin{alignedat}{2}
		0&<\Re[u_j-u_i+\pi\Delta_I]<2\pi\,,&\qquad -2\pi&<\Re[u_j-u_i-\pi\Delta_I]<0\,,\\
		0&<\Re[u_i+\pi\Delta_{q}]<2\pi\,,&\qquad 0&<\Re[-u_i+\pi\Delta_{\tq}]<2\pi\,.
	\end{alignedat}\label{ADHM:range}
\end{equation}
Note that the large $N$ BAE solution (\ref{ADHM:BAE:sol:largeN}) does not determine $v(t)$. 

Substituting the large $N$ solution (\ref{ADHM:BAE:sol:largeN}) back into the Bethe potential (\ref{ADHM:Beth:largeN}) yields
\begin{equation}
	\mathcal V=\mathrm{i}N^\fft32\fft{2\pi^2\mu}{3},\label{ADHM:Beth:largeN:onshell}
\end{equation}
which is indeed extremized at the superconformal $\Delta$-configuration (\ref{ADHM:constraints:sc}) under the constraint (\ref{ADHM:constraints}). Note that the large $N$ Bethe potential (\ref{ADHM:Beth:largeN:onshell}) is independent of $(\Delta_q,\Delta_{\tq})$ due to \eqref{ADHM:constraints}, and therefore their superconformal values are not determined by extremization. This was to be expected from the comment regarding the independence of the (modulus of the) TTI on $(\Delta_q,\Delta_{\tq})$ at the end of the previous subsection.

\medskip

One can now obtain the large $N$ limit of the ADHM TTI by substituting the BAE solution (\ref{ADHM:BAE:sol:largeN}) into the BA formula (\ref{ADHM:TTI:BA}), see \cite{Hosseini:2016ume} for details. The result reads
\begin{equation}
\begin{split}
	\log Z^\text{ADHM}_{S^1\times S^2}(N,N_f,\Delta,\mn)&=-\fft{\pi\mu}{3}N^\fft32\sum_{a=1}^4\fft{\tmn_a}{\tDelta_a}+o(N^\fft32)\\
	&=-\fft{\pi\sqrt{2N_f\tilde\Delta_1\tilde\Delta_2\tilde\Delta_3\tilde\Delta_4}}{3}N^\fft32\sum_{a=1}^4\fft{\tmn_a}{\tDelta_a}+o(N^\fft32)\,.\label{ADHM:TTI:N32}
\end{split}
\end{equation}
where $\mu$ is given in (\ref{ADHM:mu}) and we have defined $\tDelta_a$ and $\tmn_a$ as
\begin{equation}
\begin{split}
	\tilde\Delta_a&=\bigg(\Delta_1,\Delta_2,\fft{2-\Delta_q-\Delta_{\tq}}{2}-\fft{\Delta_m}{N_f},\fft{2-\Delta_q-\Delta_{\tq}}{2}+\fft{\Delta_m}{N_f}\bigg)\,,\\
	\tilde\mn_a&=\bigg(\mn_1,\mn_2,\fft{2-\mn_q-\mn_{\tq}}{2}+\fft{\mathfrak t}{N_f},\fft{2-\mn_q-\mn_{\tq}}{2}-\fft{\mathfrak t}{N_f}\bigg)\,. \label{ADHM:tilde}
\end{split}
\end{equation}
%
%%%%%
\subsubsection{Numerical approach for the all-order $1/N$ expansion: results}\label{sec:TTI:ADHM:num-result}
%%%%%
In this subsection we improve on the known analytic result for the large $N$ ADHM TTI (\ref{ADHM:TTI:N32}) by providing an analytic expression for its all-order perturbative $1/N$ expansion that we infer from a detailed numerical study. Our final result reads
\begin{equation}
\begin{split}
	&\log Z^\text{ADHM}_{S^1\times S^2}(N,N_f,\Delta,\mn)\\
	&=-\fft{\pi\sqrt{2N_f\tilde\Delta_1\tilde\Delta_2\tilde\Delta_3\tilde\Delta_4}}{3}\sum_{a=1}^4\tilde\mn_a\left[\fft{1}{\tilde\Delta_a}(\hat N_{N_f,\tilde\Delta})^\fft32+\bigg(\mathfrak c_a(\tilde\Delta)N_f+\fft{\mathfrak d_a(\tilde\Delta)}{N_f}\bigg)(\hat N_{N_f,\tilde\Delta})^\fft12\right]\\
	&\quad-\fft12\log\hat N_{N_f,\tilde\Delta}+\hat f_0(N_f,\tilde\Delta,\tilde\mn)+\hat f_\text{np}(N,N_f,\tilde\Delta,\tilde\mn)\\
	&\quad+\fft{\mathrm{i}N\pi\left(N_f(2-\mn_q-\mn_{\tq}-(\mn_{\tq}-\mn_q)\Delta_m)+\mkt(\Delta_{\tq}-\Delta_q)-2\right)}{2}\,,
\end{split}\label{ADHM:TTI:all}
\end{equation}
where we have defined
\begin{equation}
	\hat N_{N_f,\tilde\Delta}=N-\fft{2-\Delta_q-\Delta_{\tq}}{\Delta_3}\fft{N_f}{24}+\fft{N_f}{12}\bigg(\fft{1}{\tilde\Delta_1}+\fft{1}{\tilde\Delta_2}\bigg)+\fft{1}{12N_f}\bigg(\fft{1}{\tilde\Delta_3}+\fft{1}{\tilde\Delta_4}\bigg)\,,\label{ADHM:N:shifted}
\end{equation}
and also introduced the rational functions $\mathfrak c_a(\tilde\Delta)$ and $\mathfrak d_a(\tilde\Delta)$ as
\begin{subequations}
	\begin{align}
		\mathfrak c_a(\tilde\Delta)&=\bigg(\!\!-\fft{1}{\tilde\Delta_1}\fft{(\tilde\Delta_2+\tilde\Delta_3+\tilde\Delta_4)(\tilde\Delta_1+\tilde\Delta_2)}{8\tilde\Delta_1\tilde\Delta_2},-\fft{1}{\tilde\Delta_2}\fft{(\tilde\Delta_1+\tilde\Delta_3+\tilde\Delta_4)(\tilde\Delta_1+\tilde\Delta_2)}{8\tilde\Delta_1\tilde\Delta_2},\nn\\
		&\quad~-\fft{\tilde\Delta_3+\tilde\Delta_4}{8\tilde\Delta_1\tilde\Delta_2},-\fft{\tilde\Delta_3+\tilde\Delta_4}{8\tilde\Delta_1\tilde\Delta_2}\bigg)\,,\\
		\mathfrak d_a(\tilde\Delta)&=\bigg(\!\!-\fft{(\tilde\Delta_1+\tilde\Delta_2)(\tilde\Delta_2+\tilde\Delta_3+\tilde\Delta_4)(\tilde\Delta_1+\tilde\Delta_3+\tilde\Delta_4)}{8\tilde\Delta_1\tilde\Delta_2\tilde\Delta_3\tilde\Delta_4},\nn\\
		&\quad~-\fft{(\tilde\Delta_1+\tilde\Delta_2)(\tilde\Delta_2+\tilde\Delta_3+\tilde\Delta_4)(\tilde\Delta_1+\tilde\Delta_3+\tilde\Delta_4)}{8\tilde\Delta_1\tilde\Delta_2\tilde\Delta_3\tilde\Delta_4},\nn\\
		&\quad~-\fft{1}{\tilde\Delta_3}\fft{(\tilde\Delta_3+\tilde\Delta_4)((\tDelta_1+\tDelta_2)(\tDelta_2+\tDelta_3)(\tDelta_3+\tDelta_1)+(\tDelta_1\tDelta_2+\tDelta_2\tDelta_3+\tDelta_3\tDelta_1)\tDelta_4)}{8\tilde\Delta_1\tilde\Delta_2\tilde\Delta_3\tilde\Delta_4},\nn\\
		&\quad~-\fft{1}{\tilde\Delta_4}\fft{(\tilde\Delta_3+\tilde\Delta_4)((\tDelta_1+\tDelta_2)(\tDelta_2+\tDelta_4)(\tDelta_4+\tDelta_1)+(\tDelta_1\tDelta_2+\tDelta_2\tDelta_4+\tDelta_4\tDelta_1)\tDelta_3)}{8\tilde\Delta_1\tilde\Delta_2\tilde\Delta_3\tilde\Delta_4}\bigg)\,.
	\end{align}\label{ADHM:TTI:all:coeffi}%
\end{subequations}
We will present numerical evidence for the all-order $1/N$ expansion (\ref{ADHM:TTI:all}) in the following subsection. Before doing so, we collect some general remarks on the result:
\begin{itemize}
	\item As mentioned in the beginning of Section~\ref{sec:TTI:ADHM} the ADHM TTI with $N_f=1$ matches the ABJM TTI with $k=1$ obtained in~\cite{Bobev:2022jte,Bobev:2022eus}, provided we map $(\tDelta_a,\tmn_a)$ defined in (\ref{ADHM:tilde}) to the flavor chemical potentials and flavor magnetic fluxes of the ABJM theory $(\Delta_a,\mn_a)$. For example, it is straightforward to check that
	\begin{equation}
	\label{eq:ADHM-ABJM}
	\begin{split}
		\tDelta_a\big(\mathfrak c_a(\tilde\Delta)+\mathfrak d_a(\tilde\Delta)\big)=-\fft{\prod_{b\neq a}(\tDelta_a+\tDelta_b)}{8\tDelta_1\tDelta_2\tDelta_3\tDelta_4}\sum_{b\neq a}\tDelta_b \, ,
	\end{split}
	\end{equation}
	under the constraint (\ref{ADHM:constraints}). Here the RHS corresponds to the $\hat N_\Delta^\fft12$ coefficient of the ABJM TTI up to an overall prefactor, see~\cite{Bobev:2022jte,Bobev:2022eus}.
	
	\item We were unable to obtain the closed-form expression for the $N$-independent constant $\hat f_0(N_f,\Delta,\mn)$, although its numerical value can be obtained with great precision for various $N_f$ and $(\Delta,\mn)$-configurations, see Appendix~\ref{App:ADHM}. We leave a complete analytic determination of $\hat f_0(N_f,\Delta,\mn)$ for future research.
	
	\item We numerically confirmed that the non-perturbative correction $\hat f_\text{np}(N,N_f,\Delta,\mn)$ is indeed exponentially suppressed when focusing on the superconformal configuration (\ref{ADHM:constraints:sc}), see the following subsection for details.
	
	\item The real part of the all-order $1/N$ expansion (\ref{ADHM:TTI:all}) is independent of $(\Delta_q,\Delta_{\tq},\mn_q,\mn_{\tq})$ provided they satisfy the constraints (\ref{ADHM:constraints}) for a given $\Delta_3$ and $\mn_3$. This is consistent with our remarks at the end of subsection \ref{sec:TTI:ADHM:BA}. The imaginary part of (\ref{ADHM:TTI:all}) has been confirmed exactly for various configurations of flavor chemical potentials and magnetic fluxes listed in Appendix \ref{App:ADHM}.
\end{itemize}
%

%%%%%
\subsubsection{Numerical approach for the all-order $1/N$ expansion: derivation}\label{sec:TTI:ADHM:num-derive}
%%%%%
In this subsection we explain how we obtained the all-order $1/N$ expansion of the ADHM TTI (\ref{ADHM:TTI:all}) from a numerical analysis. The method parallels the analysis for the ABJM TTI studied in \cite{Bobev:2022eus}, but for completeness we repeat the procedure here. We collect the relevant numerical data in Appendix \ref{App:ADHM}.

\medskip

To begin with, we construct a numerical solution of the BAE (\ref{ADHM:BAE}) with the left-hand side given by (\ref{ADHM:integer}) for a given $N$, $N_f$ and $\Delta$-configuration. We do so using \texttt{FindRoot} in \textit{Mathematica} at \texttt{WorkingPrecision} $=200$ and using the leading order solution (\ref{ADHM:BAE:sol:largeN}) as initial conditions\footnote{Since $v(t)$ is not determined in the large $N$ BAE solution (\ref{ADHM:BAE:sol:largeN}), we set the initial condition for $v_i$ based on the ranges (\ref{ADHM:range}) as~\cite{PandoZayas:2020iqr}
	\begin{equation}
		v_i\big|_\text{initial condition}=\fft{\pi(\Delta_{\tq}-\Delta_q)}{2} \, . \nonumber
	\end{equation}
}. Substituting the numerical solution into the BA formula (\ref{ADHM:TTI:BA}) with a given $\mn$-configuration, we obtain the numerical value of the ADHM TTI for the chosen set of $N$, $N_f$, and $(\Delta,\mn)$. After repeating this process for $N=101\sim301~(\text{step}=10)$, we fit the resulting data with respect to $N$ using \texttt{LinearModelFit} in \textit{Mathematica}. As a result, we obtain the following numerical $1/N$ expansion of the ADHM TTI,
\begin{equation}
	\begin{split}
		&\Re\log Z^\text{ADHM}_{S^1\times S^2}(N,N_f,\Delta,\mn) \\
		&=f^\text{(lmf)}_{3/2}(N_f,\Delta,\mn)N^\fft32+f^\text{(lmf)}_{1/2}(N_f,\Delta,\mn)N^\fft12+f^\text{(lmf)}_\text{log}(N_f,\Delta,\mn)\log N \\
		&\quad+f^\text{(lmf)}_0(N_f,\Delta,\mn)+\sum_{s=1}^L f^\text{(lmf)}_{-s/2}(N_f,\Delta,\mn)N^{-\fft{s}{2}}\,,
	\end{split}
\label{ADHM:TTI:expansion}
\end{equation}
with numerical coefficients $f^\text{(lmf)}_X(N_f,\Delta,\mn)$ for a given $N_f$ and $(\Delta,\mn)$-configuration. The superscript ``(lmf)'' in the expansion coefficients emphasizes that they are numerical coefficients obtained by \texttt{LinearModelFit}. The upperbound $L$ for fitting is chosen as $L=16$ to minimize standard errors in estimating the coefficients $f^\text{(lmf)}_X(N_f,\Delta,\mn)$. 

Next, repeating the \texttt{LinearModelFit} (\ref{ADHM:TTI:expansion}) with five different $\mn$-configurations satisfying the constraint (\ref{ADHM:constraints}), namely
\begin{equation}
	\begin{split}
		\mathfrak n&=(\mn_1,\mn_2,\mn_3,\mn_q,\mn_{\tq},\mkt)\\
		&\in\bigg\{(\fft12,\fft12,1,\fft12,\fft12,0),(\fft14,\fft12,\fft54,\fft14,\fft12,\fft12),(\fft12,\fft34,\fft34,\fft14,1,\fft12),\\
		&\qquad(\fft14,\fft58,\fft98,\fft38,\fft12,\fft14),(\fft12,\fft58,\fft78,\fft14,\fft78,\fft14)\bigg\}\,,
	\end{split}\label{ADHM:nas}
\end{equation}
we obtain the numerical $1/N$ expansion of the ADHM TTI that is linear in the flavor magnetic fluxes $\mn$. The result reads
\begin{equation}
	\begin{split}
		&\Re\log Z^\text{ADHM}_{S^1\times S^2}(N,N_f,\Delta,\mn) \\
		&=\sum_{a=1}^4\left(f^\text{(lmf)}_{3/2,a}(N_f,\Delta)N^\fft32+f^\text{(lmf)}_{1/2,a}(N_f,\Delta)N^\fft12\right)\tmn_a+f^\text{(lmf)}_{\log}(N_f,\Delta,\mn)\log N\\
		&\quad+f^\text{(lmf)}_0(N_f,\Delta,\mn)+\sum_{s=1}^{L}f^\text{(lmf)}_{-s/2}(N_f,\Delta,\mn)N^{-\fft{s}{2}}\,.
	\end{split}\label{ADHM:TTI:expansion:linear}
\end{equation}
Note that the five $\mn$-configurations in (\ref{ADHM:nas}) are enough to determine the numerical coefficients in (\ref{ADHM:TTI:expansion:linear}) since the logarithm of the ADHM TTI is expected to be linear in the four magnetic fluxes $(\mn_I,\mkt)$; recall that the ADHM TTI is independent of the other two magnetic fluxes $(\mn_q,\mn_{\tq})$ as we discussed in Section~\ref{sec:TTI:ADHM:BA}. In (\ref{ADHM:TTI:expansion:linear}), we did not expand the numerical coefficients of the part linear in the magnetic fluxes beyond $N^\fft12$ as it will not be needed in what follows.

The $N^\fft32$ leading order coefficient derived in~\cite{Hosseini:2016ume} and summarized in (\ref{ADHM:TTI:N32}) is reproduced accurately by our numerical analysis where we find
\begin{equation}
	f^\text{(lmf)}_{3/2,a}(N_f,\Delta)\simeq-\fft{\pi\sqrt{2N_f\tDelta_1\tDelta_2\tDelta_3\tDelta_4}}{3}\fft{1}{\tDelta_a}\, ,\label{ADHM:f:known}
\end{equation}
for various $N_f$ and $(\Delta,\mn)$-configurations. Going to subleading orders, we found that the $\log N$ term has a universal coefficient
\begin{equation}
	f^\text{(lmf)}_\text{log}(N_f,\Delta,\mn)\simeq-\fft12\,,\label{ADHM:f:log}
\end{equation}
which is independent of $N_f$ and $(\Delta,\mn)$, just as in the ABJM TTI \cite{Liu:2017vll}. Then, by inspection of the negative integer powers of $N$ in the numerical $1/N$ expansion (\ref{ADHM:TTI:expansion:linear}) for various $N_f$ and $(\Delta,\mn)$-configurations, we observe that they can be resummed together with the universal $\log N$ contribution as
\begin{equation}
	\begin{split}
		&f^\text{(lmf)}_\text{log}(N_f,\Delta,\mn)\log N+\sum_{s=1}^\infty f^\text{(lmf)}_{-s}(N_f,\Delta,\mn)N^{-s}\\
		&\simeq-\fft12\log\left(N-\fft{2-\Delta_q-\Delta_{\tq}}{\Delta_3}\fft{N_f}{24}+\fft{N_f}{12}\bigg(\fft{1}{\tilde\Delta_1}+\fft{1}{\tilde\Delta_2}\bigg)+\fft{1}{12N_f}\bigg(\fft{1}{\tilde\Delta_3}+\fft{1}{\tilde\Delta_4}\bigg)\right)\\
		&=-\fft12\log\hat N_{N_f,\tDelta} \, ,\label{ADHM:resum}
	\end{split}
\end{equation}
producing a simple function of the shifted $N$ parameter introduced in (\ref{ADHM:N:shifted}). The resummation (\ref{ADHM:resum}) strongly motivates implementing the \texttt{LinearModelFit} step of the analysis with respect to $\hat N_{N_f,\tDelta}$ rather than $N$. Quite remarkably, we found that such a fit then terminates at order $\mathcal O(1)$. In other words, we found that the following \texttt{LinearModelFit},
\begin{equation}
	\begin{split}
		&\Re\log Z^\text{ADHM}_{S^1\times S^2}(N,N_f,\Delta,\mn)+\fft12\log\hat N_{N_f,\tDelta} \\
		&=\sum_{a=1}^4\left(\hat f^\text{(lmf)}_{3/2,a}(N_f,\Delta)(\hat N_{N_f,\tDelta})^\fft32+\hat f^\text{(lmf)}_{1/2,a}(N_f,\Delta)(\hat N_{N_f,\tDelta})^\fft12\right)\tmn_a+\hat f^\text{(lmf)}_0(N_f,\Delta,\mn) \, ,
	\end{split}\label{ADHM:TTI:expansion:shifted}
\end{equation}
with only three fitting functions -- $(\hat N_{N_f,\tDelta})^\fft32$, $(\hat N_{N_f,\tDelta})^\fft12$, and a constant -- yields numerical coefficients with much lower standard errors compared to the one in (\ref{ADHM:TTI:expansion:linear}). Note that in (\ref{ADHM:TTI:expansion:shifted}) we pulled out the universal logarithmic contribution $-\fft12\log\hat N_{N_f,\tDelta}$ analytically to estimate the remaining numerical coefficients with greater accuracy. 

Finally, investigating the numerical coefficients of the improved $1/N$ expansion for various $N_f$ and $\Delta$-configurations, we found
\begin{equation}
	\begin{split}
		\hat f^\text{(lmf)}_{3/2,a}(N_f,\Delta)&\simeq-\fft{\pi\sqrt{2N_f\tDelta_1\tDelta_2\tDelta_3\tDelta_4}}{3}\fft{1}{\tDelta_a}\,,\\
		\hat f^\text{(lmf)}_{1/2,a}(N_f,\Delta)&\simeq-\fft{\pi\sqrt{2N_f\tilde\Delta_1\tilde\Delta_2\tilde\Delta_3\tilde\Delta_4}}{3}\bigg(\mathfrak c_a(\tilde\Delta)N_f+\fft{\mathfrak d_a(\tilde\Delta)}{N_f}\bigg)\,,
	\end{split}\label{ADHM:TTI:expansion:shifted:coeffi}
\end{equation}
with $\mathfrak c_a$ and $\mathfrak d_a$ given in (\ref{ADHM:TTI:all:coeffi}). We refer the reader to Appendix \ref{App:ADHM} for the numerical values of $\hat f^\text{(lmf)}_0(N_f,\Delta,\mn)$ for various $N_f$ and $(\Delta,\mn)$-configurations. Substituting the numerical estimates (\ref{ADHM:TTI:expansion:shifted:coeffi}) back into the expansion (\ref{ADHM:TTI:expansion:shifted}), we arrive at the all-order $1/N$ expansion given in (\ref{ADHM:TTI:all}), where we have replaced the $N$-independent numerical contribution $\hat f^\text{(lmf)}_0(N_f,\Delta,\mn)$ with the corresponding analytic symbol $\hat f_0(N_f,\Delta,\mn)$ and restored the non-perturbative corrections $\hat f_\text{np}(N,N_f,\Delta,\mn)$. The latter corrections were ignored in the above analysis since they are exponentially suppressed, as we now show.

\medskip
%%%%%
\noindent\textbf{Non-perturbative corrections to the ADHM TTI}
%%%%%
\medskip

Let us study the non-perturbative corrections in (\ref{ADHM:TTI:all}) numerically. We will focus on the superconformal and universal configuration (\ref{ADHM:constraints:sc}), in which case the quantities $\hat N_{N_f,\tDelta}$, $\mathfrak c_a$ and $\mathfrak d_a$ take a simple form. To begin with, consider the following combination of $\mathcal O(1)$ contributions and non-perturbative corrections, 
\begin{equation}
	\begin{split}
		&\Re\log Z^\text{ADHM}_{S^1\times S^2}(N,N_f)+\fft{\pi\sqrt{2N_f}}{3}\left[(\hat N_{N_f})^\fft32-\bigg(\fft{N_f}{2}+\fft{5}{2N_f}\bigg)(\hat N_{N_f})^\fft12\right]+\fft12\log\hat N_{N_f}\\
		&=\hat f_0(N_f)+\hat f_\text{np}(N,N_f)\,,
	\end{split}\label{ADHM:f0np}
\end{equation}
obtained from the all-order expression (\ref{ADHM:TTI:all}). Above we have introduced 
\begin{equation}
\hat N_{N_f}=N+\fft{7N_f}{24}+\fft{1}{3N_f} \, ,
\end{equation}
obtained from (\ref{ADHM:N:shifted}) at the superconformal point. Then, using the $N$-independence of $\hat f_0(N_f)$, one can extract $\hat f_\text{np}(N,N_f)$ from (\ref{ADHM:f0np}). To be specific, we evaluate (\ref{ADHM:f0np}) for $N=101\sim301~(\text{step}=10)$ with fixed $N_f$ and then subtract the results with adjacent $N$ values. As a result, we gather numerical data for
\begin{equation}
	\log\left|\hat f_\text{np}(N+10,N_f)-\hat f_\text{np}(N,N_f)\right|\simeq\log\left|\hat f_\text{np}(N,N_f)\right|+\mathcal O(N^0) \, ,\label{ADHM:fnp:approx}
\end{equation}
for $N=101\sim291~(\text{step}=10)$. The approximation in (\ref{ADHM:fnp:approx}) will be justified a posteriori. Then we use \texttt{LinearModelFit} to fit the resulting data with respect to $N$,
\begin{equation}
	\log\left|\hat f_\text{np}(N,N_f)\right|=\hat f^\text{(lmf)}_{\text{np},1/2}(N_f)N^\fft12+\hat f^\text{(lmf)}_{\text{np},\log}(N_f)\log N+\sum_{s=0}^L\hat f^\text{(lmf)}_{\text{np},-s}(N_f)N^{-s/2}\,,\label{ADHM:fnp:expansion}
\end{equation}
where the upperbound $L$ is chosen as $L=16$ to minimize standard errors in estimating the numerical coefficients $\hat f^\text{(lmf)}_{\text{np},X}(N_f)$. Repeating this process for $N_f\in\{1,2,3,4\}$, we found
\begin{equation}
	\hat f^\text{(lmf)}_{\text{np},1/2}(k)\simeq-2\pi\sqrt{\fft{2}{N_f}}.\label{ADHM:fnp:expansion:coeffi}
\end{equation}
Substituting the estimate (\ref{ADHM:fnp:expansion:coeffi}) back into the expansion (\ref{ADHM:fnp:expansion}), we obtain the non-perturbative behavior 
\begin{equation}
	\hat f_\text{np}(N,N_f)=e^{-2\pi\sqrt{2N/N_f}\,+\,\mathcal O(\log N)}\,.\label{ADHM:fnp}
\end{equation}
Note that the non-perturbative behavior (\ref{ADHM:fnp}) justifies the approximation in (\ref{ADHM:fnp:approx}). The following table provides error-ratios for the leading order estimate (\ref{ADHM:fnp:expansion:coeffi}), namely
\begin{equation}
	R_{\text{np},1/2}=\fft{\hat f^\text{(lmf)}_{\text{np},1/2}(N_f)-(-2\pi\sqrt{2/N_f})}{-2\pi\sqrt{2/N_f}}\,,\label{ADHM:Rdelta}
\end{equation}
for $N_f\in\{1,2,3,4\}$:
\begin{center}
	\footnotesize
	\begin{tabular}{ |c||c| } 
		\hline
		& $R_{\text{np},1/2}$  \\
		\hline\hline
		$N_f=1$ & $7.372{\times}10^{-10}$ \\
		\hline
		$N_f=2$ & $-1.059{\times}10^{-9}$ \\
		\hline
		$N_f=3$ & $-5.557{\times}10^{-11}$ \\ 
		\hline
		$N_f=4$ & $1.907{\times}10^{-11}$ \\ 
		\hline
	\end{tabular}
\end{center}

Most importantly, the exponentially suppressed non-perturbative behavior (\ref{ADHM:fnp}) confirms that the $1/N$ expansion of the ADHM TTI (\ref{ADHM:TTI:all}) indeed captures all the perturbative contributions to the TTI for the superconformal/universal configuration (\ref{ADHM:constraints:sc}). We leave the analysis of non-perturbative corrections to the ADHM TTI with generic $(\Delta,\mn)$-configurations for future research. It would be most interesting to understand the physical origin of these non-perturbative corrections as coming from appropriate instanton contributions, similar to the case of the $S^3$ partition function in the ABJM theory~\cite{Drukker:2010nc,Drukker:2011zy,Hatsuda:2012dt}.

%%%%%
\subsection{$\mathcal N=3$ SCFT dual to AdS$_4\times N^{0,1,0}/\mathbb Z_k$}
\label{sec:TTI:N010}
%%%%%

We now consider the TTI of the $\mathcal N=3$ SCFT dual to AdS$_4\times N^{0,1,0}/\mathbb Z_k$~\cite{Gaiotto:2009tk,Hohenegger:2009as,Hikida:2009tp,Cheon:2011th}. This theory can be constructed as follows: we start from the ABJM quiver which consists of the gauge group U$(N)_k\times$U$(N)_{-k}$ with the corresponding vector multiplets and 2 pairs of bi-fundamental and anti-bi-fundamental chiral multiplets $A_{1,2}$ and $B_{1,2}$ We then add $r_{1,2}$ pairs of fundamental and anti-fundamental chiral multiplets $\psi^{(s)}_{1,2}$ and $\tpsi^{(s)}_{1,2}~(s=1,\ldots,r_{1,2})$ to each gauge node with opposite CS levels $\pm k$ with $r_1+r_2=k$~\cite{Cheon:2011th}. We find it illustrative to keep $r = r_1 + r_2$ independent from $k$ in the following formulas\footnote{In the IIA string theory point of view, the independent $(k,r)$ configuration is crucial to explore the holographic dual description by $r$ D6 branes on top of the AdS$_4\times\mathbb{CP}_3$ background \cite{Gaiotto:2009tk,Hohenegger:2009as,Hikida:2009tp}.}, but the constraint $r=k$ should be kept in mind throughout. The superpotential for this model reads
\begin{equation}
	W=\Tr\Bigg[\bigg(A_1B_2-A_2B_1-\sum_{s=1}^{r_1}\tpsi_1^{(s)}\psi_1^{(s)}\bigg)^2-\bigg(A_1B_2-A_2B_1+\sum_{s=1}^{r_2}\tpsi_2^{(s)}\psi_2^{(s)}\bigg)^2\,\Bigg]\,.\label{N010:W}
\end{equation}
For $r=1$ theory has $\SU(3)$ flavor symmetry in addition to the $\SO(3)$ R-symmetry. We refer to the topologically twisted index of this theory as the $N^{0,1,0}$ TTI.

To evaluate the $N^{0,1,0}$ TTI, we first briefly introduce its BA formulation in Section~\ref{sec:TTI:N010:BA}. Then, in Section~\ref{sec:TTI:N010:analytic}, we review the known analytic calculation of the $N^{0,1,0}$ TTI in the large $N$ limit. Finally, in Section~\ref{sec:TTI:N010:num}, we provide an analytic expression for the $N^{0,1,0}$ TTI to all orders in the perturbative $1/N$ expansion.

%%%%%
\subsubsection{Bethe Ansatz formulation}\label{sec:TTI:N010:BA}
%%%%%
Following the general formula (\ref{general:TTI}), the $S^1\times S^2$ $N^{0,1,0}$ TTI is given by the matrix model
\begin{align}
\label{N010:TTI:1} 
		&Z^{N^{0,1,0}}_{S^1\times S^2}(N,k,r,\Delta,\mn) \nonumber \\
		&=\fft{1}{(N!)^2}\sum_{\mm,\tmm\in\mathbb Z^N}\int_{\mathcal C}\prod_{i=1}^N\fft{dx_i}{2\pi \mathrm{i}x_i}\fft{d\tx_i}{2\pi \mathrm{i}\tx_i}x_i^{k\mm_i}\tx_i^{-k\tmm_i}(-1)^{(N+1)(\mm_i-\tmm_i)}\prod_{i\neq j}^N\left(1-\fft{x_i}{x_j}\right)\left(1-\fft{\tx_i}{\tx_j}\right) \nonumber \\
		&\quad\times\prod_{i,j=1}^N\prod_{a=1,2}\left(\fft{\sqrt{\fft{x_i}{\tx_j}y_a}}{1-\fft{x_i}{\tx_j}y_a}\right)^{\mm_i-\tmm_j-\mn_a+1}\prod_{b=3,4}\left(\fft{\sqrt{\fft{\tx_j}{x_i}y_b}}{1-\fft{\tx_j}{x_i}y_b}\right)^{\tmm_j-\mm_i-\mn_b+1}\nonumber \\
		&\quad\times\prod_{i=1}^N\left(\fft{\sqrt{x_iy_{q_1}}}{1-x_iy_{q_1}}\right)^{r_1(\mm_i-\mn_{q_1}+1)}\left(\fft{\sqrt{\fft{1}{x_i}y_{\tq_1}}}{1-\fft{1}{x_i}y_{\tq_1}}\right)^{r_1(-\mm_i-\mn_{\tq_1}+1)}\\
		&\quad\times\prod_{i=1}^N\left(\fft{\sqrt{\tx_iy_{q_2}}}{1-\tx_iy_{q_2}}\right)^{r_2(\tmm_i-\mn_{q_2}+1)}\left(\fft{\sqrt{\fft{1}{\tx_i}y_{\tq_2}}}{1-\fft{1}{\tx_i}y_{\tq_2}}\right)^{r_2(-\tmm_i-\mn_{\tq_2}+1)}\, , \nonumber 
\end{align}
where we have assigned the same fugacities $y_{q_{1,2}}$ and $y_{\tilde q_{1,2}}$ and magnetic fluxes $\mn_{q_{1,2}}$ and $\mn_{\tilde q_{1,2}}$ for the $r_{1,2}$ pairs of fundamental and anti-fundamental multiplets for simplicity, as in the ADHM case above. Note that we have also introduced the fugacities $(-1)^{N+1}$ associated with the U(1) topological symmetries for later convenience. From the marginality of the $N^{0,1,0}$ superpotential (\ref{N010:W}) and its invariance under global symmetries, the chemical potentials and magnetic fluxes are constrained according to
\begin{equation}
\begin{alignedat}{2}
	1&=\mn_1+\mn_4=\mn_2+\mn_4\,,&\qquad 1&=\mn_{q_1}+\mn_{\tq_1}=\mn_{q_2}+\mn_{\tq_2}\,,\\
	1&=\Delta_1+\Delta_4=\Delta_2+\Delta_3\,,&\qquad1&=\Delta_{q_1}+\Delta_{\tq_1}=\Delta_{q_2}+\Delta_{\tq_2}\,,
\end{alignedat}\label{N010:constraints}
\end{equation}
where we have relabeled the fugacities in (\ref{general:TTI}) as
\begin{equation}
	y_{\Psi_{(a,b)}}\to y_a=e^{\mathrm{i}\pi\Delta_a}\,,\qquad y_{\Psi_a}\to y_{q_n}=e^{\mathrm{i}\pi\Delta_{q_n}}\,,\qquad
	y_{\tPsi_a}\to y_{\tq_n}=e^{\mathrm{i}\pi\Delta_{\tq_n}}\,.\label{N010:Delta:simple}
\end{equation}
The superconformal $\Delta$-configuration and the universal $\mn$-configuration correspond to
\begin{equation}
	\Delta_a=\Delta_{q_n}=\Delta_{\tq_n}=\fft12\,,\qquad \mn_a=\mn_{q_n}=\mn_{\tq_n}=\fft12\,,\label{N010:constraints:sc}
\end{equation}
as can be derived from matching the $S^3$ free energy with the above trial $R$-charges and the holographically dual regularized on-shell action~\cite{Santamaria:2010dm}.
%The superconformal $\Delta$-configuration in (\ref{N010:constraints:sc}) can be confirmed in principle by extremizing the large $N$ Beth potential with respect to chemical potentials as in the ADHM case but the Beth potential with generic chemical potentials is more involved for the $N^{0,1,0}$ TTI so we skip the explicit confirmation of (\ref{N010:constraints:sc}). The superconformal $\Delta$-configuration in (\ref{N010:constraints:sc}) can be supported instead by the precise match between the $S^3$ free energy with the trial $R$-charges chosen as (\ref{N010:constraints:sc}) and the holographic dual gravitational on-shell action \cite{Santamaria:2010dm}. 
For convenience, we write $\Delta$ and $\mn$ to collectively represent all the chemical potentials and the magnetic fluxes as
\begin{equation}
	\Delta=(\Delta_a,\Delta_{q_n},\Delta_{\tq_n})\,,\qquad \mn=(\mn_a,\mn_{q_n},\mn_{\tq_n})\,.\label{N010:Delta:mn}
\end{equation}

\medskip

To present the BA formulation of the $S^1\times S^2$ $N^{0,1,0}$ TTI, we first rewrite the matrix model (\ref{N010:TTI:1}) as
\begin{equation}
	\begin{split}
		Z^{N^{0,1,0}}_{S^1\times S^2}(N,k,r,\Delta,\mn)&=\fft{1}{(N!)^2}\int_{\mathcal C}\prod_{i=1}^N\fft{dx_i}{2\pi \mathrm{i}x_i}\fft{d\tx_i}{2\pi \mathrm{i}\tx_i}\prod_{i\neq j}^N\left(1-\fft{x_i}{x_j}\right)\left(1-\fft{\tx_i}{\tx_j}\right)\\
		&\quad\times\prod_{i,j=1}^N\prod_{a=1,2}\left(\fft{\sqrt{\fft{x_i}{\tx_j}y_a}}{1-\fft{x_i}{\tx_j}y_a}\right)^{1-\mn_a}\prod_{b=3,4}\left(\fft{\sqrt{\fft{\tx_j}{x_i}y_b}}{1-\fft{\tx_j}{x_i}y_b}\right)^{1-\mn_b}\\
		&\quad\times\prod_{i=1}^N\left(\fft{\sqrt{x_iy_{q_1}}}{1-x_iy_{q_1}}\right)^{r_1(1-\mn_{q_1})}\left(\fft{\sqrt{\fft{1}{x_i}y_{\tq_1}}}{1-\fft{1}{x_i}y_{\tq_1}}\right)^{r_1(1-\mn_{\tq_1})}\\
		&\quad\times\prod_{i=1}^N\left(\fft{\sqrt{\tx_iy_{q_2}}}{1-\tx_iy_{q_2}}\right)^{r_2(1-\mn_{q_2})}\left(\fft{\sqrt{\fft{1}{\tx_i}y_{\tq_2}}}{1-\fft{1}{\tx_i}y_{\tq_2}}\right)^{r_2(1-\mn_{\tq_2})}\\
		&\quad\times\prod_{i=1}^N\fft{(e^{\mathrm{i}B_i})^M}{e^{\mathrm{i}B_i}-1}\prod_{j=1}^N\fft{(e^{\mathrm{i}\tB_j})^M}{e^{\mathrm{i}\tB_j}-1}\,,\label{N010:TTI:2}
	\end{split}
\end{equation}
in terms of a large integer cut-off $M$ $(\mm_i\leq M-1,~\tmm_j\geq1-M)$ and using the BA operators
\begin{equation}
	\begin{split}
		e^{\mathrm{i}B_i}&=(-1)^{N+1}\sigma_ix_i^k\left(\fft{\sqrt{x_iy_{q_1}}}{1-x_iy_{q_1}}\right)^{r_1}\left(\fft{\sqrt{\fft{1}{x_i}y_{\tq_1}}}{1-\fft{1}{x_i}y_{\tq_1}}\right)^{-r_1}\prod_{j=1}^N\fft{(1-\fft{\tx_j}{x_i}y_3)(1-\fft{\tx_j}{x_i}y_4)}{(1-\fft{\tx_j}{x_i}y_1^{-1})(1-\fft{\tx_j}{x_i}y_2^{-1})}\,,\\
		e^{\mathrm{i}\tB_j}&=(-1)^{N+1}\tilde\sigma_j\tx_j^k\left(\fft{\sqrt{\tx_jy_{q_2}}}{1-\tx_jy_{q_2}}\right)^{-r_2}\left(\fft{\sqrt{\fft{1}{\tx_j}y_{\tq_2}}}{1-\fft{1}{\tx_j}y_{\tq_2}}\right)^{r_2}\prod_{i=1}^N\fft{(1-\fft{\tx_j}{x_i}y_3)(1-\fft{\tx_j}{x_i}y_4)}{(1-\fft{\tx_j}{x_i}y_1^{-1})(1-\fft{\tx_j}{x_i}y_2^{-1})}\,.
	\end{split}\label{N010:B}
\end{equation}
The sign ambiguities $\sigma_i,\tilde\sigma_j$ can be fixed to
\begin{equation}
\label{eq:N010-sign-fix}
\sigma_i = \tilde\sigma_j = (-1)^N \, ,
\end{equation}
as explained in Appendix \ref{App:Q111}.
%given explicitly for the particular BAE solutions discussed in subsections \ref{sec:TTI:N010:analytic} and \ref{sec:TTI:N010:num} as
%%
%\begin{equation}
%\begin{split}
%	\sigma_i\equiv\prod_{j=1}^N\fft{\sqrt{\fft{x_i}{\tx_j}y_1}}{-\fft{x_i}{\tx_j}y_1}\fft{\sqrt{\fft{x_i}{\tx_j}y_2}}{-\fft{x_i}{\tx_j}y_2}\fft{1}{\sqrt{\fft{\tx_j}{x_i}y_3}}\fft{1}{\sqrt{\fft{\tx_j}{x_i}y_4}}\to(-1)^N,\\
%	\tilde\sigma_j\equiv\prod_{i=1}^N\fft{\sqrt{\fft{x_i}{\tx_j}y_1}}{-\fft{x_i}{\tx_j}y_1}\fft{\sqrt{\fft{x_i}{\tx_j}y_2}}{-\fft{x_i}{\tx_j}y_2}\fft{1}{\sqrt{\fft{\tx_j}{x_i}y_3}}\fft{1}{\sqrt{\fft{\tx_j}{x_i}y_4}}\to(-1)^N.
%\end{split}\label{sign}
%\end{equation}
%%
%From here on we fix the sign ambiguities as (\ref{sign}) and focus on the BAE solutions of interest, see \cite{Bobev:2022eus} for related discussion. 
Evaluating the integrals in (\ref{N010:TTI:2}) by residues and using the constraints (\ref{N010:constraints}), we obtain the following BA formula for the $S^1\times S^2$ $N^{0,1,0}$ TTI,
\begin{align}
\label{N010:TTI:BA}
		&Z^{N^{0,1,0}}_{S^1\times S^2}(N,k,r,\Delta,\mn) \nonumber \\
		&=\prod_{a=1}^2y_{q_a}^{\fft12Nr_a(1-\mn_{q_a})}y_{\tq_a}^{\fft12Nr_a(1-\mn_{\tq_a})}\times\prod_{a=1}^4y_a^{-\fft{N^2}{2}\mn_a} \nonumber \\
		&\quad\times\sum_{\{x_i,\tx_j\}\in\text{BAE}}\left[\fft{1}{\det\mathbb B}\fft{\prod_{i=1}^Nx_i^N\tx_i^N\prod_{i\neq j}^N(1-\fft{x_i}{x_j})(1-\fft{\tx_i}{\tx_j})}{\prod_{i,j=1}^N\prod_{a=1,2}(\tx_j-x_iy_a)^{1-\mn_a}\prod_{3,4}(x_i-\tx_jy_a)^{1-\mn_a}}\right.\\
		&\kern4em\left.\times\prod_{i=1}^N\fft{x_i^{\fft12r_1}}{(1-x_iy_{q_1})^{r_1(1-\mn_{q_1})}(x_i-y_{\tq_1})^{r_1(1-\mn_{\tq_1})}}\fft{\tx_i^{\fft12r_2}}{(1-\tx_iy_{q_2})^{r_2(1-\mn_{q_2})}(\tx_i-y_{\tq_2})^{r_2(1-\mn_{\tq_2})}}\right] \, , \nonumber
\end{align}
where the BAE are $e^{\mathrm{i}B_i}=e^{\mathrm{i}\tB_j}=1$. Introducing $x_i=e^{\mathrm{i}u_i}$ and $\tx_j=e^{\mathrm{i}\tu_j}$, the latter can be written as
\begin{align}
\label{N010:BAE}
		2\pi n_i&=\pi+ku_i+\mathrm{i}\sum_{j=1}^N\bigg[\sum_{a=3,4}\text{Li}_1(e^{\mathrm{i}(\tu_j-u_i+\pi\Delta_a)})-\sum_{a=1,2}\text{Li}_1(e^{\mathrm{i}(\tu_j-u_i-\pi\Delta_a)})\bigg] \nonumber \\
		&\quad+\mathrm{i}r_1\bigg[\text{Li}_1(e^{\mathrm{i}(-u_i+\pi\Delta_{\tq_1})})-\text{Li}_1(e^{\mathrm{i}(-u_i-\pi\Delta_{q_1})})+\fft{\mathrm{i}\pi}{2}(\Delta_{q_1}+\Delta_{\tq_1}-2)\bigg]\quad(n_i\in\mathbb Z)\,,\nonumber\\
		2\pi\tn_j&=\pi+k\tu_j+\mathrm{i}\sum_{i=1}^N\bigg[\sum_{a=3,4}\text{Li}_1(e^{\mathrm{i}(\tu_j-u_i+\pi\Delta_a)})-\sum_{a=1,2}\text{Li}_1(e^{\mathrm{i}(\tu_j-u_i-\pi\Delta_a)})\bigg]\\
		&\quad-\mathrm{i}r_2\bigg[\text{Li}_1(e^{\mathrm{i}(-\tu_j+\pi\Delta_{\tq_2})})-\text{Li}_1(e^{\mathrm{i}(-\tu_j-\pi\Delta_{q_2})})+\fft{\mathrm{i}\pi}{2}(\Delta_{q_2}+\Delta_{\tq_2}-2)\bigg]\quad(\tn_j\in\mathbb Z)\, . \nonumber
\end{align}
% 
%One can check that the BAE (\ref{N010:BAE}) can be obtained by differentiating the following Bethe potential with respect to $u_i$ and $\tu_j$:
%%
%\begin{equation}
%	\begin{split}
%		\mathcal V&=\sum_{i=1}^N\left[\fft{k}{2}(\tu_i^2-u_i^2)-\pi\left(2\tn_i-1\right)\tu_i+\pi\left(2n_i-1\right)u_i\right]\\
%		&\quad+\sum_{i,j=1}^N\bigg[\sum_{a=3,4}\text{Li}_2(e^{i(\tu_j-u_i+\pi\Delta_a)})-\sum_{a=1,2}\text{Li}_2(e^{i(\tu_j-u_i-\pi\Delta_a)})\bigg]\\
%		&\quad+r_1\sum_{i=1}^N\bigg[\text{Li}_2(e^{i(-u_i+\pi\Delta_{\tq_1})})-\text{Li}_2(e^{i(-u_i-\pi\Delta_{q_1})})+\fft\pi2u_i(\Delta_{q_1}+\Delta_{\tq_1}-2)\bigg]\\
%		&\quad+r_2\sum_{j=1}^N\bigg[\text{Li}_2(e^{i(-\tu_j+\pi\Delta_{\tq_2})})-\text{Li}_2(e^{i(-\tu_j-\pi\Delta_{q_2})})+\fft\pi2\tu_j(\Delta_{q_2}+\Delta_{\tq_2}-2)\bigg].
%	\end{split}\label{N010:Beth}
%\end{equation}
%%
In the BA formula (\ref{N010:TTI:BA}) we have also introduced the Jacobian matrix $\mathbb B$ as
\begin{equation}
	\begin{split}
		\mathbb B&=\fft{\partial(e^{\mathrm{i}B_1},\cdots,e^{\mathrm{i}B_N},e^{\mathrm{i}\tB_1},\cdots,e^{\mathrm{i}\tB_N})}{\partial(\log x_1,\cdots,\log x_N,\log\tx_1,\cdots,\log\tx_N)}\,,\\
		\mathbb B\big|_\text{BAE}&=\fft{\partial(B_1,\cdots,B_N,\tB_1,\cdots,\tB_N)}{\partial(u_1,\cdots, u_N,\tu_1,\cdots,\tu_N)}\, ,
	\end{split}\label{N010:Jacobian}
\end{equation}
in analogy with the ADHM case.
%The components of the Jacobian matrix (\ref{N010:Jacobian}) are given explicitly as
%%
%\begin{equation}
%	\begin{split}
%		\mathbb B\big|_\text{BAE}&=\begin{pmatrix}
%			\delta_{jl}\Big(k-\sum_{m=1}^NG_{jm}+r_1x_j(\fft{1}{x_j-y_{\tq_1}}-\fft{1}{x_j-y_{\tq_1}^{-1}})\Big) & G_{jl}\\
%			-G_{lj} & \delta_{jl}\Big(k+\sum_{m=1}^NG_{mj}+r_2\tx_j(\fft{1}{\tx_j-y_{\tq_2}}-\fft{1}{\tx_j-y_{\tq_2}^{-1}})\Big)\end{pmatrix}.
%	\end{split}
%\end{equation}
%%
%in terms of the $N\times N$ square matrix $G_{ij}$ defined as
%%
%\begin{equation}
%	G_{ij}\equiv\fft{\partial\log D(z)}{\partial\log z}\bigg|_{z=\tx_j/x_i},\qquad D(z)\equiv\fft{(1-zy_3)(1-zy_4)}{(1-zy_1^{-1})(1-zy_2^{-1})}.\label{def:G}
%\end{equation}
%%
%To evaluate the $N^{0,1,0}$ TTI using the BA formula (\ref{N010:TTI:BA}), one must find solutions of the BAE (\ref{N010:BAE}) first and then substitute the BAE solution into the BA formula (\ref{N010:TTI:BA}). 
In view of the BA formula \eqref{N010:TTI:BA}, it is clear that the first three remarks we have spelled out for the ADHM TTI below \eqref{eq:BB} apply to the $N^{0,1,0}$ TTI as well.

%%%%%
\subsubsection{Analytic approach for the large $N$ limit}\label{sec:TTI:N010:analytic}
%%%%%

The large $N$ limit of the $N^{0,1,0}$ TTI was studied analytically in \cite{Hosseini:2016tor,Hosseini:2016ume}, following the same method we reviewed in the ADHM case. We skip the derivation and simply quote the final result,
\begin{equation}
	\log Z^{N^{0,1,0}}_{S^1\times S^2}(N,k,r)=-\fft{2\pi(k+r)}{3\sqrt{2k+r}}\,N^\fft32+o(N^\fft32) \, ,\label{N010:TTI:N32}
\end{equation}
where we have focused on the superconformal point for simplicity. Even at large $N$, the full dependence on $(\Delta,\mn)$ is complicated.\footnote{An expression for the case $k=r=1$, $\Delta_3 = \Delta_4$, and $\mn_3 = \mn_4$ can be found in~\cite{Hosseini:2016ume}} Because of this difficulty, we will only focus on the superconformal point when investigating the all-order $1/N$ expansion of the $N^{0,1,0}$ TTI. Recall also that the large $N$ TTI of the SCFT holographically dual to AdS$_4\times N^{0,1,0}/\mathbb Z_k$ is given by (\ref{N010:TTI:N32}) with $k=r$~\cite{Cheon:2011th}. For $r=0$, (\ref{N010:TTI:N32}) reduces to the large $N$ ABJM TTI~\cite{Benini:2015eyy} as expected from the quiver description.

%%%%%
\subsubsection{Numerical approach for the all-order $1/N$ expansion}\label{sec:TTI:N010:num}
%%%%%
Now we improve on the leading order analytic result (\ref{N010:TTI:N32}) by providing an analytic expression for the all-order perturbative $1/N$ expansion of the $N^{0,1,0}$ TTI. Our result reads
\begin{equation}
	\begin{split}
		\log Z^{N^{0,1,0}}_{S^1\times S^2}(N,k,r)&=-\fft{2\pi(k+r)}{3\sqrt{2k+r}}\left((\hat N_{k,r})^\fft32-\left(\fft{r}{4}+\fft{3k+2r}{(k+r)^2}\right)(\hat N_{k,r})^\fft12\right)\\
		&\quad-\fft12\log\hat N_{k,r}+\hat f_0(k,r)+\hat f_\text{np}(N,k,r)+\fft{\mathrm{i} Nr\pi}{2}\, ,
	\end{split}\label{N010:TTI:all}
\end{equation}
where we have chosen $r_1=r_2=r/2$ and defined
\begin{equation}
	\hat N_{k,r}=N+\fft{7r-2k}{48}+\fft{2}{3(k+r)} \, .\label{N010:N:shifted}
\end{equation}
The numerical analysis leading to (\ref{N010:TTI:all}) proceeds just as in the ADHM case and we refer to Appendix \ref{App:N010} for the relevant numerical data. We close this section with a few remarks:
\begin{itemize}	
	\item We have been unable to obtain the closed-form expression for the $N$-independent constants $\hat f_0(k,r)$ in~\eqref{N010:TTI:all}, although their numerical values can be obtained with high precision for various $k$ and $r$, see Appendix \ref{App:N010}.
	
	\item The imaginary part of the $N^{0,1,0}$ TTI can be determined analytically as in (\ref{N010:TTI:all}) from the BA formula (\ref{N010:TTI:BA}) and the property of a numerical BAE solution $\{x_i^\star,\tx_i^\star\}$ under complex conjugation, namely $\overline{x_i^\star}=\tx_i^\star$, which we have also numerically confirmed.
	
	\item We have been able to numerically deduce the leading non-perturbative behavior of the $N^{0,1,0}$ TTI following the procedure described for the ADHM TTI in Section~\ref{sec:TTI:ADHM:num-derive} to find
	\begin{equation}
		\hat f_\text{np}(N,k,r)=e^{-\fft{4\pi}{(1+r/k)\sqrt{2k+r}}\sqrt{N}+\mathcal O(\log N)}\quad(r\geq 1)\,,\label{N010:fnp}
	\end{equation}
	although the $N^{0,1,0}$ TTI involves larger numerical errors compared to the ADHM TTI and the outlying case $(k,r)=(2,4)$ comes with especially large numerical errors. We leave a complete numerical confirmation of the exponentially suppressed behavior (\ref{N010:fnp}) and the corresponding physical interpretation for future research. 
\end{itemize}
%

%%%%%
\subsection{$\mathcal N=2$ SCFT dual to AdS$_4\times V^{5,2}/\mathbb Z_{N_f}$}
\label{sec:TTI:V52}
%%%%%

We now consider the TTI of the $\mathcal N=2$ SCFT dual to AdS$_4\times V^{5,2}/\mathbb Z_{N_f}$, which has been constructed in two different ways~\cite{Martelli:2009ga,Jafferis:2009th} dubbed as Model I and Model II respectively in~\cite{Hosseini:2016ume}. 
%For simplicity, we call the TTI of this theory simply the $V^{5,2}$ TTI from here on.

In Model I, the theory has a U$(N)_{N_f}\times$U$(N)_{-N_f}$ gauge group and consists of two pairs of bi-fundamental $A_{1,2}$ and anti-bi-fundamental $B_{1,2}$ chiral multiplets together with two adjoint chiral multiplets $\Psi_{1,2}$ for each gauge node. The superpotential reads
\begin{equation}
	W=\Tr\Big[\Psi_1^3+\Psi_2^3+\Psi_1(A_1B_2+A_2B_1)+\Psi_2(B_2A_1+B_1A_2)\Big].\label{V52:Model1:W}
\end{equation}

In Model II, the theory has gauge group U$(N)$ with a vanishing CS level and consists of three adjoint chiral multiplets $\Psi_I~(I=1,2,3)$ and $N_f$ pairs of fundamental and anti-fundamental chiral multiplets $\psi_q$ and $\tpsi_q~(q=1,\ldots,N_f)$. The superpotential reads
\begin{equation}
	W=\Tr\Bigg[\sum_{q=1}^{N_f}\tpsi_q(\Psi_1\Psi_2+\Psi_2\Psi_1-\Psi_3^2)\psi_q+\Psi_3[\Psi_1,\Psi_2]\Bigg].\label{V52:Model2:W}
\end{equation}
Note that Model II has exactly the same matter content as the ADHM theory studied in Section~\ref{sec:TTI:ADHM} but the different superpotential breaks the supersymmetry to $\mathcal{N}=2$. We will use Model II to study the $V^{5,2}$ TTI. Note that for $N_f=1$ the theory has an $\SO(5)$ flavor symmetry in addition to the $\U(1)$ R-symmetry.

%%%%%
\subsubsection{Bethe Ansatz formulation}\label{sec:TTI:V52:BA}
%%%%%
Since Model II has the same field content as the ADHM theory, we can simply use the BA formula for the ADHM TTI spelled out in Section~\ref{sec:TTI:ADHM:BA} to study the $V^{5,2}$ TTI. The constraints on the chemical potentials and magnetic fluxes are different in the two SCFTs, however, since the two theories have different superpotentials. For the SCFT dual to AdS$_4\times V^{5,2}/\mathbb Z_{N_f}$, we have the following constraints
\begin{equation}
	\Delta_1+\Delta_2=\fft43\,,\quad\Delta_3=\fft23\,,\quad\Delta_q+\Delta_{\tq}=\fft23\,,\quad \mn_1+\mn_2=\fft43\,,\quad\mn_3=\fft23\,,\quad\mn_q+\mn_{\tq}=\fft23\,,\label{V52:constraints}
\end{equation}
from the marginality of the $V^{5,2}$ superpotential (\ref{V52:Model2:W}). The $\Delta$-configuration and the universal $\mn$-configuration are also different from the ADHM case and read
\begin{equation}
	\Delta_1=\Delta_2=\fft23\,,\qquad\Delta_m=0\,,\qquad \mn_1=\mn_2=\fft23\,,\qquad\mkt=0\,.\label{V52:constraints:sc}
\end{equation}
These values can be obtained by extremizing the large $N$ Bethe potential (\ref{ADHM:Beth:largeN:onshell}) with respect to the chemical potentials under the constraint (\ref{V52:constraints}).

%%%%%
\subsubsection{Analytic approach for the large $N$ limit}\label{sec:TTI:V52:analytic}
%%%%%
Since the BA formula for the $V^{5,2}$ TTI is the same as the one for the ADHM TTI, the large $N$ limit of the $V^{5,2}$ TTI is readily obtained from (\ref{ADHM:TTI:N32}). The result is~\cite{Hosseini:2016ume}
\begin{equation}
\begin{split}
	\log Z^{V^{5,2}}_{S^1\times S^2}(N,N_f,\Delta,\mn)&=-\fft{\pi\mu}{3}N^\fft32\sum_{a=1}^4\fft{\tmn_a}{\tDelta_a}+o(N^\fft32)\\
	&=-\fft{\pi\sqrt{N_f\tilde\Delta_1\tilde\Delta_2\tilde\Delta_3\tilde\Delta_4}}{3}N^\fft32\sum_{a=1}^4\fft{\tmn_a}{\tDelta_a}+o(N^\fft32)\,,\label{V52:TTI:N32}
\end{split}
\end{equation}
where $\mu$ is given in (\ref{ADHM:mu}) and we have defined $\tDelta_a$ and $\tmn_a$ in (\ref{ADHM:tilde}). The only difference with the ADHM case is that chemical potentials and magnetic fluxes satisfy the set of constraints (\ref{V52:constraints}).

%%%%%
\subsubsection{Numerical approach for the all-order $1/N$ expansion}\label{sec:TTI:V52:num}
%%%%%
Just as in the previous cases, a numerical analysis allows us to propose an analytic expression for the all-order perturbative $1/N$ expansion of the $V^{5,2}$ TTI:
\begin{equation}
	\begin{split}
		&\Re\log Z^{V^{5,2}}_{S^1\times S^2}(N,N_f,\Delta,\mn)\\
		&=-\fft{\pi\sqrt{N_f\tilde\Delta_1\tilde\Delta_2\tilde\Delta_3\tilde\Delta_4}}{3}\sum_{a=1}^4\fft{\tilde\mn_a}{\tilde\Delta_a}(\hat N_{N_f,\tDelta})^\fft32\\
		&\quad-\fft{\pi\sqrt{N_f\tilde\Delta_1\tilde\Delta_2\tilde\Delta_3\tilde\Delta_4}}{3}\left(\sum_{I=1}^2(\mathfrak{a}_IN_f+\fft{\mathfrak{b}_I}{N_f})\mn_I+\fft{\tDelta_3-\tDelta_4}{3\tDelta_3^2\tDelta_4^2}\fft{\mkt}{N_f^2}\right)(\hat N_{N_f,\tDelta})^\fft12\\
		&\quad-\fft12\log\hat N_{N_f,\tDelta}+\hat f_0(N_f,\tilde\Delta,\tilde\mn)+\hat f_\text{np}(N,N_f,\tilde\Delta,\tilde\mn)\,,
	\end{split}\label{V52:TTI:all}
\end{equation}
where we have introduced the rational functions $\mathfrak a_I(\tilde\Delta)$, $\mathfrak b_I(\tilde\Delta)$ (with $I=1,2$) as
\begin{subequations}
	\begin{align}
		\mathfrak a_I(\tilde\Delta)&=-\fft{1}{\tilde\Delta_I}\fft{2-(\fft23-\tilde\Delta_I)}{4\tilde\Delta_1\tilde\Delta_2}\,,\\
		\mathfrak b_I(\tilde\Delta)&=-\fft{2}{3\tilde\Delta_1\tilde\Delta_2\tilde\Delta_3\tilde\Delta_4}-\fft{3}{4\tDelta_3\tDelta_4}-\fft{(\tDelta_3-\tDelta_4)^2}{8\tDelta_3^2\tDelta_4^2}\,,
	\end{align}\label{V52:TTI:all:coeffi}%
\end{subequations}
and the shifted $N$ parameter $\hat N_{N_f,\tDelta}$ is defined in \eqref{ADHM:N:shifted}. We only focused on the real part of $\log Z^{V^{5,2}}_{S^1\times S^2}$ since the imaginary part is the same as (\ref{ADHM:TTI:all}). The numerical analysis leading to this result follows the same logic as that in the ADHM case. The only difference is that the set of $\mn$-configurations used for the \texttt{LinearModelFit} is chosen to satisfy  (\ref{V52:constraints}),
\begin{equation}
	\begin{split}
		\mathfrak n&=(\mn_1,\mn_2,\mn_3,\mn_q,\mn_{\tq},\mkt)\\
		&\in\bigg\{(\fft23,\fft23,\fft23,\fft13,\fft13,0),(\fft12,\fft56,\fft23,\fft14,\fft{5}{12},\fft12),(\fft25,\fft{14}{15},\fft23,\fft12,\fft16,\fft13),\\
		&\qquad(\fft14,\fft{13}{12},\fft23,\fft38,\fft{7}{24},\fft14),(\fft13,1,\fft23,\fft12,\fft16,\fft25)\bigg\}\, ,
	\end{split}\label{V52:nas}
\end{equation}
in contrast with (\ref{ADHM:nas}). As before, we were unable to obtain the analytic form of the $N$-independent term $\hat f_0(N_f,\tilde\Delta,\tilde\mn)$, although we could compute it to high precision for various values of $N_f$ and $(\tilde\Delta,\tilde\mn)$. We refer the reader to Appendix \ref{App:V52} for the relevant numerical data. The first three remarks summarized in Section~\ref{sec:TTI:ADHM:num-result} also apply to the $V^{5,2}$ TTI.

\medskip
%%%%%
\noindent\textbf{Non-perturbative corrections to the $V^{5,2}$ TTI}
%%%%%
\medskip

Here we numerically explore the non-perturbative corrections in (\ref{V52:TTI:all}), focusing on the superconformal and universal configuration (\ref{V52:constraints:sc}). Just as in the ADHM case, the combination of the $N$-independent contribution and the non-perturbative corrections is given by
\begin{align}
\label{V52:f0np}
		&\Re\log Z^{V^{5,2}}_{S^1\times S^2}(N,N_f)+\fft{16\pi\sqrt{N_f}}{27}\left[(\hat N_{N_f})^\fft32-\bigg(\fft{9N_f}{16}+\fft{27}{16N_f}\bigg)(\hat N_{N_f})^\fft12\right]+\fft12\log\hat N_{N_f} \nonumber\\
		&=\hat f_0(N_f)+\hat f_\text{np}(N,N_f) \, ,
\end{align}
where we have omitted $(\Delta,\mn)$ in the argument to lighten the notation and introduced the quantity $\hat N_{N_f}=N+\fft{N_f}{6}+\fft{1}{4N_f}$ based on (\ref{ADHM:N:shifted}) and \eqref{V52:constraints:sc}. We then follow the same procedure as described in Section~\ref{sec:TTI:ADHM:num-derive}, using \texttt{LinearModelFit} to fit the non-perturbative correction $\hat f_\text{np}(N,N_f)$ with respect to $N$ and with $L=16$:
\begin{equation}
	\log\left|\hat f_\text{np}(N,N_f)\right|=\hat f^\text{(lmf)}_{\text{np},1/2}(N_f)N^\fft12+\hat f^\text{(lmf)}_{\text{np},\log}(N_f)\log N+\sum_{s=0}^L\hat f^\text{(lmf)}_{\text{np},-s}(N_f)N^{-s/2} \, .\label{V52:fnp:expansion}
\end{equation}
Repeating this process for $N_f\in\{1,2,3,4,5\}$, we found 
\begin{equation}
	\hat f^\text{(lmf)}_{\text{np},1/2}(N_f)\simeq-2\pi\sqrt{\fft{1}{N_f}}\, .\label{V52:fnp:expansion:coeffi}
\end{equation}
Substituting the estimate (\ref{V52:fnp:expansion:coeffi}) back into the expansion (\ref{V52:fnp:expansion}), we obtain the non-perturbative behavior 
\begin{equation}
	\hat f_\text{np}(N,N_f)=e^{-2\pi\sqrt{N/N_f}\,+\,\mathcal O(\log N)} \, .\label{V52:fnp}
\end{equation}
The following table provides error-ratios for the leading order estimate (\ref{V52:fnp:expansion:coeffi}), namely
\begin{equation}
	R_{\text{np},1/2} = \fft{\hat f^\text{(lmf)}_{\text{np},1/2}(N_f)-(-2\pi\sqrt{1/N_f})}{-2\pi\sqrt{1/N_f}},
\end{equation}
for $N_f\in\{1,2,3,4,5\}$:
\begin{center}
	\footnotesize
	\begin{tabular}{ |c||c| } 
		\hline
		& $R_{\text{np},1/2}$  \\
		\hline\hline
		$N_f=1$ & $-9.786{\times}10^{-10}$ \\
		\hline
		$N_f=2$ & $2.603{\times}10^{-12}$ \\
		\hline
		$N_f=3$ & $-1.162{\times}10^{-11}$ \\ 
		\hline
		$N_f=4$ & $6.036{\times}10^{-9}$ \\ 
		\hline
		$N_f=5$ & $-6.007{\times}10^{-8}$ \\ 
		\hline
	\end{tabular}
\end{center}

Most importantly, the exponentially suppressed non-perturbative behavior (\ref{V52:fnp}) confirms that the $1/N$ expansion (\ref{V52:TTI:all}) indeed captures all the perturbative contributions to the TTI at the superconformal point (\ref{V52:constraints:sc}). We leave a more general analysis of non-perturbative corrections to the $V^{5,2}$ TTI with generic $(\Delta,\mn)$-configurations, as well as their physical interpretation, for future research.

%%%%%
\subsection{$\mathcal N=2$ SCFT dual to AdS$_4\times Q^{1,1,1}/\mathbb Z_{N_f}$}
\label{sec:TTI:Q111}
%%%%%

As our last example, we study the TTI of the $\mathcal N=2$ SCFT dual to AdS$_4\times Q^{1,1,1}/\mathbb Z_{N_f}$~\cite{Franco:2008um,Franco:2009sp,Benini:2009qs,Cremonesi:2010ae}. In the conventions of~\cite{Benini:2009qs,Cremonesi:2010ae}, the theory has a U$(N)\times$U$(N)$ gauge group with vanishing CS levels together with two pairs of bi-fundamental $A_{1,2}$ and anti-bi-fundamental $B_{1,2}$ chiral multiplets and $2N_f$ pairs of fundamental $\psi^{(s)}_{1,2}$ and anti-fundamental chiral multiplets $\tpsi^{(s)}_{1,2}~(s=1,\ldots,N_f)$. Here $\psi^{(s)}_a$ and $\tpsi^{(s)}_a$ are fundamental and anti-fundamental with respect to the second and first U$(N)$ factors, so that they yield a single trace operator together with $A_a~(a=1,2)$. The superpotential is given by
\begin{equation}
	W=\Tr\Bigg[A_1B_1A_2B_2-A_1B_2A_2B_1+\sum_{s=1}^{N_f}\tpsi^{(s)}_1A_1\psi^{(s)}_1+\sum_{s=1}^{N_f}\tpsi^{(s)}_2A_2\psi^{(s)}_2\Bigg]\, .\label{Q111:W}
\end{equation}
For $N_f=1$ the theory has $\SU(2)^3$ flavor symmetry in addition to the $\U(1)$ R-symmetry.
%For simplicity, we call the TTI of this theory simply the $Q^{1,1,1}$ TTI from here on.

%To evaluate the $Q^{1,1,1}$ TTI, we first briefly introduce the BA formula for the $Q^{1,1,1}$ TTI in subsection \ref{sec:TTI:Q111:BA}. Then in subsection \ref{sec:TTI:Q111:analytic} we review the analytic calculation of the $Q^{1,1,1}$ TTI in the large $N$ limit based on the BA formula. Finally in subsection \ref{sec:TTI:Q111:num} we provide an analytic expression for the $Q^{1,1,1}$ TTI to all orders in the perturbative $\fft1N$-expansion.

%%%%%
\subsubsection{Bethe Ansatz formulation}\label{sec:TTI:Q111:BA}
%%%%%
According to the general formula (\ref{general:TTI}), the $Q^{1,1,1}$ TTI takes the form of a matrix model,
\begin{align}
\label{Q111:TTI:1}
		&Z^{Q^{1,1,1}}_{S^1\times S^2}(N,N_f,\Delta,\mn) \nonumber \\
		&=\fft{1}{(N!)^2}\sum_{\mm,\tilde{\mm}\in\mathbb Z^N}\oint_{\mathcal C}\prod_{i=1}^N\fft{dx_i}{2\pi \mathrm{i}x_i}\fft{d\tx_i}{2\pi \mathrm{i}\tx_i}(-1)^{(N+N_f-2\lfloor N_f/2\rfloor)(\mm_i-\tmm_i)}\prod_{i\neq j}^N\left(1-\fft{x_i}{x_j}\right)\left(1-\fft{\tx_i}{\tx_j}\right)\nonumber \\
		&\quad\times\prod_{i,j=1}^N\prod_{a=1,2}\left(\fft{\sqrt{\fft{x_i}{\tx_j}y_a}}{1-\fft{x_i}{\tx_j}y_a}\right)^{\mm_i-\tilde{\mm}_j-\mn_a+1}\prod_{a=3,4}\left(\fft{\sqrt{\fft{\tx_j}{x_i}y_a}}{1-\fft{\tx_j}{x_i}y_a}\right)^{-\mm_i+\tilde{\mm}_j-\mn_a+1}\\
		&\quad\times\prod_{i=1}^N\prod_{n=1,2}\left(\fft{\sqrt{\fft{1}{x_i}y_{\tq_n}}}{1-\fft{1}{x_i}y_{\tq_n}}\right)^{N_f(-\mm_i-\mn_{\tq_n}+1)}\times\prod_{j=1}^N\prod_{n=1,2}\left(\fft{\sqrt{\tx_jy_{q_n}}}{1-\tx_jy_{q_n}}\right)^{N_f(\tilde{\mm}_j-\mn_{q_n}+1)} \, , \nonumber 
\end{align}
where we have again assigned the same fugacities $y_{q_{1,2}}$ and $y_{\tilde q_{1,2}}$ and magnetic fluxes $\mn_{q_{1,2}}$ and $\mn_{\tilde q_{1,2}}$ for the $N_f$ pairs of fundamental and anti-fundamental multiplets for simplicity. Note that we have also included the fugacities $(-1)^{N+N_f-2\lfloor N_f/2\rfloor}$ associated with the U(1) topological symmetries for later convenience. From the marginality of the $Q^{1,1,1}$ superpotential (\ref{Q111:W}) and its invariance under global symmetries, the chemical potentials and magnetic fluxes are constrained to satisfy
\begin{equation}
	\begin{alignedat}{2}
		2&=\sum_{a=1}^4\mn_a\,,&\qquad 2&=\mn_1+\mn_{q_1}+\mn_{\tq_1}=\mn_2+\mn_{q_2}+\mn_{\tq_2}\,,\\
		2&=\sum_{a=1}^4\Delta_a\,,&\qquad 2&=\Delta_1+\Delta_{q_1}+\Delta_{\tq_1}=\Delta_2+\Delta_{q_2}+\Delta_{\tq_2}\,,
	\end{alignedat}\label{Q111:constraints}
\end{equation}
where we have used the notation
\begin{equation}
	y_{\Psi_{(a,b)}}\to y_a=e^{\mathrm{i}\pi\Delta_a},\qquad y_{\Psi_a}\to y_{q_n}=e^{\mathrm{i}\pi\Delta_{q_n}},\qquad y_{\tPsi_a}\to y_{\tq_n}=e^{\mathrm{i}\pi\Delta_{\tq_n}}.\label{Q111:Delta:simple}
\end{equation}
The superconformal $\Delta$-configuration and the universal $\mn$-configuration correspond to
\begin{equation}
\begin{alignedat}{3}
	\Delta_1&=\Delta_2\,,&\qquad\Delta_3&=\Delta_4\,,&\qquad\Delta_{q_{1,2}}&=\Delta_{\tq_{1,2}}\,,\\
	\mn_1&=\mn_2\,,&\qquad\mn_3&=\mn_4\,,&\qquad\mn_{q_{1,2}}&=\mn_{\tq_{1,2}}\,.\label{Q111:constraints:sc}
\end{alignedat}
\end{equation}
This can be obtained in principle by extremizing the large $N$ Bethe potential with respect to chemical potentials as in the ADHM case, but the dependence of the Bethe potential on $(\Delta,\mn)$ is much more involved for the $Q^{1,1,1}$ TTI. An alternative way to derive the result is to match the $S^3$ free energy with trial $R$-charges chosen as in (\ref{Q111:constraints:sc}) and the holographically dual regularized on-shell action \cite{Jafferis:2011zi}. For notational convenience, we use $\Delta$ and $\mn$ to collectively represent all the chemical potentials and the magnetic fluxes as
\begin{equation}
	\Delta=(\Delta_a,\Delta_{q_{1,2}},\Delta_{\tq_{1,2}})\,,\qquad \mn=(\mn_a,\mn_{q_{1,2}},\mn_{\tq_{1,2}})\,.\label{Q111:Delta:mn}
\end{equation}

Following the procedure detailed in Section~\ref{sec:TTI:ADHM:BA}, the BA formula for the $S^1\times S^2$ $Q^{1,1,1}$ TTI reads
\begin{equation}
	\begin{split}
		&Z^{Q^{1,1,1}}_{S^1\times S^2}(N,N_f,\Delta,\mn)\\
		&=\prod_{n=1}^2y_{q_n}^{\fft12Nk(1-\mn_{q_n})}y_{\tq_n}^{\fft12Nk(1-\mn_{\tq_n})}\times\prod_{a=1}^4y_a^{-\fft{N^2}{2}\mn_a}\\
		&\quad\times\sum_{\{x_i,\tx_j\}\in\text{BAE}}\left[\fft{1}{\det\mathbb B}\fft{\prod_{i=1}^Nx_i^{N}\tx_i^{N}\prod_{i\neq j}^N(1-\fft{x_i}{x_j})(1-\fft{\tx_i}{\tx_j})}{\prod_{i,j=1}^N\prod_{a=1,2}(\tx_j-x_iy_a)^{1-\mn_a}\prod_{a=3,4}(x_i-\tx_jy_a)^{1-\mn_a}}\right.\\
		&\kern7em~\left.\times\prod_{i=1}^N\prod_{n=1}^2\fft{x_i^{N_f(1-\mn_{\tq_n})/2}}{(x_i-y_{\tq_n})^{N_f(1-\mn_{\tq_n})}}\fft{\tx_i^{N_f(1-\mn_{q_n})/2}}{(1-\tx_iy_{q_n})^{N_f(1-\mn_{q_n})}}\right]\,,
	\end{split}\label{Q111:TTI:BA}
\end{equation}
in terms of the Jacobian matrix $\mathbb B$ given in (\ref{N010:Jacobian}). The explicit BAE can be written as follows in terms of the variables $u_i$ and $\tu_j$ introduced via $x_i=e^{\mathrm{i}u_i}$ and $\tx_j=e^{\mathrm{i}\tu_j}$,
\begin{align}
\label{Q111:BAE}
		2\pi\left(n_i-\fft{N_f}{2}+\left\lfloor\fft{N_f}{2}\right\rfloor\right)&=\mathrm{i}\sum_{j=1}^N\bigg[\sum_{a=3,4}\text{Li}_1(e^{\mathrm{i}(\tu_j-u_i+\pi\Delta_a)})-\sum_{a=1,2}\text{Li}_1(e^{\mathrm{i}(\tu_j-u_i-\pi\Delta_a)})\bigg] \nonumber \\
		&\quad+\mathrm{i}N_f\sum_{n=1}^2\text{Li}_1(e^{\mathrm{i}(-u_i+\pi\Delta_{\tq_n})})-\fft{N_f\pi}{2}\sum_{n=1}^2\Delta_{\tq_n}+N_fu_i\quad(n_i\in\mathbb Z)\,, \nonumber\\
		2\pi\left(\tn_j-\fft{N_f}{2}+\left\lfloor\fft{N_f}{2}\right\rfloor\right)&=\mathrm{i}\sum_{i=1}^N\bigg[\sum_{a=3,4}\text{Li}_1(e^{\mathrm{i}(\tu_j-u_i+\pi\Delta_a)})-\sum_{a=1,2}\text{Li}_1(e^{\mathrm{i}(\tu_j-u_i-\pi\Delta_a)})\bigg]\\
		&+\mathrm{i}N_f\sum_{n=1}^2\text{Li}_1(e^{\mathrm{i}(-\tu_j-\pi\Delta_{q_n})})+\fft{N_f\pi}{2}\sum_{n=1}^2(\Delta_{q_n}-2)+N_f\tu_j\quad(\tn_j\in\mathbb Z) \, . \nonumber
\end{align}
Once again, the remarks we have spelled out in the ADHM case below \eqref{eq:BB} apply \textit{mutatis mutandis} to the BA formula for the $Q^{1,1,1}$ TTI.

\subsubsection{Analytic approach for the large $N$ limit}\label{sec:TTI:Q111:analytic}
%%%%%

The large $N$ limit of the $Q^{1,1,1}$ TTI follows the same logic as in the other cases. In the derivation, we do however correct a couple of subtle points in the intermediate steps of the Bethe potential analysis conducted in~\cite{Hosseini:2016tor,Hosseini:2016ume}. To streamline the presentation, we include some details in Appendix \ref{App:Q111}. These corrections do not affect the final large $N$ result. For the superconformal $\Delta$-configuration in \eqref{Q111:constraints:sc} and arbitrary $\mn$, it reads~\cite{Hosseini:2016ume}
\begin{equation}
	\log Z^{Q^{1,1,1}}_{S^1\times S^2}(N,N_f,\Delta,\mn) =-\fft{4\pi\sqrt{N_f}}{3\sqrt{3}}\,N^\fft32+o(N^\fft32) \, .\label{Q111:TTI:N32}
\end{equation}
Observe that, because of \eqref{Q111:constraints} and \eqref{Q111:constraints:sc}, the left hand side can only depend on the chemical potential $\Delta_1$ at the superconformal point. The upshot of \eqref{Q111:TTI:N32} is that $\Delta_1$ and $\mn$ are in fact flat directions for the large $N$ TTI. We will see below that this remains true in the all-order $1/N$ expansion. See~\cite{Jafferis:2011zi} for a similar phenomenon in the $S^3$ partition function of the $Q^{1,1,1}$ theory.

Similarly to the $N^{0,1,0}$ case discussed in Section~\ref{sec:TTI:N010}, the full dependence on $\Delta$ is unwieldy even at large $N$. A partial result obtained by setting some of the chemical potentials to be equal can be found in~\cite{Hosseini:2016ume}. Because of this complication, we will only study the all-order $1/N$ expansion of the $Q^{1,1,1}$ TTI at the superconformal point, but we will keep the magnetic fluxes $\mn$ arbitrary modulo the constraint \eqref{Q111:constraints}.

%%%%%
\subsubsection{Numerical approach for the all-order $1/N$ expansion}\label{sec:TTI:Q111:num}
%%%%%

To go beyond the analytic result for the large $N$ $Q^{1,1,1}$ TTI (\ref{Q111:TTI:N32}), we use the same numerical method as in the other cases. This leads us to an analytic expression for the all-order perturbative $1/N$ expansion of the TTI. At the superconformal point where the only independent chemical potential is $\Delta_1$, the result reads
\begin{equation}
	\begin{split}
		\Re\log Z^{Q^{1,1,1}}_{S^1\times S^2}(N,N_f,\Delta_1,\mn)&=-\fft{4\pi\sqrt{N_f}}{3\sqrt3}\left((\hat N_{N_f})^\fft32-\bigg(\fft{N_f}{4}+\fft{3}{4N_f}\bigg)(\hat N_{N_f})^\fft12\right)\\
		&\quad-\fft12\log\hat N_{N_f}+\hat f_0(N_f)+\hat f_\text{np}(N,N_f) \, ,
	\end{split}\label{Q111:TTI:all}
\end{equation}
where we have defined
\begin{equation}
	\hat N_{N_f}=N+\fft{N_f}{6}\,.\label{Q111:N:shifted}
\end{equation}
Once again, we have focused on the real part of the TTI, see the comment below \eqref{N010:N:shifted}. Our numerical investigations confirm that the $Q^{1,1,1}$ TTI is in fact independent of $(\Delta_1,\mn)$.
%
%Numerical derivation of the all-order $\fft1N$-expansion of the $Q^{1,1,1}$ TTI (\ref{Q111:TTI:all}) is exactly parallel to the ADHM case described in subsection \ref{sec:TTI:ADHM:num-derive}\footnote{Since $v(t)+\tv(t)$ is not determined in the large $N$ BAE solution (\ref{N010:BAE:sol:largeN}), we set the initial condition for $v_i+\tv_i$ based on the ranges (\ref{Q111:range}) as \cite{PandoZayas:2020iqr}
%	%
%	\begin{equation}
%		v_i+\tv_i\big|_\text{initial condition}=\pi(2-\Delta_1-\Delta_{q_1}-\Delta_{q_2})=0,\label{BAE:Q111:sol:v}
%	\end{equation}
%	%
%where the second equation is for the superconformal $\Delta$-configuration (\ref{Q111:constraints:sc}).}. 

The configurations of magnetic fluxes used in the numerical fits have been chosen to satisfy the constraints (\ref{Q111:constraints}),
\begin{align}
\label{Q111:nas}
		&(\mn_1,\mn_2,\mn_3,\mn_4,\mn_{q_1},\mn_{\tq_1},\mn_{q_2},\mn_{\tq_2}) \nonumber \\
		&\in\bigg\{(\fft12,\fft12,\fft12,\fft12,\fft34,\fft34,\fft34,\fft34),(\fft14,\fft12,\fft23,\fft{7}{12},\fft12,\fft54,\fft12,1),(\fft14,\fft23,\fft38,\fft{17}{24},\fft23,\fft{13}{12},\fft12,\fft56),\\
		&\qquad(\fft13,\fft25,\fft12,\fft{23}{30},\fft25,\fft{19}{15},\fft34,\fft{17}{20}),(\fft13,\fft34,\fft78,\fft{1}{24},\fft78,\fft{19}{24},\fft13,\fft{11}{12}),(\fft58,\fft23,\fft14,\fft{11}{24},\fft45,\fft{23}{40},\fft13,1)\bigg\} \, . \nonumber
\end{align}
We refer the reader to Appendix \ref{App:Q111} for the relevant numerical data used in the derivation of (\ref{Q111:TTI:all}). As in the other cases, we have not yet found the closed-form expression for the $N$-independent constant $\hat f_0(N_f)$. The numerical values can be obtained with a high precision, see Appendix \ref{App:Q111} where we show they are independent of $(\Delta_1,\mn)$. The numerical analysis of the non-perturbative corrections parallels the ADHM case. We found the following behavior at the superconformal point,
	\begin{equation}
		\hat f_\text{np}(N,N_f) = e^{-2\pi\sqrt{\fft{N}{3N_f}}\,+\,\mathcal O(\log N)} \, ,\label{Q111:fnp}
	\end{equation}
	for $N_f\in\{1,2,3,5\}$. Numerical errors in the non-perturbative corrections for $N_f=4$ are however too large to confirm (\ref{Q111:fnp}) in that case. We leave a physical interpretation of the above non-perturbative behavior and further numerical study for the future. 

%%%%%%%%%%%%%%%%%
\section{Sphere partition functions}
\label{sec:S3}
%%%%%%%%%%%%%%%%%

In this section we study another observable for the 3d $\mathcal{N}\geq 2$ SCFTs introduced in Section~\ref{sec:TTI}, namely their $S^3_b$ partition functions with real mass deformations. In Section~\ref{sec:S3:ADHM} we focus on the perturbative part of the $S^3_b$ partition function of the ADHM theory with an $\mathcal N=4$ symmetry breaking mass parameter $\mu$.
%, see subsection \ref{sec:TTI:ADHM} for detailed introduction to the ADHM theory. 
Then in Section~\ref{sec:S3:others} we briefly comment on the $S^3_b$ partition function of other $\mathcal N\geq2$ holographic SCFTs with real mass deformations, highlighting some obstacles that remain to obtain the all-order expression for the partition functions in an $1/N$ expansion.

%%%%%
\subsection{$\mathcal N=4$ ADHM theory}\label{sec:S3:ADHM}
%%%%%

The sphere partition function of the ADHM theory can be computed using supersymmetric localization, which yields a matrix model~\cite{Mezei:2013gqa,Hatsuda:2014vsa}. It is possible to introduce geometric and mass deformations to this result by squashing the sphere with a parameter $b$ (where $b=1$ corresponds to the round sphere) and turning on an $\mathcal N=4$ symmetry breaking mass parameter $\mu$~\cite{Minahan:2021pfv}. In this case, the matrix model is given as~\cite{Minahan:2021pfv,Chester:2020jay,Willett:2016adv}\footnote{We mainly follow the conventions of~\cite{Minahan:2021pfv} but for the precise normalization of $Z^\text{ADHM}_{S^3_b}$ we refer to the other two references.}
\begin{equation}
	\begin{split}
		&Z^\text{ADHM}_{S^3_b}(N,N_f,\mu)\\
		&=\fft{1}{N!}\int d^Nx\,\prod_{i<j}^N2\sinh\relax(\pi bx_{ij})\,2\sinh\relax(\pi b^{-1}x_{ij}) \prod_{i,j=1}^Ns_b(\mu-x_{ij})\\
		&\quad\times\prod_{i,j=1}^Ns_b\Bigl(\fft{\mathrm{i}(b+b^{-1})}{4}-\fft{\mu}{2}+x_{ij}\Bigr)s_b\Bigl(\fft{\mathrm{i}(b+b^{-1})}{4}-\fft{\mu}{2}-x_{ij}\Bigr)\\
		&\quad\times\Bigg[\prod_{i=1}^Ns_b\Bigl(\fft{\mathrm{i}(b+b^{-1})}{4}-\fft{\mu}{2}+x_i\Bigr)s_b\Bigl(\fft{\mathrm{i}(b+b^{-1})}{4}-\fft{\mu}{2}-x_i\Bigr)\Bigg]^{N_f} \, ,
	\end{split}\label{ADHM:S3:mm}
\end{equation}
where the double sine function $s_b(x)$ is defined in e.g. App. A of~\cite{Hatsuda:2016uqa}. For $\mu=0$ and $b=1$, the matrix model (\ref{ADHM:S3:mm}) reduces to the one for the $S^3$ partition function of the undeformed $\mathcal N=4$ ADHM theory, which has been evaluated to all orders in the perturbative $1/N$ expansion in terms of an Airy function using the ideal Fermi-gas formulation~\cite{Mezei:2013gqa,Hatsuda:2014vsa},
\begin{equation}
\begin{split}
	&Z^\text{ADHM}_{S^3}(N,N_f,0)\\
	&=\bigg(\fft{2}{\pi^2 N_f}\bigg)^{-\fft13}e^{\fft12(\mathcal{A}(N_f)+\mathcal {A}(1)N_f^2)}\text{Ai}\Bigg[\bigg(\fft{2}{\pi^2 N_f}\bigg)^{-\fft13}\bigg(N+\fft{N_f}{8}-\fft{1}{2N_f}\bigg)\Bigg]+\mathcal O(e^{-\sqrt{N}}) \, .\label{ADHM:S3}
\end{split}
\end{equation}
The function $\mathcal{A}$ above is the constant map contribution to the partition function of ABJM theory on the round $S^3$~\cite{Marino:2011eh}. Its explicit expression will not be needed in what follows.

The matrix model (\ref{ADHM:S3:mm}) has an interesting property for the special value of the $\mathcal N=4$ symmetry breaking mass parameter $\mu=\pm\fft{\mathrm{i}(b-b^{-1})}{2}$, namely \cite{Minahan:2021pfv}
\begin{equation}
	Z^\text{ADHM}_{S^3_b}(N,N_f,\pm\fft{\mathrm{i}(b-b^{-1})}{2})=Z^\text{ADHM}_{S^3}(N,N_f,0) \, .\label{ADHM:S3:constraint}
\end{equation}
Combining the relation (\ref{ADHM:S3:constraint}) with the Airy formula (\ref{ADHM:S3}) shows that the perturbative part of the $S^3_b$ ADHM partition function with $\mu=\pm\fft{\mathrm{i}(b-b^{-1})}{2}$ is also given in terms of an Airy function:
\begin{equation}
\begin{split}
	&Z^\text{ADHM}_{S_b^3}(N,N_f,\pm\fft{\mathrm{i}(b-b^{-1})}{2})\\
	&=\bigg(\fft{2}{\pi^2 N_f}\bigg)^{-\fft13}e^{\fft12(\mathcal{A}(N_f)+\mathcal {A}(1)N_f^2)}\text{Ai}\Bigg[\bigg(\fft{2}{\pi^2 N_f}\bigg)^{-\fft13}\bigg(N+\fft{N_f}{8}-\fft{1}{2N_f}\bigg)\Bigg]+\mathcal O(e^{-\sqrt{N}}) \, . \label{ADHM:Sb3mu:special}
\end{split}
\end{equation}

\medskip

Motivated by the observation (\ref{ADHM:Sb3mu:special}), we conjecture that the perturbative part of the $S^3_b$ ADHM partition function with generic mass parameter $\mu$ is given by an Airy function as
\begin{equation}
\begin{split}
	Z^\text{ADHM}_{S_b^3}(N,N_f,\mu)=C_b(N_f,\mu)^{-\fft13}e^{\mathcal{A}_b(N_f,\mu)}\text{Ai}\bigg[C_b(N_f,\mu)^{-\fft13}\big(N-B_b(N_f,\mu)\big)\bigg]+\mathcal O(e^{-\sqrt{N}})\, ,\label{ADHM:Sb3mu}
\end{split}
\end{equation}
where the functions $C_b$ and $B_b$ read
\begin{equation}
\label{eq:Cb-Bb}
\begin{split}
	C_b(N_f,\mu)&=\fft{2}{\pi^2N_f}\fft{1}{(\fft{b+b^{-1}}{2}+\mathrm{i}\mu)^2(\fft{b+b^{-1}}{2}-\mathrm{i}\mu)^2} \, \\[1mm]
	B_b(N_f,\mu)&=N_f\left(\fft{1}{24}-\fft{\fft{(b+b^{-1})^2(b^2+b^{-2})}{48}+\fft{b^2+4+b^{-2}}{12}\mu^2}{(\fft{b+b^{-1}}{2}+\mathrm{i}\mu)^2(\fft{b+b^{-1}}{2}-\mathrm{i}\mu)^2}\right) \\
&\quad -\fft{1}{N_f}\fft{\fft{(b+b^{-1})^2(b^2-8+b^{-2})}{48}+\fft{b^2-4+b^{-2}}{12}\mu^2}{(\fft{b+b^{-1}}{2}+\mathrm{i}\mu)^2(\fft{b+b^{-1}}{2}-\mathrm{i}\mu)^2} \, .
\end{split}
\end{equation}
At present, we do not have a conjecture for the $N$-independent prefactor $\mathcal{A}_b(N_f,\mu)$. Our Airy proposal (\ref{ADHM:Sb3mu}) passes various non-trivial consistency checks, listed below:
\begin{itemize}
	\item It reduces to the known result (\ref{ADHM:Sb3mu:special}) for the special value of the $\mathcal N=4$ symmetry breaking mass parameter $\mu=\pm\fft{\mathrm{i}(b-b^{-1})}{2}$.
	
	\item When expanding the Airy function at large $N$, it is consistent with the leading $\mathcal{O}(N^\fft32)$ and subleading $\mathcal{O}(N^\fft12)$ terms of the $\mu=0$ squashed sphere partition function that were obtained in~\cite{Bobev:2021oku} in the dual bulk supergravity theory.
	
	\item The Airy formula (\ref{ADHM:Sb3mu}) with $N_f=1$ becomes equivalent to the Airy formula for the $S^3_b$ partition function of real mass deformed ABJM theory conjectured in~\cite{Bobev:2022jte,Bobev:2022eus} where the CS level and the real masses are set to $k=1$ and $(m_1,m_2,m_3)=(0,0,\mu)$, respectively.
	
	\item For  $\mu=0$ and $b^2=3$ it was shown in \cite{Hatsuda:2016uqa} that the supersymmetric localization matrix model admits a Fermi gas description and its perturbative part in the large $N$ expansion can be resummed into an Airy function. It is easy to check that our conjecture in \eqref{ADHM:Sb3mu} and \eqref{eq:Cb-Bb} reduces to the results in \cite{Hatsuda:2016uqa} for $\mu=0$ and $b^2=3$.
\end{itemize}
In addition, one can use our Airy proposal to compute the coefficient $c_T$ of the stress tensor 2-point function in the undeformed ADHM theory following~\cite{Closset:2012ru,Chester:2014fya},
\begin{equation}
		c_T = -\fft{32}{\pi^2}\fft{\partial^2\log Z^\text{ADHM}_{S^3_b}(N,N_f,0)}{\partial b^2}\Bigg|_{b=1} = -\fft{32}{\pi^2}\fft{\partial^2\log Z^\text{ADHM}_{S^3}(N,N_f,\mu)}{\partial\mu^2}\Bigg|_{\mu=0} \, .
\end{equation}
Upon taking the derivatives, we obtain 
\begin{equation}
\begin{split}
		c_T&=-\fft{64N_f^\fft12(2\pi)^\fft23\left(N+\fft{3N_f}{8}+\fft{3}{4N_f}\right)\text{Ai}'\left[\left(\fft{2}{\pi^2N_f}\right)^{-\fft13}\left(N+\fft{N_f}{8}-\fft{1}{2N_f}\right)\right]}{3\pi^2\text{Ai}\left[\left(\fft{2}{\pi^2N_f}\right)^{-\fft13}\left(N+\fft{N_f}{8}-\fft{1}{2N_f}\right)\right]}\\
		&\quad+(N\text{\,-\,independent constant}) + \mathcal O(e^{-\sqrt{N}}) \, ,
\end{split}
\end{equation}
which matches the result of~\cite{Chester:2020jay} derived using the ideal Fermi-gas approach. \\

On top of the above consistency checks, the conjecture (\ref{ADHM:Sb3mu}) displays an interesting property. In the $b\to0$ limit (or equivalently $b\to\infty$ since there is a $b \leftrightarrow b^{-1}$ symmetry apparent in the matrix model \eqref{ADHM:S3:mm}) with fixed $\fft{\mu}{b+b^{-1}}$, the argument of the Airy function in (\ref{ADHM:Sb3mu}) becomes
\begin{equation}
	\begin{split}
		N-B_b(N_f,\mu) \; \xrightarrow[b\to0]{\fft{\mu}{b+b^{-1}} \; \mathrm{fixed}}\;\; &N-\fft{N_f}{24}+\fft{N_f}{12}\left(\fft{1}{\fft12+\fft{\mathrm{i}\mu}{b+b^{-1}}}+\fft{1}{\fft12-\fft{\mathrm{i}\mu}{b+b^{-1}}}\right)\\
		&\quad+\fft{1}{12N_f}\left(\fft{1}{\fft12+\fft{\mathrm{i}\mu}{b+b^{-1}}}+\fft{1}{\fft12-\fft{\mathrm{i}\mu}{b+b^{-1}}}\right) \, .
	\end{split}
\end{equation}
The RHS is now identical to the shifted $N$ parameter for the $S^1\times\Sigma_{\mg}$ ADHM TTI (\ref{ADHM:N:shifted}) provided we identify the mass $\mu$ and the chemical potentials $\tDelta$ as
\begin{equation}
	\tDelta_1=\tDelta_3=\fft12-\fft{\mathrm{i}\mu}{b+b^{-1}}\,,\qquad \tDelta_2=\tDelta_4=\fft12+\fft{\mathrm{i}\mu}{b+b^{-1}}\,.\label{tDelta:mu}
\end{equation}
Note that this identification is consistent with the remark around \eqref{eq:ADHM-ABJM}: the flavor chemical potentials $\tDelta$ in the ADHM theory are identified with the chemical potentials $\Delta$ in the ABJM theory based on the duality between the two theories at $k=N_f=1$, and the ABJM $\Delta$ are given precisely by (\ref{tDelta:mu}) for $(m_1,m_2,m_3)=(0,0,\mu)$ as described in~\cite{Bobev:2022jte,Bobev:2022eus}. The correspondence between the shifted $N$ for the $S^3_b$ partition function and the shifted $N$ for the $S^1\times\Sigma_{\mg}$ TTI upon taking the $b\to0$ limit is not specific to the ADHM model and can also be deduced for the ABJM theory. Using the results in~\cite{Bobev:2022jte,Bobev:2022eus} we find the following analogous expressions for the ABJM theory
\begin{equation}
	\begin{split}
		B_b(k,\Delta) &=\fft{k}{24}+\fft{1}{12k}\Bigg[\fft{4-\sum_{a=1}^4\Delta_a^2}{(b+b^{-1})^2\Delta_1\Delta_2\Delta_3\Delta_4}-\sum_{a=1}^4\fft{1}{\Delta_a}\Bigg]\,,\\
		N-B_b(k,\Delta) \; &\xrightarrow[b\to0]{\Delta \;\; \mathrm{fixed}}\;\; \hat N_\Delta=N-\fft{k}{24}+\fft{k}{12}\sum_{a=1}^4\fft{1}{\Delta_a}\, ,
	\end{split}
\end{equation}
where $N-B_b(k,\Delta)$ is the shift for the $S^3$ partition function and $\hat N_\Delta$ is the shift for $S^1\times\Sigma_{\mg}$ TTI. It would be interesting to understand the physics behind this relation between different observables in these holographic SCFTs in more detail.\\

Considering that the $S^3_b$ partition function of the ABJM theory with generic real mass deformations also admits an Airy form~\cite{Bobev:2022jte,Bobev:2022eus}, and that the ADHM theory with $N_f=1$ is dual to the ABJM theory with $k=1$, we expect that the $S^3_b$ partition function of the ADHM theory deformed by the real masses of the adjoint/fundamental hypermultiplets and by the FI parameter in addition to $\mu$ will also admit an Airy form. The latter should then be valid to all orders in the perturbative $1/N$ expansion, although we have been unable to formulate a precise conjecture at this stage. 

%%%%%
\subsection{$\mathcal N=2,3$ holographic SCFTs}\label{sec:S3:others}
%%%%%

In this subsection we briefly present some of the difficulties one faces when trying to derive or conjecture an Airy formula for the $S^3_b$ partition function of $\mathcal N=2,3$ holographic SCFTs with real mass deformations. We focus on the theories introduced in Sections~\ref{sec:TTI:N010},~\ref{sec:TTI:V52}, and~\ref{sec:TTI:Q111}.

For the $\mathcal N=3$ SCFT dual to AdS$_4\times N^{0,1,0}/\mathbb Z_k$, the Airy formula for the undeformed $S^3$ partition function with $r_1=r_2=r/2$ was derived  in~\cite{Marino:2011eh} and is given by
\begin{equation}
\begin{split}
	Z^{N^{0,1,0}}_{S^3}(N,k,r)&=C(k,r)^{-\fft13}e^{\mathcal A(k,r)}\text{Ai}\left[C(k,r)^{-\fft13}(N-B(k,r))\right]+\mathcal O(e^{-\sqrt{N}})\,,\\[1mm]
	C(k,r)&=\fft{2k+r}{\pi^2(k+r)^2}\,,\\
	B(k,r)&=\fft{2k-3r}{48}+\fft{k}{3(k+r)^2}\,,
\end{split}\label{N010:S3}
\end{equation}
where $r$ is kept independent from $k$ to keep the discussion general, but one should ultimately identify $r=k$. Generalizing the Airy formula (\ref{N010:S3}) to the $S^3_b$ partition function with real mass deformations is non-trivial, mainly due to the lack of a relation of the form (\ref{ADHM:S3:constraint}) in the $\mathcal N=3$ SCFT of interest. Without such a constraint, it is difficult to conjecture an Airy formula for the $S^3_b$ partition function of the $N^{0,1,0}$ theory with real mass deformations. Furthermore, the duality between the ADHM theory with $N_f=1$ and the ABJM theory with $k=1$ was crucial in formulating our conjecture (\ref{ADHM:Sb3mu}). In contrast, the $\mathcal N=3$ SCFT dual to AdS$_4\times N^{0,1,0}/\mathbb Z_k$ does not exhibit such a duality as far as we are aware.\\

The situation is even worse for the $\mathcal N=2$ SCFTs dual to AdS$_4\times V^{5,2}/\mathbb Z_{N_f}$ and AdS$_4\times V^{5,2}/\mathbb Z_{k}$ since there are no known Airy formula even for their undeformed $S^3$ partition functions. The matrix models for their $S^3$ partition functions can be written explicitly based on~\cite{Kapustin:2009kz}, but the ideal Fermi-gas formulation of~\cite{Marino:2011eh} has not yet been successfully applied to obtain the all-order perturbative $1/N$ expansion of the $S^3$ partition functions from the corresponding matrix models.

%%%%%
\section{Intermezzo: The mABJM theory}\label{sec:mABJM}
%%%%%

We now discuss another example of a holographic 3d $\mathcal{N}=2$ SCFT for which the TTI and the $S^3$ partition function can be computed in the large $N$ limit via supersymmetric localization. The theory in question is a close cousin of the ABJM model at $k=1,2$ and can be obtained from it by adding a superpotential mass term for one of the four bi-fundamental chiral superfields of the ABJM theory.\footnote{To form a gauge invariant superpotential one needs to multiply the bi-fundamental chiral superfield with a monopole operator. To have a relevant superpotential deformation the monopole operator has to have a sufficiently low conformal dimension. This is possible for $k=1$ and $k=2$, see \cite{Benna:2008zy} for a detailed discussion. Thus, in this Section, all formulas apply only for the values $k=1,2$.} In the IR the RG flow ends at a strongly interacting 3d $\mathcal{N}=2$ SCFT which we will refer to as the mABJM theory, see \cite{Benna:2008zy,Jafferis:2011zi,Bobev:2018uxk,Bobev:2018wbt} for more details. In addition to the $\U(1)$ R-symmetry the theory also enjoys $\SU(3)$ flavor symmetry. Its holographic dual description is provided by an AdS$_4$ vacuum of 11d supergravity, see \cite{Corrado:2001nv}, which can be found by uplifting one of the non-trivial vacua of the 4d $\mathcal{N}=8$ $\SO(8)$ gauged supergravity found in \cite{Warner:1983vz}.

We conjecture that the $S^3$ partition function and the TTI of the mABJM theory to all orders in the $1/N$ expansion can be obtained from the corresponding results in the ABJM model after a particular specialization of the real mass parameters, the electric fugacities, and the background magnetic fluxes.

We start with the $S^3$ partition function. For the ABJM theory in the presence of the three real masses, $m_i$, and the supersymmetric preserving squashing deformation it was conjectured in \cite{Bobev:2022jte,Bobev:2022eus,Hristov:2022lcw} that the $S^3$ partition function takes the form
\begin{equation}
\label{eq:Z}
	Z_{S^3_b} = C_b^{-\frac{1}{3}}\,e^{\mathcal{A}_b(k)}\,\text{Ai}\bigl[C_b^{-\frac{1}{3}}\bigl(N - B_b\bigr)\bigr] + \mathcal{O}(e^{-\sqrt{N}}) \, ,
\end{equation}
where $\text{Ai}$ is the Airy function and the expression above captures all terms perturbative in $1/N$ and receives non-trivial instanton corrections. The parameters in \eqref{eq:Z} are defined as follows:
\begin{equation}
	C_b(k,\Delta) = \frac{2}{\pi^2k}\frac{(b + b^{-1})^{-4}}{\prod_{a=1}^4 \Delta_a} \, , \qquad B_b(k,\Delta) = \frac{k}{24} + \frac{\alpha(\Delta,b)}{k} \, ,\label{CB:general}
\end{equation}
where $b$ is the squashing parameter and we have introduced the constrained quantities
\begin{equation}\label{eq:DeltaaABJM}
\begin{split}
	\Delta_1 =&\; \frac12 - \ri\,\frac{m_1 + m_2 + m_3}{b+b^{-1}} \, , \qquad \Delta_2 = \frac12 - \ri\,\frac{m_1 - m_2 - m_3}{b+b^{-1}} \, , \\
	\Delta_3 =&\; \frac12 + \ri\,\frac{m_1 + m_2 - m_3}{b+b^{-1}} \, , \qquad \Delta_4 = \frac12 + \ri\,\frac{m_1 - m_2 + m_3}{b+b^{-1}} \, ,
\end{split}
\end{equation}
such that $\sum_a \Delta_a = 2$, and
\begin{equation}\label{eq:alphabetaABJM}
	\alpha(\Delta,b) = -\frac1{12}\,\sum_{a=1}^4\Delta_a^{-1} + \frac{1 - \frac14\sum_a\Delta_a^2}{3(b+b^{-1})^2\prod_{a=1}^4 \Delta_a} \, .
\end{equation}
The function $\mathcal{A}_b(k)$ in \eqref{eq:Z} for general $(m_i,b)$ is currently not known but its form has been obtained in various limits. See \cite{Marino:2011eh,Hatsuda:2012dt,Nosaka:2015iiw,Hatsuda:2016uqa,Chester:2021gdw} for more details on the calculation of the ABJM $S^3$ partition function for different values of the parameters $(m_i,b)$.

Once we introduce the superpotential mass term and flow to the IR, the $\SO(8)$ global symmetry of the ABJM theory is broken to $\SU(3)\times \U(1)$ which in turn restricts the allowed values of the real mass parameters in the following way
\begin{equation}\label{eq:mABJMdelta}
\Delta_1=1\,, \qquad \Delta_2+\Delta_3+\Delta_4=1\,.
\end{equation}
Fixing $\Delta_1=1$ is a simple manifestation of the fact that this real mass parameter is no longer present once we integrate out one of the chiral superfield. The other two mass parameters are associated with the $\SU(3)$ flavor symmetry of the mABJM theory in the IR. We conjecture that the all-order $1/N$ expansion of the mABJM $S^3$ partition function can be obtained from \eqref{eq:Z}-\eqref{eq:alphabetaABJM} by restricting the parameters $\Delta_a$ as in \eqref{eq:mABJMdelta}. Note that the mABJM theory preserves the full $\mathcal{N}=2$ superconformal invariance when the real masses vanish, i.e. when $\Delta_{2}=\Delta_{3}=\Delta_{4}=1/3$. These values of $\Delta_a$ can be obtained either by a direct inspection of the superpotential or by using $F$-maximization.

We now proceed with the TTI. For the ABJM TTI the all-order $1/N$ expression was found in \cite{Bobev:2022jte,Bobev:2022eus} and reads
\begin{align}
\label{TTI:exact}
	F_{S^1\times \Sigma_\mathfrak{g}} &\!= \fft{\pi\sqrt{2k\Delta_1\Delta_2\Delta_3\Delta_4}}{3}\sum_{a=1}^4\fft{\mn_a}{\Delta_a}\Bigl(\hat{N}_{\Delta}^\fft32 -\fft{\mathfrak c_a}{k}\hat{N}_{\Delta}^\fft12\Bigr) \nonumber \\
	&\;\;\;+\frac{1-\mathfrak{g}}{2}\log\hat{N}_{\Delta} - \hat f_0(k,\Delta,\mn)  \, ,
\end{align}
where $\mathfrak c_a$ are given by
\begin{equation}
	\mathfrak c_a=\fft{\prod_{b\neq a}(\Delta_a + \Delta_b)}{8\Delta_1\Delta_2\Delta_3\Delta_4}\,\sum_{b\neq a}\Delta_b \, , 
\end{equation}
and $\hat f_0$ is a function that does not depend on $N$ and was computed numerically for some value of the parameters in \cite{Bobev:2022jte,Bobev:2022eus}. Note that we have introduced the quantity
\begin{equation}\label{eq:Ndeltadef}
	\hat{N}_{\Delta} \equiv N-\fft{k}{24}+\fft{1}{12k}\,\sum_{a=1}^4 \Delta_a^{-1} \, ,
\end{equation}
in terms of which the TTI admits a nice compact form. This expression for the TTI was obtained numerically in \cite{Bobev:2022jte,Bobev:2022eus} where it was extensively checked with high accuracy and was shown to be valid up to $\mathcal{O}(e^{-\sqrt{N}})$ corrections. 

The TTI for the mABJM theory can be obtained from the expression in~\eqref{TTI:exact} by taking $k=1,2$ and restricting the fugacities $\Delta_a$ to obey \eqref{eq:mABJMdelta} and in addition constraining the magnetic fluxes to obey
\begin{equation}\label{eq:mnamABJM}
\mn_1=1-\mg\,, \qquad \mn_2+\mn_3+\mn_4=1-\mg\,.
\end{equation}
Note that for the ABJM theory there is a single constraint on the magnetic fluxes determining the TTI in \eqref{TTI:exact}, namely~$\sum_{a}\mn_a=2(1-\mg)$, which is now modified since the flavor symmetry is partially broken in the mABJM model which in turn fixes $\mn_1=(1-\mg)$. The superconformal configuration for the mABJM TTI is obtained by setting
\begin{equation}
\Delta_{2}=\Delta_{3}=\Delta_{4}=\frac{1}{3}\,, \qquad \mn_2=\mn_3=\mn_4=\frac{(1-\mg)}{3}\,.
\end{equation}
It is possible to obtain these expression for the mABJM TTI by simply taking limits of the ABJM TTI in~\eqref{TTI:exact} because the supersymmetric localization formula for the TTI depends only on the matter content, R-charges, and flavor fluxes of the theory and is thus to a large extent similar for two SCFTs related by a supersymmetric RG flow.

We end our discussion of the mABJM theory by pointing out that based on the recent results in \cite{Bobev:2022wem} we can also obtain some explicit results for the SCI of this model. It was shown in \cite{Bobev:2022wem} that the ABJM SCI in the Cardy-like limit of small angular fugacity $\omega$ takes the following explicit form\footnote{See \cite{Bobev:2022wem} for mode details on the definitions and conventions used to define the SCI and the corresponding Cardy-like limit.}
\begin{align}\label{eq:SCImABJM}
		&\log\mathcal{I}_\text{ABJM}(N,k,\omega,\Delta,\mn)\nn\\
		&=-\fft{2}{\omega}\left[\fft{\pi\sqrt{2k\Delta_1\Delta_2\Delta_3\Delta_4}}{3}\hat N_{k,\Delta}^\fft32+\hat g_0(k,\Delta)\right]\\
		&\quad+\left[-\fft{\pi\sqrt{2k\Delta_1\Delta_2\Delta_3\Delta_4}}{3}\sum_{a=1}^4\fft{\mn_a}{\Delta_a}\left(\hat N_{k,\Delta}^\fft32-\fft{\mathfrak{c}_a(\Delta)}{k}\hat N_{k,\Delta}^\fft12\right)-\fft12\log\hat N_{k,\Delta}+\hat f_0(k,\Delta,\mn)\right]\nn\\
		&\quad+\mathcal O(e^{-\sqrt{N}})+\mathcal O(\omega)\,.\nn
\end{align}
Based on the discussion above we the SCI for the mABJM theory can be obtained from this expression by fixing $k=1,2$ and restricting $\Delta_a$ and $\mn_a$ as in \eqref{eq:mABJMdelta} and~\eqref{eq:mnamABJM}, respectively.

The expressions for the partition functions of the mABJM theory presented above can be compared with available results in the literature. The leading $N^{\fft32}$ behavior of both the TTI and the $S^3$ partition function for general values of $\Delta_a$ and $\mn_a$ agree with the holographic calculations in \cite{Jafferis:2011zi,Bobev:2018uxk,Bobev:2018wbt}. For the SCI, the holographic result is only available for the so-called universal limit corresponding to the superconformal configuration, see \cite{Bobev:2019zmz}, and it agrees with the corresponding limit of~\eqref{eq:SCImABJM}. From the perspective of the matrix model for the TTI and SCI it is clear from the analysis of \cite{Bobev:2022jte,Bobev:2022eus,Bobev:2022wem} that the results for the mABJM theory can be obtained as a limit of the ABJM theory results. Finally, as we discuss in Section~\ref{sec:holo:SUGRAtoSCFT} below, the $N^{\fft12}$ terms in the expansion of all three partition functions of the mABJM theory for the superconformal configuration are consistent with the 4d minimal gauged supergravity higher-derivative holographic analysis of \cite{Bobev:2020egg,Bobev:2021oku}.

%%%%%
\section{Holography}\label{sec:holo}
%%%%%
In this section we discuss some holographic implications of our all-order perturbative $1/N$ expansions for the TTI of the various holographic SCFTs studied above. In Section \ref{sec:holo:SCFTtoSUGRA} we discuss the holographic dual description of the all-order TTI in terms of the Euclidean M-theory path integral and the BPS magnetic Reissner-Nordstr\"om AdS$_4$ black hole entropy. In Section~\ref{sec:holo:SUGRAtoSCFT} we explain how our TTI results can be combined with the recent analysis of the 4-derivative corrections to $\mathcal N=2$ minimal gauged supergravity \cite{Bobev:2020egg,Bobev:2021oku} to yield results for the subleading corrections to other types of SCFT partition functions in the large $N$ limit.

%%%%%
\subsection{Holographic duals of the TTI}\label{sec:holo:SCFTtoSUGRA}
%%%%%
For simplicity, we will first focus on the superconformal and universal configurations of flavor chemical potentials and magnetic fluxes to discuss holographic duals of the TTI. In this special case, the real part of the all-order TTI for all examples studied above simplify to
\begin{align}
\label{TTI:sc}
	\fft{\log Z^\text{ADHM}_{S^1\times\Sigma_{\mg}}(N,N_f)}{\mg-1}&=\fft{\pi\sqrt{2N_f}}{3}\Bigg[\bigg(N+\fft{7N_f}{24}+\fft{1}{3N_f}\bigg)^\fft32-\bigg(\fft{N_f}{2}+\fft{5}{2N_f}\bigg)\bigg(N+\fft{7N_f}{24}+\fft{1}{3N_f}\bigg)^\fft12\Bigg]\nn\\
	&\quad+\fft12\log\bigg(N+\fft{7N_f}{24}+\fft{1}{3N_f}\bigg)-\hat f_0(N_f)-\hat f_\text{np}(N,N_f)\,,\nn\\[1mm]
	\fft{\log Z^{N^{0,1,0}}_{S^1\times\Sigma_{\mg}}(N,k)}{\mg-1}&=\fft{4\pi\sqrt{k}}{3\sqrt3}\Bigg[\bigg(N+\fft{5k}{48}+\fft{1}{3k}\bigg)^\fft32-\bigg(\fft{k}{4}+\fft{5}{4k}\bigg)\bigg(N+\fft{5k}{48}+\fft{1}{3k}\bigg)^\fft12\Bigg]\nn\\
	&\quad+\fft12\log\bigg(N+\fft{5k}{48}+\fft{1}{3k}\bigg)-\hat f_0(k)-\hat f_\text{np}(N,k)\,,\\[1mm]
	\fft{\log Z^{V^{5,2}}_{S^1\times\Sigma_{\mg}}(N,N_f)}{\mg-1}&=\fft{16\pi\sqrt{N_f}}{27}\Bigg[\bigg(N+\fft{N_f}{6}+\fft{1}{4N_f}\bigg)^\fft32-\bigg(\fft{9N_f}{16}+\fft{27}{16N_f}\bigg)\bigg(N+\fft{N_f}{6}+\fft{1}{4N_f}\bigg)^\fft12\Bigg]\nn\\
	&\quad+\fft12\log\bigg(N+\fft{N_f}{6}+\fft{1}{4N_f}\bigg)-\hat f_0(N_f)-\hat f_\text{np}(N,N_f)\,,\nn\\[1mm]
	\fft{\log Z^{Q^{1,1,1}}_{S^1\times\Sigma_{\mg}}(N,N_f)}{\mg-1}&=\fft{4\pi\sqrt{N_f}}{3\sqrt3}\Bigg[\bigg(N+\fft{N_f}{6}\bigg)^\fft32-\bigg(\fft{N_f}{4}+\fft{3}{4N_f}\bigg)\bigg(N+\fft{N_f}{6}\bigg)^\fft12\Bigg]\nn\\
	&\quad+\fft12\log\bigg(N+\fft{N_f}{6}\bigg)-\hat f_0(N_f) - \hat f_\text{np}(N,N_f)\,.\nn
\end{align}
Note that we have fixed $r=k$ for the $N^{0,1,0}$ TTI  and have used the relation (\ref{ADHM:S2:Riemann}) to generalize the previous results from $S^1\times S^2$ to $S^1\times\Sigma_{\mg}$. Recall that we do not consider the case $\mg = 1$, which should be treated separately. Our goal now is to discuss a holographic interpretation for the results in (\ref{TTI:sc}). \\

The AdS/CFT correspondence in the context of M2-branes leads to a duality between a 3d CFT describing the low-energy physics of the M2-branes at the tip of the cone over a 7d internal manifold $Y_7$ on one hand, and M-theory on a dual asymptotically AdS$_4\times_\text{f} Y_7$ background\footnote{The notation ``$\times_\text{f}$'' emphasizes that the 11d solution is not necessarily a direct product but may in general involve a non-trivial fibration.} obtained in the near horizon limit of the supergravity solution sourced by the M2-branes, on the other. The 3d SCFTs whose TTI has been analyzed in Section~\ref{sec:TTI} correspond to the low-energy effective worldvolume theories of $N$ M2 branes on the cone over the 7d Sasaki-Einstein manifolds $S^7/\mathbb{Z}_{N_f}$, $N^{0,1,0}/\mathbb{Z}_k$, $V^{5,2}/\mathbb{Z}_{N_f}$ and $Q^{1,1,1}/\mathbb{Z}_{N_f}$. On the M-theory side, the dual 11d Euclidean backgrounds have the following metric and 4-form flux, see for example \cite{Gauntlett:2007ma,Bobev:2022eus}:
\begin{equation}
	\begin{split}
		ds^2_{11}&=\fft{L^2}{4}ds^2_4+L^2ds^2_6+L^2\left(d\psi+\sigma+\fft14A\right)^2 \, ,\\
		G&=\fft{3L^3}{8}\,\text{vol}_4-\fft{L^3}{4}\,J\wedge *_4 F \, ,
	\end{split}\label{11d:to:4d:E}
\end{equation}
where $ds^2_6$ is a locally K\"ahler-Einstein metric such that $R^{(6)}_{\mu\nu}=8g^{(6)}_{\mu\nu}$ and $d\sigma = 2J$ with the K\"ahler form $J$ normalized as $J^3 = 3!\,\text{vol}_6$. In (\ref{11d:to:4d:E}) we have kept the original convention of the Hodge star in Lorentzian signature. Some known facts about the relevant 7d Sasaki-Einstein geometries are collected in Appendix~\ref{App:SE}.

The holographic dual of the TTI with the superconformal configuration \eqref{TTI:sc} is given by the Euclidean M-theory path integral around the 11d background (\ref{11d:to:4d:E}) where the 4d external space corresponds to the Euclidean Romans background of $\mathcal{N}=2$ minimal gauged supergravity \cite{Romans:1991nq,BenettiGenolini:2019jdz,Bobev:2020pjk}. The field strength $F=dA$ and the metric for this background take the form\footnote{We have $(ds_4^2)^\text{here}=L^{-2}(ds_4^2)^\text{there}$ and $(r,\tau,Q,A)^\text{here}=L^{-1}(r,\tau,q,A)^\text{there}$ compared to the convention of \cite{Bobev:2022eus}.}
\begin{equation}
	\begin{split}
		ds_4^2&=U(r)d\tau^2+U(r)^{-1}dr^2+r^2ds^2_{\Sigma_{\mg}}\,,\\
		U(r)&=\left(r+\fft{\kappa}{2r}\right)^2-\fft{Q^2}{4r^2}\,,\\
		F&=\fft{Q}{r^2}d\tau\wedge dr\pm\kappa\,\text{vol}_{\Sigma_{\mg}}\,.
	\end{split}\label{ext-4d:E}%
\end{equation}
Here $\kappa=1,0,-1$ is the normalized curvature of the Riemann surface $\Sigma_\mg$ with $\mg=0$, $\mg=1$, and $\mg>1$, respectively. The Euclidean solution above admits a Lorentzian black hole interpretation upon Wick rotating and taking $Q \rightarrow 0$. This limit produces a regular Lorentzian solution only when $\kappa = -1$~\cite{Bobev:2020pjk}. In this case, one obtains a magnetically charged asymptotically AdS$_4$ supersymmetric black hole with fixed charge and a hyperbolic horizon.

To leading order in $N$, the Euclidean M-theory path integral around the 11d background \eqref{11d:to:4d:E} can be approximated in terms of the regularized two-derivative on-shell action of the 4d background \eqref{ext-4d:E}. The latter can be computed in terms of the scale $L$ and the 4d Newton constant~\cite{Azzurli:2017kxo}
\begin{equation}
\label{eq:grav-2der}
	I_{2\partial}^{\text{reg}} = (1 - \mg)\,\frac{\pi L^2}{2 G_N} \, .
\end{equation}
Notably this result is independent of the electric charge $Q$, or equivalently of the periodicity $\beta$ of the coordinate $\tau$, and is valid both for the family of Euclidean solutions in \eqref{ext-4d:E} and the Lorentzian supersymmetric black hole solution. We can now use the standard holographic dictionary to relate the gravitational quantities to the field theory parameters. To leading two-derivative order, this yields
\begin{equation}
\label{eq:I-SCFT}
	I_{2\partial}^{\text{reg}} = (1-\mg)\sqrt{\fft{2\pi^6}{27\,\text{vol}[Y_7]}}\,N^\fft32 + o(N^\fft32) \, .
\end{equation}
Using the volumes computed in Appendix~\ref{App:SE}, we find perfect agreement between the TTI \eqref{TTI:sc} and the Euclidean M-theory path integral in the large $N$ limit for the various Sasaki-Einstein spaces $Y_7$ 
\begin{equation}
	\log Z^\text{SCFT}_{S^1\times\Sigma_{\mg}}=(\mg-1)\sqrt{\fft{2\pi^6}{27\,\text{vol}[Y_7]}}\,N^\fft32 + o(N^\fft32)= - I_{2\partial}^{\text{reg}} \, .\label{TTI=M-PI}
\end{equation}
Note that since the imaginary part of $\log Z^\text{SCFT}_{S^1\times\Sigma_{\mg}}$ is defined modulo $2\pi\mathbb Z$ it does not affect the leading term in the large $N$ expansion.

At finite $N$, the field theory TTI receives perturbative $1/N$ corrections as studied in detail in Section~\ref{sec:TTI}. On the gravity side, the Euclidean M-theory path integral around the 11d background \eqref{11d:to:4d:E} receives higher-derivative and loop corrections to the saddle point approximation \eqref{TTI=M-PI}. In addition, the holographic dictionary receives quantum corrections. These quantum gravitational effects should combine to account for the subleading $o(N^\fft32)$ terms in \eqref{eq:I-SCFT}. Therefore, our field theory results for the TTI in \eqref{TTI:sc} can be interpreted as a prediction for the full set of perturbative $1/N$ corrections to the M-theory path integral. It is worth noting that there has been recent progress in computing the first perturbative correction, at order $N^{\fft12}$, directly in supergravity by including four-derivative terms. This was explained for both the ABJM and ADHM theories in~\cite{Bobev:2020egg,Bobev:2021oku}. In Section~\ref{sec:holo:SUGRAtoSCFT} below we discuss how the same four-derivative supergravity analysis can be extended to more general $\mathcal N=2,3$ SCFTs. The universal $\frac12\log N$ logarithmic correction to the TTI has been obtained from a one-loop bulk computation in~\cite{Liu:2017vbl,Liu:2017vll} for the ABJM theory, i.e. for $S^7$ as internal manifold. It would of course be very interesting to understand how to include higher-derivative couplings and higher loop orders in the 11d supergravity path integral in order to reproduce the all-order perturbative $1/N$ expansions in \eqref{TTI:sc}. This appears to be a formidable task at present. Finally, we would like to point out that the universal relation between the sphere free energy and the TTI for 3d $\mathcal{N}=2$ holographic SCFTs observed in~\cite{Azzurli:2017kxo,Bobev:2017uzs} for the leading term in the large $N$ expansion is no longer valid when one includes also the subleading terms in~\eqref{TTI:sc}.

As mentioned below \eqref{ext-4d:E}, the Euclidean Romans background admits a continuation to a Lorentzian magnetic BPS AdS$_4$ charged black hole for a suitable choice of parameters. In this context, our results \eqref{TTI:sc} can be used to obtain all the perturbative corrections to the Bekenstein-Hawking entropy of the black hole. Since this background is a solution of minimal gauged supergravity there is no need to change ensembles or invoke $\mathcal{I}$-extremization in order to relate the Euclidean on-shell action (or equivalently the TTI) to the black hole entropy, see \cite{Benini:2015eyy,Benini:2016rke}. We therefore conclude that AdS/CFT dictates the following result for the black hole entropy $\mathcal{S}$ 
\begin{equation}
	 \mathcal{S} = \Re\log Z^\text{SCFT}_{S^1\times\Sigma_{\mg}} \, , \label{Z=S:TTI}
\end{equation}
where the perturbative part of the right sand side is given by \eqref{TTI:sc} to all orders in the $1/N$ expansion. Through the holographic dictionary, this translates to an infinite set of perturbative corrections to the Bekenstein-Hawking entropy contained in the left hand side. \\

To go beyond the superconformal values of TTI parameters, corresponding to minimal 4d $\mathcal{N}=2$ gauged supergravity, one has to turn on suitable deformations in the SCFTs that are captured by supergravity vector and hyper multiplets. In the holographic context we are then faced with the hard problem of either finding appropriate 4d consistent truncations of 11d supergravity that include these additional multiplets or construct the corresponding supergravity backgrounds directly in 11d. For the four 11d AdS$_4$ vacua discussed in this work we do not know how to solve this problem in general. From the 4d perspective there are no known consistent truncations that capture all relevant deformations, see e.g.~\cite{Gauntlett:2007ma,Cassani:2011fu,Cassani:2012pj}, while constructing 11d generalizations of the background in \eqref{11d:to:4d:E} and \eqref{ext-4d:E} to include these additional deformations is prohibitively hard since one has to solve coupled system of PDEs. This state of affairs should be contrasted with the situation for the ABJM theory where there is a truncation of 11d supergravity on $S^7$ to 4d maximal gauged supergravity which can be exploited to construct supersymmetric Lorentzian black hole solutions with non-trivial profiles for fields in vector and hyper multiplets, see for example \cite{Benini:2015eyy,Benini:2016rke,Bobev:2018uxk}, as well as more general Euclidean supersymmetric solutions that are dual to the general ABJM TTI to leading order in the large $N$ limit, see \cite{Bobev:2020pjk}. Note also that the continuation of Euclidean supergravity backgrounds to Lorentzian black hole solutions that are free of pathologies is subtle in matter-coupled gauged supergravity theories. As shown in \cite{Bobev:2018uxk,Bobev:2020pjk}, in the context of 4d $\mathcal{N}=8$ gauged supergravity the space of Euclidean supergravity solutions that capture the $(\Delta,\mn)$-dependent TTI is larger than the space of known supersymmetric Lorentzian black holes. It is therefore not immediately clear which 11d black holes can be studied using the TTI of the 3d $\mathcal{N}\geq2$ SCFTs discussed above away from the superconformal point.

Even if these difficulties are overcome and more general supergravity backgrounds dual to the $(\Delta,\mn)$-dependent TTI for the four holographic SCFTs we studied are constructed, one should consider the additional subtlety of specifying a choice of ensemble. The entropy of Lorentzian black holes is usually formulated in the microcanonical ensemble of fixed charges while the TTI is written in the ensemble with fixed flavor chemical potentials. One must therefore further implement an inverse Laplace transform on the $(\Delta,\mn)$-dependent TTIs with respect to $\Delta$ in order to obtain the perturbative corrections to the Bekenstein-Hawking entropy of the dual black holes. At the leading $\mathcal{O}(N^\fft32)$ order, this change of ensemble has been implemented for the ABJM TTI and dubbed ``$\mathcal{I}$-extremization'', and the resulting expression matches the dual AdS$_4$ black hole entropy precisely~\cite{Benini:2015eyy,Benini:2016rke}.\footnote{See~\cite{Bobev:2018uxk,Zaffaroni:2019dhb} for a more general discussion of this change of ensemble in the large $N$ limit.} Going beyond this leading order implementation of the inverse Laplace transform is still an open problem. Nevertheless, we emphasize that the TTI results presented in this paper should provide a very useful guide to understand how to overcome part (or all) of the difficulties summarized above in computing the corrections to the semi-classical entropy of AdS$_4$ black holes in M-theory using holography.

%%%%%
\subsection{Subleading corrections to SCFT partition functions from holography}
\label{sec:holo:SUGRAtoSCFT}
%%%%%

The interplay between four-derivative couplings in 4d $\mathcal N=2$ minimal gauged supergravity and subleading corrections to the large $N$ limit of SCFT partition functions has been recently used to great efficacy in the ABJM and ADHM theories~\cite{Bobev:2020egg,Bobev:2021oku}. Here we extend the same analysis to more general $\mathcal N=2,3$ holographic SCFTs. We also discuss the application of these results to partition functions on the so-called spindles, see \cite{Ferrero:2020twa} and references thereof. \\

The main result of the four-derivative supergravity analysis in~\cite{Bobev:2020egg,Bobev:2021oku} is that the large $N$ partition function of a 3d holographic SCFT on any compact 3d manifold $M_3$ can be written in the form
\begin{equation}
	-\log Z^\text{SCFT}_{M_3}=\pi\mF\left(AN^\fft32+BN^\fft12\right)-\pi(\mF-\chi)CN^\fft12 + o(N^\fft12) \,,\label{Z:M3}
\end{equation}
where $\mF$ and $\chi$ are the two-derivative regularized on-shell action\footnote{We normalize this action by dividing it by $\pi L^2/(2G_N)$ so that for empty Euclidean AdS$_4$ with an $S^3$ boundary we have $\mathcal{F}=1$.} and the Euler number of the dual 4d Euclidean background whose conformal boundary corresponds to $M_3$. We list some examples of $\mF$ and $\chi$ in Table~\ref{table:Fchi}. Note that the metric for the Kerr-Newman-AdS black hole can be found in~\cite{Caldarelli:1998hg}, that of the AdS$_4$-Taub-NUT space is presented in Section 3.2.3 of \cite{Bobev:2021oku}, and the data for the black hole with spindle horizon can be found in~\cite{Cassani:2021dwa}.\footnote{Compared to~\cite{Cassani:2021dwa}, we have rescaled $\omega^\text{there} = -2\pi \mathrm{i}\omega^\text{here}$ and used the lower sign in their result for $\mF$.}
\begin{table}
	\centering
	\renewcommand*{\arraystretch}{1.3}
	\begin{tabular}{|c||c|c|}
		\hline
		4d Euclidean geometry & $\mF$ & $\chi$ \\\hhline{|=||=|=|}
		AdS$_4$ w. $S^3$ bdry & $1$ & $1$ \\\hline
		Taub-NUT-AdS$_4$ w. $S^3_b$ bdry & $\frac{1}{4}\left(b+b^{-1}\right)^2$ & $1$ \\\hline
		Romans \eqref{ext-4d:E} & $1-\mg$ & $2(1-\mg)$  \\\hline
		KN-AdS & $\fft{(\omega+1)^2}{2\omega}$ & $2$  \\\hline
		BH with spindle horizon & $\fft{1}{8\omega}\bigl[\bigl(\fft{1}{n_+}+\fft{1}{n_-}\bigr)\,\omega+2\bigr]^2+\fft{\omega}{8}\bigl[\fft{1}{n_+}-\fft{1}{n_-}\bigr]^2$ & $\fft{1}{n_+}+\fft{1}{n_-}$  \\\hline
	\end{tabular}
	\caption{The quantities $\mF$ and $\chi$ for various Euclidean 4d minimal gauged supergravity backgrounds. The spindle black hole is characterized by two integers $n_\pm$ that specify the topology of the horizon. For $n_+ = n_- = 1$, we recover the KN-AdS black hole.}
	\label{table:Fchi}
\end{table}

The information about the details of the holographic SCFT enters the result in \eqref{Z:M3} through the $(A,B,C)$ coefficients. For a given SCFT, these coefficients can be deduced from the large $N$ expansion of any two distinct physical observables that can be written as in \eqref{Z:M3}. From this, one can deduce the first subleading correction to the SCFT partition function on any 3-manifold using \eqref{Z:M3} and the quantities $\mF$ and $\chi$ computed from the bulk supergravity. This strategy was used in~\cite{Bobev:2020egg,Bobev:2021oku} where the coefficients $(A,B,C)$ for the ABJM and ADHM theories were fixed using the known large $N$ expansions of the partition functions on the round and squashed 3-spheres.

Using the all-order results for the TTI we have obtained in this paper, we can now extend this analysis to other SCFTs. For the $N^{0,1,0}$ $\mathcal{N}=3$ theory, we can combine the large $N$ expansion of \eqref{TTI:sc} and the $S^3$ partition function in \eqref{N010:S3} to read off the $(A,B,C)$ coefficients. For the mABJM theory discussed in Section~\ref{sec:mABJM} we can use the $S^3$ partition function and the TTI in a similar fashion to find $(A,B,C)$. Note that these values are compatible with \eqref{Z:M3} and the SCI for the superconformal configuration of the mABJM theory computed using \eqref{eq:SCImABJM}. The results for $(A,B,C)$ for these SCFTs are summarized in Table~\ref{table:ABC}. 

As explained in Section~\ref{sec:S3:others}, we currently do not have access to the subleading corrections in the $S^3$ partition functions of the $V^{5,2}$ and $Q^{1,1,1}$ theories, so we cannot apply the above strategy to uniquely fix their $(A,B,C)$ coefficients. It would be most interesting to derive such corrections and combine them with our \eqref{TTI:sc} to obtain the coefficients entering \eqref{Z:M3} for these $\mathcal{N}=2$ theories. In the absence of such a calculation we note that for $V^{5,2}$ and $Q^{1,1,1}$ we can use the results for the TTI in \eqref{TTI:sc} to find the linear combination $B+C$ by reading off the coefficient of the $N^{\fft12}$ term in the large $N$ expansion and using \eqref{Z:M3} with $\chi=2\mathcal{F}=2(1-\mathfrak{g})$. This results in the following values 
\begin{equation}
\begin{split}
V^{5,2}:& \qquad~~~ A = \frac{16}{27}\sqrt{N_f}\,, \qquad B+C = -\frac{5N_f^2+21}{27\sqrt{N_f}} \,,\\
Q^{1,1,1}:& \qquad~~~A = \frac{4}{3\sqrt{3}}\sqrt{N_f}\,, \qquad B+C = -\frac{1}{\sqrt{3N_f}}\,.
\end{split}
\end{equation}

It was recently observed that the Cardy-like expansion $\omega \rightarrow \mathrm{i}0^+$ of the SCI for the ABJM and ADHM theories is controlled by the so-called Bethe potential and the TTI~\cite{Bobev:2022wem}. It is highly likely that more general $\mathcal N=2$ SCIs have the same property, and we plan to study this generalization soon \cite{SCI-WIP}. At this stage, it is worth pointing out that the $N^{\frac{1}{2}}$ term in the $N^{0,1,0}$ SCI obtained by substituting the $(A,B,C)$ coefficients and the $(\mF,\chi)$ for the dual KN-AdS background into the formula \eqref{Z:M3} is already in perfect agreement with this expectation.

Finally, we note that our results lead to a prediction for the $N^{\frac{1}{2}}$ term in the partition function of the holographic SCFTs discussed here when they are placed on the recently discussed spindle geometries, see \cite{Ferrero:2020twa} and references thereof. To obtain these results one has to use the expression in \eqref{Z:M3} with the appropriate $(\mathcal{F},\chi)$ characterizing the spindle geometry, see Table~\ref{table:Fchi}, and the values of $(A,B,C)$ from Table~\ref{table:ABC} for the SCFT of interest. It will be most interesting to reproduce this supergravity prediction for the $N^{\frac{3}{2}}$ and $N^{\frac{1}{2}}$ terms in the spindle free energy by an independent calculation in the dual SCFTs based on the recently proposed localization formula in~\cite{Inglese:2023wky}.

\begin{table}
	\centering
	\renewcommand*{\arraystretch}{1.3}
	\begin{tabular}{|c||c|c|c|c|}
		\hline
		SCFT & SUSY &$A$ & $B/A$ & $C/A$ \\\hhline{|=||=|=|=|=|}
		ABJM & $\mathcal{N}=6$ & $\fft{\sqrt{2k}}{3}$ & $-\fft{k}{16}-\fft{1}{2k}$ & $-\fft{3}{2k}$ \\\hline
		ADHM & $\mathcal{N}=4$ & $\fft{\sqrt{2N_f}}{3}$ & $\fft{3N_f}{16}-\fft{3}{4N_f}$ & $-\fft{N_f}{4}-\fft{5}{4N_f}$ \\\hline
		$N^{0,1,0}$ & $\mathcal{N}=3$ & $\fft{2(k+r)}{3\sqrt{2k+r}}$ & $\fft{3r-2k}{32}-\fft{k}{2(k+r)^2}$ & $-\fft{r}{8}-\fft{3k+2r}{2(k+r)^2}$ \\\hline
		%deformed ABJM & $\fft{4\sqrt{2k\Delta_1\Delta_2\Delta_3\Delta_4}}{3}$ & $-\fft{k}{16}+\fft{1}{8k}\Big(\sum_{a=1}^4\fft{1}{\Delta_a}-\fft{\sum_{a\neq b}\Delta_a\Delta_b}{4\Delta_1\Delta_2\Delta_3\Delta_4}\Big)$ & $-\fft{\sum_{a\neq b}\Delta_a\Delta_b}{32k\Delta_1\Delta_2\Delta_3\Delta_4}$ \\\hline
		mABJM & $\mathcal{N}=2$ & $\fft{4\sqrt{2k}}{9\sqrt3}$ & $-\fft{k}{16}-\fft1k$ & $-\fft{9}{4k}$ \\\hline
	\end{tabular}
	\caption{The coefficients $(A,B,C)$ entering \eqref{Z:M3} for various $\mathcal{N}\geq 2$ SCFTs. As explained in Section~\ref{sec:TTI:N010} one should take $r=k$ for the $N^{0,1,0}$ theory.}
	\label{table:ABC}
\end{table}
%
%%%%%%%%%%%%%%%%%
\section{Discussion}
\label{sec:discussion}
%%%%%%%%%%%%%%%%%

In this work we found closed form expressions for the all-order perturbative $1/N$ expansion for the TTI of $\mathcal{N}\geq2$ holographic SCFTs arising from M2-branes in M-theory. We focused on five examples corresponding to the four Sasaki-Einstein orbifolds $Y_7\in\{S^7/\mathbb Z_{N_f},N^{0,1,0}/\mathbb Z_k,V^{5,2}/\mathbb Z_{N_f},Q^{1,1,1}/\mathbb Z_{N_f}\}$ together with the mABJM theory which arises from a superpotential mass-deformation of the ABJM model. In all these examples the $1/N$ expansion for the TTI resums into a simple compact expression that takes a similar form to the one we recently found for the ABJM theory in \cite{Bobev:2022jte,Bobev:2022eus}. We also conjectured a closed form expression in terms of an Airy function for the large $N$ $S^3$ partition function of the ADHM theory on the squashed sphere in the presence of arbitrary real mass deformation. Our results suggest several possible generalizations and questions for future exploration some of which we discuss below. 

\begin{itemize}\setlength\itemsep{0.1em}

\item While we exhibited very strong numerical evidence for the validity of the TTI expressions presented in this work  it is very important to establish these results through more direct analytical methods. In addition to providing insights into how to calculate the TTI for large $N$ holographic SCFTs, such an analytic understanding will also shed light on how to extend our results to configurations with general values for the real mass parameters some of which we set to specific values in the examples treated above. A better analytic control of our TTI calculations will also allow to understand the role of any other solutions of the BAE. In essence, we have focused on the solution to the BAE equation that dominates in the large $N$ limit and have ignored the contributions of any other possible BAE solutions, see the discussion around \eqref{eq:calB1calB2} above. It will be most interesting to investigate the space of solutions of the TTI BAE for the holographic SCFTs studied here and to uncover their possible role in holography and black hole physics. Since such a study of the BAE seems quite non-trivial in general, focusing on gauge groups of low ranks could be helpful in order to delineate the structure of the BAE solution space, as in the case for the SCI of the 4d $\mathcal N=4$ Super-Yang-Mills theory~\cite{Lezcano:2021qbj,Benini:2021ano}.

\item To this end it is also desirable to understand the non-perturbative contributions to the TTI. Our numerical approach was sufficiently accurate to estimate the leading non-perturbative correction to the TTI but further insights will require other methods that will hopefully lead to better analytic control over the infinite series of such corrections and its possible resummation.

\item Given the conjecture for the squashed sphere partition function of the ADHM and mABJM models with real mass deformation presented in this work and the analogous results for the ABJM theory in \cite{Bobev:2022jte,Bobev:2022eus}, it is clearly desirable to develop analytic or numerical methods to compute these large $N$ observables from first principle for 3d $\mathcal{N}=2$ holographic SCFTs. Perhaps the analytical techniques studied in \cite{Gang:2019jut} together with the results in \cite{Chester:2020jay,Chester:2021gdw} can prove useful in this regard. We are also currently exploring various numerical methods to study this problem \cite{Airy-WIP}.

\item The results for the sphere partition function, TTI, and SCI discussed here and in our previous work suggest that there are closed form expressions for the large $N$ partition functions of holographic SCFTs on other compact Euclidean 3-manifolds. It will be desirable to systematically derive such results and find an underlying organizing principle. Our work, together with the results in \cite{Hosseini:2019iad,Choi:2019dfu,Hristov:2021qsw,Hristov:2022lcw,Hristov:2022plc}, should provide useful guidance in this pursuit.

\item An important feature of all results for the large $N$ partition functions discussed above is that they are naturally written in terms of a ``shifted $N$''. This strongly suggest that there is an underlying mechanism in M-theory that makes this ``shifted $N$'' the natural large $N$ expansion parameter. For the ABJM round sphere partition function a partial explanation of this shift in $N$ can be attributed to a modification of the M2-brane charge quantization prescription in the presence of the well-known topological 8-derivative term in the M-theory effective action, see \cite{Bergman:2009zh}. It is clear that this cannot be the full answer since this modified charge quantization condition does not account for the full shift in $N$ and, as discussed above, the shift in $N$ can depend on continuous parameters which suggest that it has a non-topological origin. It will be very interesting to further clarify this aspect of our results.

\item Our focus in this work was on 3d $\mathcal N=2$ holographic SCFTs in the M-theory limit. It will be interesting to explore whether similar explicit closed form expressions can be found for the partition functions of other holographic SCFTs with type IIB or IIA gravitational dual descriptions. A first step in this direction could be the careful exploration of the type IIA limit of the large $N$ observables we discussed above.
\item Our results pose several open questions in the holographic context. The most important one is to reproduce the all-order $1/N$ expansion of the TTI, SCI, and sphere partition functions by using the M-theory path integral. This would constitute a very stringent precision test of AdS/CFT beyond the large $N$ limit. This appears to be a daunting task and has currently been achieved only for the $N^\fft32$, $N^\fft12$, and $\log N$ terms in some of the examples discussed above. Given our incomplete understanding of M-theory, a promising avenue to extend these results further is to use the 4d gauged supergravity localization framework put forward in~\cite{Dabholkar:2014wpa,Hristov:2018lod,Hristov:2019xku}. We hope that our results above will provide a valuable benchmark in pursuing this open problem and will ultimately lead to a precise OSV-type conjecture for asymptotically AdS gravitational solutions. Lastly, we note that if there exist other BAE solutions that contribute to the TTI, they should also be understood holographically. Such solutions should correspond to new asymptotically AdS$_4$ saddle points whose contribution to the M-theory path integral are exponentially suppressed compared to the backgrounds considered in Section~\ref{sec:holo:SCFTtoSUGRA}. Recently the authors of~\cite{BenettiGenolini:2023rkq} identified a family of such saddle points for the ABJM SCI in the generalized Cardy-like limit, and it would be most interesting to understand the relation between these saddle points and more general BAE solutions for the TTI exploiting the map between the SCI and the TTI used in~\cite{Bobev:2022wem}. 
\end{itemize}
We hope to explore some of these questions in the near future.

%%%%%%%%%%%%%%%%%%%%%%
\section*{Acknowledgments}
%%%%%%%%%%%%%%%%%%%%%%

We are grateful to Anthony Charles, Sunjin Choi, Pieter-Jan De Smet, Kiril Hristov, Sameer Murthy, and Xuao Zhang for valuable discussions. This research is supported by the FWO
projects G003523N and G094523N. NB and JH are also supported in part by the KU Leuven C1 grant ZKD1118 C16/16/005 and by Odysseus grant G0F9516N from the FWO. VR is supported by a public grant as part of the Investissement d'avenir project, reference ANR-11-LABX-0056-LMH, LabEx LMH. NB is grateful to KITP Santa Barbara for the warm hospitality during the final stages of this project. Research at KITP is supported in part by the National Science Foundation under Grant No. NSF PHY-1748958

%%%%%%%%%%%%%%%%%%%%%%

%%%%%%%%%%%%%%%%%%
\appendix
%%%%%%%%%%%%%%%%%%

%%%%%
\section{ADHM topologically twisted index}
\label{App:ADHM}
%%%%%

Here we provide the numerical data that supports the all-order $1/N$ expansion of the ADHM TTI given in (\ref{ADHM:TTI:all}). The list of $N_f$ and $\Delta$-configurations for which we confirmed (\ref{ADHM:TTI:all}) with the $\mn$-configurations (\ref{ADHM:nas}) is given as follows:

\smallskip

\noindent\textbf{Case 1.} $\Delta=(\Delta_1,1-\Delta_1,1,\fft12,\fft12,0)$
\begin{equation}
\begin{alignedat}{3}
	N_f&\in\{1,2,3,4\}&\quad&\&&\quad \Delta_1&=\fft12\\
	N_f&\in\{1,2,3\}&\quad&\&&\quad \Delta_1&\in\bigg\{\fft38,\fft25,\fft{5}{12},\fft37\bigg\}
\end{alignedat}\label{ADHM:case1}
\end{equation}
\noindent\textbf{Case 2.} $\Delta=(\Delta_1,1-\Delta_1,1,\Delta_q,1-\Delta_q,0)$
\begin{equation}
	N_f=1\quad\&\quad(\Delta_1,\Delta_q)\in\bigg\{(\fft38,\fft25),(\fft25,\fft{5}{12}),(\fft{5}{12},\fft37),(\fft37,\fft38)\bigg\}\label{ADHM:case2}
\end{equation}
\noindent\textbf{Case 3.} $\Delta=(\Delta_1,1-\Delta_1,1,\fft12,\fft12,\Delta_m)$
\begin{equation}
	\begin{alignedat}{3}
		N_f&\in\{1,2,3\}&\quad&\&&\quad (\Delta_1,\Delta_m)&=\bigg\{(\fft38,\fft{N_f}{8}),(\fft38,\fft{N_f}{10})\bigg\}\\
		N_f&\in\{2,3\}&\quad&\&&\quad (\Delta_1,\Delta_m)&\in\bigg\{(\fft25,\fft{N_f}{10}),(\fft25,\fft{N_f}{12}),(\fft{5}{12},\fft{N_f}{12}),\\
		&&&&&\qquad(\fft{5}{12},\fft{N_f}{14}),(\fft37,\fft{N_f}{14}),(\fft37,\fft{N_f}{16})\bigg\}
	\end{alignedat}\label{ADHM:case3}
\end{equation}
\noindent\textbf{Case 4.} $\Delta=(\Delta_1,\Delta_2,2-\Delta_1-\Delta_2,\Delta_q,\Delta_1+\Delta_2-\Delta_q,\Delta_m)$
\begin{equation}
	\begin{alignedat}{3}
		N_f&=3&\quad&\&&\quad (\Delta_1,\Delta_2,\Delta_q,\Delta_m)&=(\fft47,\fft47,\fft37,-\fft{N_f}{7})\\
		N_f&\in\{2,3\}&\quad&\&&\quad (\Delta_1,\Delta_2,\Delta_q,\Delta_m)&\in\bigg\{(\fft25,\fft25,\fft25,\fft{N_f}{5}),(\fft{3}{10},\fft{5}{10},\fft15,\fft{N_f}{5})\\
		&&&&&\qquad(\fft{4}{10},\fft{5}{10},\fft{3}{10},\fft{3N_f}{20}),(\fft38,\fft{17}{40},\fft25,\fft{3N_f}{40})\bigg\}\\
		N_f&\in\{1,2,3\}&\quad&\&&\quad (\Delta_1,\Delta_2,\Delta_q,\Delta_m)&=(\fft1\pi,\fft2\pi,\fft{3}{2\pi},N_f(1-\fft{3}{\pi}))\\
		N_f&\in\{3,4\}&\quad&\&&\quad (\Delta_1,\Delta_2,\Delta_q,\Delta_m)&=(\fft1\pi,\fft2\pi,\fft{e}{2\pi},N_f(1-\fft{3}{\pi}))
	\end{alignedat}\label{ADHM:case4}
\end{equation}
For all the above configurations, we compared numerical estimates
\begin{equation}
\begin{split}
	\hat f_{3/2}^\text{(lmf)}(N_f,\Delta,\mn)&=\sum_{a=1}^4\hat f_{3/2,a}^\text{(lmf)}(N_f,\Delta,\mn)\,\tmn_a\,,\\
	\hat f_{1/2,a}^\text{(lmf)}(N_f,\Delta,\mn)&=\sum_{a=1}^4\hat f_{1/2,a}^\text{(lmf)}(N_f,\Delta,\mn)\,\tmn_a\,,
\end{split}
\end{equation}
obtained from the \texttt{LinearModelFit} (\ref{ADHM:TTI:expansion:shifted}) with the corresponding analytic expressions given in the RHS of (\ref{ADHM:TTI:expansion:shifted:coeffi}), namely
\begin{equation}
\begin{split}
	\hat f_{3/2}(N_f,\Delta,\mn)&=-\fft{\pi\sqrt{2N_f\tDelta_1\tDelta_2\tDelta_3\tDelta_4}}{3}\sum_{a=1}^4\fft{\tmn_a}{\tDelta_a}\,,\\
	\hat f_{1/2}(N_f,\Delta,\mn)&=-\fft{\pi\sqrt{2N_f\tilde\Delta_1\tilde\Delta_2\tilde\Delta_3\tilde\Delta_4}}{3}\sum_{a=1}^4\bigg(\mathfrak c_a(\tilde\Delta)N_f+\fft{\mathfrak d_a(\tilde\Delta)}{N_f}\bigg)\,\tmn_a\,,
\end{split}
\end{equation}
where $\tDelta_a,\tmn_a$ are defined in (\ref{ADHM:tilde}) and $\mathfrak{c}_a,\mathfrak{d}_a$ are defined in (\ref{ADHM:TTI:all:coeffi}). To estimate the precision of our results, we checked that the error ratios
\begin{equation}
	R_{X}(N_f,\Delta,\mn) = \fft{\hat f_X^\text{(lmf)}(N_f,\Delta,\mn)-\hat f_X(N_f,\Delta,\mn)}{\hat f_X(N_f,\Delta,\mn)}\qquad(X\in\{3/2,1/2\}) \, ,\label{ADHM:error}
\end{equation}
are very small for all the configurations listed in (\ref{ADHM:case1},\ref{ADHM:case2},\ref{ADHM:case3},\ref{ADHM:case4}). Below we provide some examples to show how small the error ratios (\ref{ADHM:error}) are, and also present numerical estimates for the constant term $\hat f_0^\text{(lmf)}(N_f,\Delta,\mn)$ in the \texttt{LinearModelFit} (\ref{ADHM:TTI:expansion:shifted}) together with the corresponding standard error $\sigma_0$. 

\medskip

\noindent\textbf{Case 1.} $\Delta_1=\fft12$ \& $\mn=(\fft12,\fft12,1,\fft12,\fft12,0)$
\begin{center}
	\footnotesize
	\begin{tabular}{ |c||c|c|c|c| } 
		\hline
		& $R_{3/2}$ & $R_{1/2}$ & $\hat f_0^\text{(lmf)}$ & $\sigma_0$ \\
		\hline\hline
		$N_f=1$ & $2.436{\times}10^{-39}$ & $5.319{\times}10^{-37}$ & $-3.0459513105331823845$ & $7.834{\times}10^{-36}$  \\
		\hline
		$N_f=2$ & $-6.565{\times}10^{-29}$ & $-1.915{\times}10^{-26}$ & $-2.8393059176753911173$ & $2.859{\times}10^{-25}$  \\
		\hline
		$N_f=3$ & $4.214{\times}10^{-25}$ & $1.188{\times}10^{-22}$ & $-3.3892805274389775678$ & $2.159{\times}10^{-21}$  \\ 
		\hline
		$N_f=4$ & $1.620{\times}10^{-22}$ & $4.069{\times}10^{-20}$ & $-4.3655284762174631267$ & $9.220{\times}10^{-19}$  \\ 
		\hline
	\end{tabular}
\end{center}
\noindent\textbf{Case 2.} $N_f=1$ \& $(\Delta_1,\Delta_q)=(\fft38,\fft25)$
\begin{center}
	\footnotesize
	\begin{tabular}{ |c||c|c|c|c| } 
		\hline
		& $R_{3/2}$ & $R_{1/2}$ & $\hat f_0^\text{(lmf)}$ & $\sigma_0$ \\
		\hline\hline
		\text{1st in }(\ref{ADHM:nas}) & $-3.594{\times}10^{-33}$ & $-7.605{\times}10^{-31}$ & $-3.1653887845699785480$ & $1.122{\times}10^{-29}$  \\
		\hline
		\text{2nd in }(\ref{ADHM:nas}) & $8.694{\times}10^{-33}$ & $1.792{\times}10^{-30}$ & $-3.1419129712303989770$ & $2.612{\times}10^{-29}$  \\
		\hline
		\text{3rd in }(\ref{ADHM:nas}) & $2.285{\times}10^{-33}$ & $4.750{\times}10^{-31}$ & $-3.1592865927006754811$ & $6.990{\times}10^{-30}$  \\ 
		\hline
		\text{4th in }(\ref{ADHM:nas}) & $1.188{\times}10^{-32}$ & $2.427{\times}10^{-30}$ & $-3.1388618752957474436$ & $3.522{\times}10^{-29}$  \\ 
		\hline
		\text{5th in }(\ref{ADHM:nas}) & $-6.909{\times}10^{-34}$ & $-1.449{\times}10^{-31}$ & $-3.1623376886353270145$ & $2.113{\times}10^{-30}$  \\ 
		\hline
	\end{tabular}
\end{center}
\noindent\textbf{Case 3.} $N_f=2$ \& $(\Delta_1,\Delta_m)=(\fft{5}{12},\fft{N_f}{14})$
\begin{center}
	\footnotesize
	\begin{tabular}{ |c||c|c|c|c| } 
		\hline
		& $R_{3/2}$ & $R_{1/2}$ & $\hat f_0^\text{(lmf)}$ & $\sigma_0$ \\
		\hline\hline
		\text{1st in }(\ref{ADHM:nas}) & $8.370{\times}10^{-24}$ & $2.381{\times}10^{-21}$ & $-2.9294387406755353595$ & $3.385{\times}10^{-20}$  \\
		\hline
		\text{2nd in }(\ref{ADHM:nas}) & $1.364{\times}10^{-23}$ & $4.159{\times}10^{-21}$ & $-2.6816709273752314462$ & $5.589{\times}10^{-20}$  \\
		\hline
		\text{3rd in }(\ref{ADHM:nas}) & $1.359{\times}10^{-23}$ & $3.802{\times}10^{-21}$ & $-3.0545474608660212803$ & $5.577{\times}10^{-20}$  \\ 
		\hline
		\text{4th in }(\ref{ADHM:nas}) & $4.598{\times}10^{-24}$ & $1.351{\times}10^{-21}$ & $-2.7562653234541241733$ & $1.832{\times}10^{-20}$  \\ 
		\hline
	\end{tabular}
\end{center}
\noindent\textbf{Case 4.} $N_f=3$ \& $(\Delta_1,\Delta_2,\Delta_q,\Delta_m)=(\fft1\pi,\fft2\pi,\fft{3}{2\pi},N_f(1-\fft3\pi))$
\begin{center}
	\footnotesize
	\begin{tabular}{ |c||c|c|c|c| } 
		\hline
		& $R_{3/2}$ & $R_{1/2}$ & $\hat f_0^\text{(lmf)}$ & $\sigma_0$ \\
		\hline\hline
		\text{1st in }(\ref{ADHM:nas}) & $-2.023{\times}10^{-18}$ & $-4.727{\times}10^{-16}$ & $-4.4813859779853576433$ & $8.868{\times}10^{-15}$  \\
		\hline
		\text{2nd in }(\ref{ADHM:nas}) & $-1.879{\times}10^{-18}$ & $-4.836{\times}10^{-16}$ & $-3.5203924078852770102$ & $7.756{\times}10^{-15}$  \\
		\hline
		\text{3rd in }(\ref{ADHM:nas}) & $-2.948{\times}10^{-18}$ & $-6.613{\times}10^{-16}$ & $-4.7777616683409756638$ & $1.286{\times}10^{-14}$  \\ 
		\hline
		\text{4th in }(\ref{ADHM:nas}) & $-1.710{\times}10^{-19}$ & $-4.255{\times}10^{-17}$ & $-3.6883756122275059505$ & $6.527{\times}10^{-16}$  \\ 
		\hline
		\text{5th in }(\ref{ADHM:nas}) & $-2.484{\times}10^{-18}$ & $-5.685{\times}10^{-16}$ & $-4.6295738231631666536$ & $1.086{\times}10^{-14}$  \\ 
		\hline
	\end{tabular}
\end{center}
%

%%%%%
\section{$N^{0,1,0}$ topologically twisted index}
\label{App:N010}
%%%%%
Here we provide numerical data that supports the all-order $1/N$ expansion of the $N^{0,1,0}$ TTI given in (\ref{N010:TTI:all}). The list of $(k,r)$-configurations for which we confirmed (\ref{N010:TTI:all}) with $r_1=r_2=r/2$ is given as follows: 
\begin{equation}
\begin{split}
	k\in\{1,2,3,4\}\,,\qquad \fft{r}{k}\in\left\{\fft12,\fft23,1,\fft32,2,3\right\}\,.
\end{split}\label{N010:case}
\end{equation}
Note that the choice $r_1=r_2=r/2$ is crucial to obtain the all-order results (\ref{N010:TTI:all}) since the subleading contribution depends on both parameters $(r_1,r_2)$ in general unlike the $N^\fft32$ leading order (\ref{N010:TTI:N32}) completely determined by their sum $r=r_1+r_2$. For the above listed $(k,r)$-configurations, we compared numerical estimates
\begin{equation}
	\hat f_{3/2}^\text{(lmf)}(k,r)\,,\qquad\hat f_{1/2}^\text{(lmf)}(k,r)\,,\label{N010:TTI:coeffi:num}
\end{equation}
obtained from the \texttt{LinearModelFit} for the $N^{0,1,0}$ TTI similar to the one for the ADHM TTI (\ref{ADHM:TTI:expansion:shifted}), namely
\begin{equation}
	\log Z^{N^{0,1,0}}_{S^1\times S^2}(N,k,r)+\fft12\log\hat N_{k,r}=\hat f^\text{(lmf)}_{3/2}(k,r)(\hat N_{k,r})^\fft32+\hat f^\text{(lmf)}_{1/2}(k,r)(\hat N_{k,r})^\fft12+\hat f^\text{(lmf)}_0(k,r)\,,\label{N010:TTI:expansion:shifted}
\end{equation}
with the corresponding analytic expressions in (\ref{N010:TTI:all}),
\begin{equation}
	\begin{split}
		\hat f_{3/2}(k,r)&=-\fft{2\pi(k+r)}{3\sqrt{2k+r}}\,,\\
		\hat f_{1/2}(k,r)&=\fft{2\pi(k+r)}{3\sqrt{2k+r}}\left(\fft{r}{4}+\fft{3k+2r}{(k+r)^2}\right)\,.
	\end{split}\label{N010:TTI:coeffi:ana}
\end{equation}
We checked that the error ratios
\begin{equation}
	R_{X}(k,r) = \fft{\hat f_X^\text{(lmf)}(k,r)-\hat f_X(k,r)}{\hat f_X(k,r)}\quad(X\in\{3/2,1/2\}) \, ,\label{N010:error}
\end{equation}
are small enough for all the configurations listed in (\ref{N010:case}). Comparing these results to the ADHM case, we have encountered larger numerical errors and have therefore increased the range of $N$ used in the \texttt{LinearModelFit} (\ref{N010:TTI:expansion:shifted}) from $101\sim301$ to $101\sim401$ (step=10) to improve precision. Below we show how small the error ratios (\ref{N010:error}) are and also present numerical estimates for the constant term $\hat f_0^\text{(lmf)}(k,r)$ in the \texttt{LinearModelFit} (\ref{N010:TTI:expansion:shifted}) together with the corresponding standard error $\sigma_0$. 
\begin{center}
	\footnotesize
	\begin{tabular}{ |c||c|c|c|c| } 
		\hline
		$(k,r)$ & $R_{3/2}$ & $R_{1/2}$ & $\hat f_0^\text{(lmf)}$ & $\sigma_0$ \\
		\hline\hline
		$(1,1/2)$ & $-1.842{\times}10^{-16}$ & $-7.982{\times}10^{-14}$ & $-2.4445122210251198105$ & $7.956{\times}10^{-13}$  \\
		\hline
		$(2,1)$ & $-1.382{\times}10^{-21}$ & $-9.963{\times}10^{-19}$ & $-1.5990156188560014468$ & $9.539{\times}10^{-18}$  \\
		\hline
		$(3,3/2)$ & $-1.529{\times}10^{-18}$ & $-1.301{\times}10^{-15}$ & $-1.4747778063088432060$ & $1.174{\times}10^{-14}$  \\ 
		\hline
		$(4,2)$ & $-1.430{\times}10^{-16}$ & $-1.249{\times}10^{-13}$ & $-1.6979156145862367914$ & $1.187{\times}10^{-12}$  \\ 
		\hline
	\end{tabular}
\end{center}
\begin{center}
	\footnotesize
	\begin{tabular}{ |c||c|c|c|c| } 
		\hline
		$(k,r)$ & $R_{3/2}$ & $R_{1/2}$ & $\hat f_0^\text{(lmf)}$ & $\sigma_0$ \\
		\hline\hline
		$(1,2/3)$ & $-3.264{\times}10^{-17}$ & $-1.557{\times}10^{-14}$ & $-2.3523428919766891206$ & $1.568{\times}10^{-13}$  \\
		\hline
		$(2,4/3)$ & $-4.072{\times}10^{-19}$ & $-3.009{\times}10^{-16}$ & $-1.6252242716209757969$ & $2.823{\times}10^{-15}$  \\
		\hline
		$(3,2)$ & $-1.090{\times}10^{-16}$ & $-8.815{\times}10^{-14}$ & $-1.6284176444001315906$ & $8.459{\times}10^{-13}$  \\ 
		\hline
		$(4,8/3)$ & $-4.916{\times}10^{-15}$ & $-3.849{\times}10^{-12}$ & $-2.0166679918730627992$ & $4.102{\times}10^{-11}$  \\ 
		\hline
	\end{tabular}
\end{center}
\begin{center}
	\footnotesize
	\begin{tabular}{ |c||c|c|c|c| } 
		\hline
		$(k,r)$ & $R_{3/2}$ & $R_{1/2}$ & $\hat f_0^\text{(lmf)}$ & $\sigma_0$ \\
		\hline\hline
		$(1,1)$ & $-3.573{\times}10^{-18}$ & $-1.959{\times}10^{-15}$ & $-2.2479735914758641588$ & $2.024{\times}10^{-14}$  \\
		\hline
		$(2,2)$ & $-4.454{\times}10^{-16}$ & $-3.269{\times}10^{-13}$ & $-1.7710441322786798897$ & $3.135{\times}10^{-12}$  \\
		\hline
		$(3,3)$ & $-3.989{\times}10^{-14}$ & $-2.834{\times}10^{-11}$ & $-2.0958128593131947881$ & $3.097{\times}10^{-10}$  \\ 
		\hline
		$(4,4)$ & $-8.360{\times}10^{-13}$ & $-5.303{\times}10^{-10}$ & $-2.9167150281215899207$ & $6.910{\times}10^{-9}$  \\ 
		\hline
	\end{tabular}
\end{center}
\begin{center}
	\footnotesize
	\begin{tabular}{ |c||c|c|c|c| } 
		\hline
		$(k,r)$ & $R_{3/2}$ & $R_{1/2}$ & $\hat f_0^\text{(lmf)}$ & $\sigma_0$ \\
		\hline\hline
		$(1,3/2)$ & $-3.293{\times}10^{-16}$ & $-2.036{\times}10^{-13}$ & $-2.2194141302255562257$ & $1.941{\times}10^{-12}$  \\
		\hline
		$(2,3)$ & $-2.845{\times}10^{-13}$ & $-1.923{\times}10^{-10}$ & $-2.1883848791741933989$ & $1.976{\times}10^{-9}$  \\
		\hline
		$(3,9/2)$ & $-1.086{\times}10^{-11}$ & $-6.292{\times}10^{-9}$ & $-3.1734834121743210878$ & $8.184{\times}10^{-8}$  \\ 
		\hline
		$(4,6)$ & $-1.205{\times}10^{-10}$ & $-5.835{\times}10^{-8}$ & $-4.9043313527433843627$ & $9.564{\times}10^{-7}$  \\ 
		\hline
	\end{tabular}
\end{center}
\begin{center}
	\footnotesize
	\begin{tabular}{ |c||c|c|c|c| } 
		\hline
		$(k,r)$ & $R_{3/2}$ & $R_{1/2}$ & $\hat f_0^\text{(lmf)}$ & $\sigma_0$ \\
		\hline\hline
		$(1,2)$ & $-3.601{\times}10^{-14}$ & $-2.338{\times}10^{-11}$ & $-2.2917317046811495268$ & $2.099{\times}10^{-10}$  \\
		\hline
		$(2,4)$ & $-1.842{\times}10^{-11}$ & $-1.112{\times}10^{-8}$ & $-2.8180835746366396624$ & $1.235{\times}10^{-7}$  \\
		\hline
		$(3,6)$ & $-4.050{\times}10^{-10}$ & $-1.949{\times}10^{-7}$ & $-4.6848075856487257662$ & $2.909{\times}10^{-6}$  \\ 
		\hline
		$(4,8)$ & $-2.996{\times}10^{-9}$ & $-1.167{\times}10^{-6}$ & $-7.6419214327213593603$ & $2.248{\times}10^{-5}$  \\ 
		\hline
	\end{tabular}
\end{center}
\begin{center}
	\footnotesize
	\begin{tabular}{ |c||c|c|c|c| } 
		\hline
		$(k,r)$ & $R_{3/2}$ & $R_{1/2}$ & $\hat f_0^\text{(lmf)}$ & $\sigma_0$ \\
		\hline\hline
		$(1,3)$ & $-2.018{\times}10^{-11}$ & $-1.291{\times}10^{-8}$ & $-2.6498077170116517266$ & $1.104{\times}10^{-7}$  \\
		\hline
		$(2,6)$ & $-2.996{\times}10^{-9}$ & $-1.441{\times}10^{-6}$ & $-4.6689712958005925271$ & $1.821{\times}10^{-5}$  \\
		\hline
		$(3,9)$ & $-3.148{\times}10^{-8}$ & $-1.126{\times}10^{-5}$ & $-8.9730889697708208865$ & $2.015{\times}10^{-4}$  \\ 
		\hline
		$(4,12)$ & $-1.397{\times}10^{-7}$ & $-3.939{\times}10^{-5}$ & $-15.329031790765612168$ & $9.246{\times}10^{-4}$  \\ 
		\hline
	\end{tabular}
\end{center}

Here we also provide some numerical errors involved in the estimate for the leading non-perturbative behavior (\ref{N010:fnp}) following the definition (\ref{ADHM:Rdelta}).
\begin{center}
	\footnotesize
	\begin{tabular}{ |c||c| } 
		\hline
		$(k,r)$ & $R_{\text{np},1/2}$  \\
		\hline\hline
		$(1,1)$ & $3.549{\times}10^{-10}$ \\
		\hline
		$(2,2)$ & $-2.420{\times}10^{-8}$ \\
		\hline
		$(3,3)$ & $4.451{\times}10^{-5}$ \\ 
		\hline
		$(4,4)$ & $-3.138{\times}10^{-4}$ \\ 
		\hline
	\end{tabular}
	\begin{tabular}{ |c||c| } 
		\hline
		$(k,r)$ & $R_{\text{np},1/2}$  \\
		\hline\hline
		$(1,2)$ & $-3.737{\times}10^{-11}$ \\
		\hline
		$(2,4)$ & $-3.968{\times}10^{-7}$ \\
		\hline
		$(3,6)$ & $-8.227{\times}10^{-6}$ \\ 
		\hline
		$(4,8)$ & $8.126{\times}10^{-5}$ \\ 
		\hline
	\end{tabular}
	\begin{tabular}{ |c||c| } 
		\hline
		$(k,r)$ & $R_{\text{np},1/2}$  \\
		\hline\hline
		$(1,3)$ & $-5.599{\times}10^{-8}$ \\
		\hline
		$(2,6)$ & $2.636{\times}10^{-8}$ \\
		\hline
		$(3,9)$ & $-3.638{\times}10^{-5}$ \\ 
		\hline
		$(4,12)$ & $1.289{\times}10^{-4}$ \\ 
		\hline
	\end{tabular}
\end{center}
%

%%%%%
\section{$V^{5,2}$ topologically twisted index}
\label{App:V52}
%%%%%

Here we provide numerical data that supports the all-order $1/N$ expansion of the $V^{5,2}$ TTI given in (\ref{V52:TTI:all}). The list of $N_f$ and $\Delta$-configurations for which we confirmed (\ref{V52:TTI:all}) with five different $\mn$-configurations (\ref{V52:nas}) is given as follows: 

\smallskip

\noindent\textbf{Case 1.} $\Delta=(\Delta_1,\fft43-\Delta_1,\fft23,\fft13,\fft13,0)$
\begin{equation}
	\begin{alignedat}{3}
		N_f&\in\{1,2,3,4,5\}&\quad&\&&\quad \Delta_1&=\fft23\\
		N_f&\in\{1,2,3\}&\quad&\&&\quad \Delta_1&\in\bigg\{\fft12,\fft59\bigg\}\\
		N_f&=1&\quad&\&&\quad\Delta_1&=\fft{7}{12}
	\end{alignedat}\label{V52:case1}
\end{equation}
\noindent\textbf{Case 2.} $\Delta=(\Delta_1,\fft43-\Delta_1,\fft23,\fft13,\fft13,\Delta_m)$
\begin{equation}
	\begin{alignedat}{3}
		N_f&\in\{1,2,3\}&\quad&\&&\quad (\Delta_1,\Delta_m)&=\bigg\{(\fft12,\fft{N_f}{6}),(\fft12,\fft{N_f}{9}),(\fft59,\fft{N_f}{9}),\\
		&&&&&\qquad(\fft59,\fft{N_f}{12}),(\fft{7}{12},\fft{N_f}{12}),(\fft{7}{12},\fft{N_f}{15})\bigg\}\\
		N_f&\in\{1,2\}&\quad&\&&\quad (\Delta_1,\Delta_m)&\in\bigg\{(\fft{5}{12},\fft{N_f}{12}),(\fft{7}{15},\fft{N_f}{15}),(\fft{13}{24},\fft{N_f}{10})\bigg\}
	\end{alignedat}\label{V52:case2}
\end{equation}
\noindent\textbf{Case 3.} $\Delta=(\Delta_1,\fft43-\Delta_1,\fft23,\Delta_q,\fft23-\Delta_q,\Delta_m)$
\begin{equation}
	\begin{alignedat}{3}
		N_f&\in\{1,2\}&\quad&\&&\quad (\Delta_1,\Delta_q,\Delta_m)&=(\fft23-\fft{1}{2\pi},\fft16,N_f(\fft23-\fft2\pi))\\
		N_f&\in\{3,4\}&\quad&\&&\quad (\Delta_1,\Delta_q,\Delta_m)&=(\fft23-\fft{1}{2\pi},\fft14,N_f(\fft23-\fft2\pi))
	\end{alignedat}\label{V52:case3}
\end{equation}
For all the above $N_f$ and $\Delta$-configurations with the $\mn$-configurations (\ref{V52:nas}), we compared the numerical estimates
\begin{equation}
	\hat f_{3/2}^\text{(lmf)}(N_f,\Delta,\mn)\,,\qquad\hat f_{1/2}^\text{(lmf)}(N_f,\Delta,\mn)\,,
\end{equation}
obtained from the \texttt{LinearModelFit} for the $V^{5,2}$ TTI with the corresponding analytic expressions in (\ref{V52:TTI:all}), namely
\begin{equation}
	\begin{split}
		\hat f_{3/2}(N_f,\Delta,\mn)&=-\fft{\pi\sqrt{N_f\tDelta_1\tDelta_2\tDelta_3\tDelta_4}}{3}\sum_{a=1}^4\fft{\tmn_a}{\tDelta_a}\,,\\
		\hat f_{1/2}(N_f,\Delta,\mn)&=-\fft{\pi\sqrt{N_f\tilde\Delta_1\tilde\Delta_2\tilde\Delta_3\tilde\Delta_4}}{3}\left(\sum_{I=1}^2(\mathfrak{a}_IN_f+\fft{\mathfrak{b}_I}{N_f})\mn_I+\fft{\tDelta_3-\tDelta_4}{3\tDelta_3^2\tDelta_4^2}\fft{\mkt}{N_f^2}\right)\,,
	\end{split}
\end{equation}
where $\tDelta_a,\tmn_a$ are defined in (\ref{ADHM:tilde}) and $\mathfrak{a}_I,\mathfrak{b}_I$ are defined in (\ref{V52:TTI:all:coeffi}). To estimate the precision, we checked that the error ratios
\begin{equation}
	R_{X}(N_f,\Delta,\mn)=\fft{\hat f_X^\text{(lmf)}(N_f,\Delta,\mn)-\hat f_X(N_f,\Delta,\mn)}{\hat f_X(N_f,\Delta,\mn)}\quad(X\in\{3/2,1/2\}) \, ,\label{V52:error}
\end{equation}
are small enough for all the configurations listed in (\ref{V52:case1},\ref{V52:case2},\ref{V52:case3}). Below we provide some examples to show how small the error ratios (\ref{V52:error}) are, and also present numerical estimates for the constant term $\hat f_0^\text{(lmf)}(N_f,\Delta,\mn)$ in the \texttt{LinearModelFit} (\ref{ADHM:TTI:expansion:shifted}) together with the corresponding standard error $\sigma_0$. 

\medskip

\noindent\textbf{Case 1.} $\Delta_1=\fft23$ \& $\mn=(\fft23,\fft23,\fft23,\fft13,\fft13,0)$
\begin{center}
	\footnotesize
	\begin{tabular}{ |c||c|c|c|c| } 
		\hline
		& $R_{3/2}$ & $R_{1/2}$ & $\hat f_0^\text{(lmf)}$ & $\sigma_0$ \\
		\hline\hline
		$N_f=1$ & $2.446{\times}10^{-28}$ & $7.122{\times}10^{-26}$ & $-2.7620858097124988759$ & $9.445{\times}10^{-25}$  \\
		\hline
		$N_f=2$ & $-4.779{\times}10^{-22}$ & $-1.594{\times}10^{-19}$ & $-3.2671366455659909955$ & $2.399{\times}10^{-18}$  \\
		\hline
		$N_f=3$ & $-7.188{\times}10^{-19}$ & $-2.104{\times}10^{-16}$ & $-4.7624014875151824187$ & $4.110{\times}10^{-15}$  \\ 
		\hline
		$N_f=4$ & $1.210{\times}10^{-16}$ & $2.994{\times}10^{-14}$ & $-7.0161650179852435476$ & $7.423{\times}10^{-13}$  \\ 
		\hline
		$N_f=5$ & $9.005{\times}10^{-15}$ & $1.895{\times}10^{-12}$ & $-9.9830275183613382542$ & $5.851{\times}10^{-11}$  \\ 
		\hline
	\end{tabular}
\end{center}
\noindent\textbf{Case 2.} $N_f=2$ \& $(\Delta_1,\Delta_q)=(\fft12,\fft{N_f}{9})$
\begin{center}
	\footnotesize
	\begin{tabular}{ |c||c|c|c|c| } 
		\hline
		& $R_{3/2}$ & $R_{1/2}$ & $\hat f_0^\text{(lmf)}$ & $\sigma_0$ \\
		\hline\hline
		\text{1st in }(\ref{V52:nas}) & $1.078{\times}10^{-16}$ & $3.448{\times}10^{-14}$ & $-3.4528096365944814244$ & $4.489{\times}10^{-13}$  \\
		\hline
		\text{2nd in }(\ref{V52:nas}) & $1.732{\times}10^{-16}$ & $5.639{\times}10^{-14}$ & $-3.3715743804211707717$ & $7.244{\times}10^{-13}$  \\
		\hline
		\text{3rd in }(\ref{V52:nas}) & $6.362{\times}10^{-17}$ & $2.046{\times}10^{-14}$ & $-3.3447537752182470037$ & $2.564{\times}10^{-13}$  \\ 
		\hline
		\text{4th in }(\ref{V52:nas}) & $-5.510{\times}10^{-17}$ & $-1.751{\times}10^{-14}$ & $-3.2966941000920532717$ & $2.195{\times}10^{-13}$  \\ 
		\hline
	\end{tabular}
\end{center}
\noindent\textbf{Case 3.} $N_f=3$ \& $(\Delta_1,\Delta_q,\Delta_m)=(\fft23-\fft{1}{2\pi},\fft14,N_f(\fft23-\fft2\pi))$
\begin{center}
	\footnotesize
	\begin{tabular}{ |c||c|c|c|c| } 
		\hline
		& $R_{3/2}$ & $R_{1/2}$ & $\hat f_0^\text{(lmf)}$ & $\sigma_0$ \\
		\hline\hline
		\text{1st in }(\ref{V52:nas}) & $1.584{\times}10^{-15}$ & $4.419{\times}10^{-13}$ & $-5.0895614683300330156$ & $7.852{\times}10^{-12}$  \\
		\hline
		\text{2nd in }(\ref{V52:nas}) & $1.597{\times}10^{-15}$ & $4.479{\times}10^{-13}$ & $-4.9454216523869612663$ & $7.691{\times}10^{-12}$  \\
		\hline
		\text{3rd in }(\ref{V52:nas}) & $6.608{\times}10^{-17}$ & $1.862{\times}10^{-14}$ & $-4.8692433262566337160$ & $2.645{\times}10^{-13}$  \\ 
		\hline
		\text{4th in }(\ref{V52:nas}) & $-1.782{\times}10^{-15}$ & $-4.984{\times}10^{-13}$ & $-4.7512952786913154264$ & $8.264{\times}10^{-12}$  \\ 
		\hline
		\text{5th in }(\ref{V52:nas}) & $-3.923{\times}10^{-16}$ & $-1.098{\times}10^{-13}$ & $-4.8145318465752665874$ & $1.898{\times}10^{-12}$ \\ 
		\hline
	\end{tabular}
\end{center}
%

%%%%%
\section{$Q^{1,1,1}$ topologically twisted index}
\label{App:Q111}
%%%%%

We begin with a review of the analytic derivation of the large $N$ result for the TTI \eqref{Q111:TTI:N32}. Along the way, we correct a few minor errors in~\cite{Hosseini:2016tor,Hosseini:2016ume}.

\medskip

To derive the BA formula for the $S^1\times S^2$ $Q^{1,1,1}$ TTI, we first rewrite the matrix model (\ref{Q111:TTI:1}) as
\begin{align}
\label{Q111:TTI:2}
		Z^{Q^{1,1,1}}_{S^1\times S^2}(N,N_f,\Delta,\mn)&=\fft{1}{(N!)^2}\sum_{\mm,\tilde{\mm}\in\mathbb Z^N}\oint_{\mathcal C}\prod_{i=1}^N\fft{dx_i}{2\pi \mathrm{i}x_i}\fft{d\tx_i}{2\pi \mathrm{i}\tx_i}\,\prod_{i\neq j}^N\left(1-\fft{x_i}{x_j}\right)\left(1-\fft{\tx_i}{\tx_j}\right) \nonumber \\
		&\quad\times\prod_{i,j=1}^N\prod_{a=1,2}\left(\fft{\sqrt{\fft{x_i}{\tx_j}y_a}}{1-\fft{x_i}{\tx_j}y_a}\right)^{1-\mn_a}\prod_{a=3,4}\left(\fft{\sqrt{\fft{\tx_j}{x_i}y_a}}{1-\fft{\tx_j}{x_i}y_a}\right)^{1-\mn_a}\\
		&\quad\times\prod_{i=1}^N\prod_{n=1,2}\left(\fft{\sqrt{\fft{1}{x_i}y_{\tq_n}}}{1-\fft{1}{x_i}y_{\tq_n}}\right)^{N_f(1-\mn_{\tq_n})}\prod_{j=1}^N\prod_{n=1,2}\left(\fft{\sqrt{\tx_jy_{q_n}}}{1-\tx_jy_{q_n}}\right)^{N_f(1-\mn_{q_n})}\nonumber\\
		&\quad\times\prod_{i=1}^N\fft{(e^{\mathrm{i}B_i})^M}{e^{\mathrm{i}B_i}-1}\times\prod_{j=1}^N\fft{(e^{\mathrm{i}\tilde{B}_j})^M}{e^{\mathrm{i}\tilde{B}_j}-1} \, , \nonumber
\end{align}
in terms of a large integer cut-off $M$ $(\mm_i\leq M-1,~\tmm_j\geq1-M)$ and using the BA operators
\begin{equation}
	\begin{split}
		e^{\mathrm{i}B_i}&=(-1)^{N+N_f-2\lfloor N_f/2\rfloor}\sigma_i\prod_{n=1,2}\left(\fft{\sqrt{\fft{1}{x_i}y_{\tq_n}}}{1-\fft{1}{x_i}y_{\tq_n}}\right)^{-N_f}\prod_{j=1}^N\fft{(1-\fft{\tx_j}{x_i}y_3)(1-\fft{\tx_j}{x_i}y_4)}{(1-\fft{\tx_j}{x_i}y_1^{-1})(1-\fft{\tx_j}{x_i}y_2^{-1})}\,,\\
		e^{\mathrm{i}\tilde B_j}&=(-1)^{N+N_f-2\lfloor N_f/2\rfloor}\tilde\sigma_j\prod_{n=1,2}\left(\fft{\sqrt{\tx_jy_{q_n}}}{1-\tx_jy_{q_n}}\right)^{-N_f}\prod_{i=1}^N\fft{(1-\fft{\tx_j}{x_i}y_3)(1-\fft{\tx_j}{x_i}y_4)}{(1-\fft{\tx_j}{x_i}y_1^{-1})(1-\fft{\tx_j}{x_i}y_2^{-1})}\,.
	\end{split}\label{Q111:B}
\end{equation}
The sign ambiguities $\sigma_i,\tilde\sigma_j$ in the BAE are defined as
\begin{equation}
\begin{split}
	\sigma_i = \prod_{j=1}^N\fft{\sqrt{\fft{x_i}{\tx_j}y_1}}{-\fft{x_i}{\tx_j}y_1}\fft{\sqrt{\fft{x_i}{\tx_j}y_2}}{-\fft{x_i}{\tx_j}y_2}\fft{1}{\sqrt{\fft{\tx_j}{x_i}y_3}}\fft{1}{\sqrt{\fft{\tx_j}{x_i}y_4}} \in \{-1,1\} \, ,\\
	\tilde\sigma_j = \prod_{i=1}^N\fft{\sqrt{\fft{x_i}{\tx_j}y_1}}{-\fft{x_i}{\tx_j}y_1}\fft{\sqrt{\fft{x_i}{\tx_j}y_2}}{-\fft{x_i}{\tx_j}y_2}\fft{1}{\sqrt{\fft{\tx_j}{x_i}y_3}}\fft{1}{\sqrt{\fft{\tx_j}{x_i}y_4}} \in \{-1,1\} \, .
\end{split}\label{sign}
\end{equation}
Anticipating the following results, we will show below that the analytic large $N$ BAE solution imposes a range on the values of $(x_i,\tx_j)$ and $y_a$ such that the ambiguities take values
\begin{equation}
\label{eq:sign-fix}
\sigma_i = \tilde\sigma_j = (-1)^N \, , 
\end{equation}
on the BAE solution. Thus, it is consistent to use this choice in our analysis. Note that the same remark holds in the $N^{0,1,0}$ theory, see \eqref{eq:N010-sign-fix}.
%From here on we fix the sign ambiguities $\sigma_i,\tilde\sigma_j$ as (\ref{sign}), which is consistent with the particular BAE solutions discussed in subsections \ref{sec:TTI:Q111:analytic} and \ref{sec:TTI:Q111:num}. 
%
%Removing contour integrals in (\ref{Q111:TTI:2}) by Cauchy's integral formula and using the constraints (\ref{Q111:constraints}), 
%
%In this subsection we briefly review how to evaluate the $Q^{1,1,1}$ TTI analytically using the BA formula (\ref{Q111:TTI:BA}) in the large $N$ limit \cite{Hosseini:2016tor,Hosseini:2016ume}.

As discussed in the ADHM case, the residues picked up by the integrals in \eqref{Q111:B} lead to the BA formula \eqref{Q111:TTI:BA} for the $Q^{1,1,1}$ TTI, where the BAE are given in (\ref{Q111:BAE}). One can check that these BAE can be obtained by differentiating the following Bethe potential with respect to $u_i$ and $\tu_j$:
\begin{align}
\label{Q111:Beth}
		\mathcal V&=\sum_{i=1}^N\left[\fft{N_f}{2}(\tu_i^2-u_i^2)-2\pi\left(\tn_i-\fft{N_f}{2}+\left\lfloor\fft{N_f}{2}\right\rfloor\right)\tu_i+2\pi\left(n_i-\fft{N_f}{2}+\left\lfloor\fft{N_f}{2}\right\rfloor\right)u_i\right] \nonumber \\
		&\quad+\sum_{i,j=1}^N\bigg[\sum_{a=3,4}\text{Li}_2(e^{\mathrm{i}(\tu_j-u_i+\pi\Delta_a)})-\sum_{a=1,2}\text{Li}_2(e^{\mathrm{i}(\tu_j-u_i-\pi\Delta_a)})\bigg]\\
		&\quad+N_f\sum_{i=1}^N\sum_{n=1}^2\bigg[\text{Li}_2(e^{\mathrm{i}(-u_i+\pi\Delta_{\tq_n})})-\text{Li}_2(e^{\mathrm{i}(-\tu_i-\pi\Delta_{q_n})})+\fft\pi2u_i\Delta_{\tq_n}+\fft\pi2\tu_i(\Delta_{q_n}-2)\bigg] \, . \nonumber
\end{align}
We note that this Bethe potential is slightly different from the one written previously in the literature~\cite{Hosseini:2016tor,Hosseini:2016ume,PandoZayas:2020iqr}. Our result is equivalent to the previous ones in the large $N$ limit, but we stress that (\ref{Q111:Beth}) is the exact Bethe potential consistent with the BA operators (\ref{Q111:B}) at finite $N$. This slight modification leads to a different set of integers (\ref{Q111:integer}) chosen to find a BAE solution instead of $(n_i,\tn_j)=(1-i,j-N)$.

The large $N$ limit of the solutions to the BAE (\ref{Q111:BAE}) has been constructed following the same procedure as described in Section~\ref{sec:TTI:N010:analytic}. This involves the construction of a large $N$ ansatz and its continuum version in the large $N$ limit, see~\cite{Hosseini:2016ume} for details. Here we only summarize the resulting large $N$ BAE solution for the superconformal $\Delta$-configuration in~(\ref{Q111:constraints:sc}):
\begin{equation}
	\begin{split}
		\rho(t)&=\fft{2\mu-k|t|}{\pi^2}\qquad(t_<<t<t_>)\,,\\
		\delta v(t)&=\pi\Delta_1-\fft{\pi\mu}{2\mu-k|t|}\qquad(t_<<t<t_>)\,,\label{Q111:BAE:sol:largeN}
	\end{split}
\end{equation}
where $\delta v(t)=\tv(t)-v(t)$ and the end points for the eigenvalue distribution are given by
\begin{equation}
	t_\ll=-\fft{\mu}{N_f}\,,\qquad t_\gg=\fft{\mu}{N_f}\,.
\end{equation}
The parameter $\mu$ is determined by the normalization (\ref{ADHM:rho:normal}) and reads
\begin{equation}
	\mu=\sqrt{\fft{N_f}{3}}\pi\,.
\end{equation}
To find the solution (\ref{Q111:BAE:sol:largeN}), the integers $n_i,\tn_j$ in the BAE (\ref{Q111:BAE}) have been chosen as
\begin{equation}
	n_i=1-i-\left\lfloor\fft{N_f}{2}\right\rfloor\,,\qquad \tn_j=j-N-\left\lfloor\fft{N_f}{2}\right\rfloor\,\label{Q111:integer}
\end{equation}
and we have also assumed
\begin{equation}
	\begin{alignedat}{2}
		0&<\Re[\tu_j-u_i+\pi\Delta_a]<2\pi\,,&\qquad -2\pi&<\Re[\tu_j-u_i-\pi\Delta_a]<0\qquad(i\geq j)\,,\\
		0&<\Re[\tu_i+\pi\Delta_{q_n}]<2\pi\,,&\qquad 0&<\Re[-u_i+\pi\Delta_{\tq_n}]<2\pi\,.
	\end{alignedat}\label{Q111:range}
\end{equation}
These ranges lead to \eqref{eq:sign-fix}. One then obtains the large $N$ limit of the $Q^{1,1,1}$ TTI by substituting (\ref{Q111:BAE:sol:largeN}) into the large $N$ limit of the BA formula. The result is given in \eqref{Q111:TTI:N32}. 

Observe that $v(t)+\tv(t)$ is not determined in the large $N$ BAE solution (\ref{Q111:BAE:sol:largeN}). For this reason, when conducting the numerical study in Section~\ref{sec:TTI:Q111:num}, we should also set the initial condition for $v_i+\tv_i$ based on the ranges (\ref{Q111:range}) as in \cite{PandoZayas:2020iqr}
	\begin{equation}
		v_i+\tv_i\big|_\text{initial condition}=\pi(2-\Delta_1-\Delta_{q_1}-\Delta_{q_2})=0,\label{BAE:Q111:sol:v}
	\end{equation}
where the second equation is for the superconformal $\Delta$-configuration (\ref{Q111:constraints:sc}). \\

Finally, we provide the numerical data that supports the all-order $1/N$ expansion of the $Q^{1,1,1}$ TTI given in (\ref{Q111:TTI:all}). The list of $N_f$ and $\Delta$-configurations satisfying the condition (\ref{Q111:constraints:sc}) for which we confirmed (\ref{Q111:TTI:all}) with the $\mn$-configurations (\ref{Q111:nas}) is given as follows:
\begin{equation}
	\begin{alignedat}{3}
		N_f&\in\{1,2,3,4,5\}&\quad&\&&\quad\Delta_1&=\fft12\,,\\
		N_f&\in\{1,2,3\}&\quad&\&&\quad\Delta_1&\in\bigg\{\fft38,\fft{5}{12},\fft37\bigg\}\,.
	\end{alignedat}\label{Q111:case}
\end{equation}
For the above $N_f$ and $\Delta$-configurations together with the six $\mn$-configurations (\ref{Q111:nas}), we compared the numerical estimates
\begin{equation}
	\hat f_{3/2}^\text{(lmf)}(N_f,\Delta,\mn)\,,\qquad\hat f_{1/2}^\text{(lmf)}(N_f,\Delta,\mn)\,,\label{Q111:TTI:coeffi:num}
\end{equation}
obtained from the \texttt{LinearModelFit} for the $Q^{1,1,1}$ TTI with the corresponding analytic expressions in (\ref{Q111:TTI:all}),
\begin{equation}
	\begin{split}
		\hat f_{3/2}(N_f,\Delta,\mn)&=-\fft{4\pi\sqrt{N_f}}{3\sqrt{3}}\,,\\
		\hat f_{1/2}(N_f,\Delta,\mn)&=\fft{4\pi\sqrt{N_f}}{3\sqrt{3}}\left(\fft{N_f}{4}+\fft{3}{4N_f}\right)\,.
	\end{split}\label{Q111:TTI:coeffi:ana}
\end{equation}
Recall that we expect $(\Delta_1,\mn)$ to be flat directions from the large $N$ analytic result \eqref{Q111:TTI:N32}, so we do not introduce a dependence on these parameters in the RHS of \eqref{Q111:TTI:coeffi:ana}. To estimate the precision of these results, we checked that the error ratios
\begin{equation}
	R_{X}(N_f,\Delta,\mn) = \fft{\hat f_X^\text{(lmf)}(N_f,\Delta,\mn)-\hat f_X(N_f,\Delta,\mn)}{\hat f_X(N_f,\Delta,\mn)}\quad(X\in\{3/2,1/2\}) \, ,\label{Q111:error}
\end{equation}
are small enough for all the configurations listed in (\ref{Q111:case}). Below we show how small the error ratios (\ref{Q111:error}) are and also present numerical estimates for the constant term $\hat f_0^\text{(lmf)}(N_f,\Delta,\mn)$ in the \texttt{LinearModelFit} (\ref{ADHM:TTI:expansion:shifted}) together with the corresponding standard error $\sigma_0$. 
\begin{center}
	\footnotesize
	\begin{tabular}{ |c||c|c|c|c| } 
		\hline
		& $R_{3/2}$ & $R_{1/2}$ & $\hat f_0^\text{(lmf)}$ & $\sigma_0$ \\
		\hline\hline
		$N_f=1$ & $1.503{\times}10^{-17}$ & $9.883{\times}10^{-15}$ & $-2.1415723730798296354$ & $6.340{\times}10^{-14}$  \\
		\hline
		$N_f=2$ & $-3.490{\times}10^{-14}$ & $-2.641{\times}10^{-11}$ & $-2.0385864384989237526$ & $1.780{\times}10^{-10}$  \\
		\hline
		$N_f=3$ & $-1.753{\times}10^{-12}$ & $-1.168{\times}10^{-9}$ & $-2.2368141361938090934$ & $9.784{\times}10^{-9}$  \\ 
		\hline
		$N_f=4$ & $5.722{\times}10^{-12}$ & $3.251{\times}10^{-9}$ & $-2.6005901148883909862$ & $2.981{\times}10^{-8}$  \\ 
		\hline
		$N_f=5$ & $2.396{\times}10^{-10}$ & $1.157{\times}10^{-7}$ & $-3.1045097958934355205$ & $1.393{\times}10^{-6}$  \\ 
		\hline
	\end{tabular}
\end{center}
The numerical estimates for the $Q^{1,1,1}$ TTI do not depend on the $\Delta_1$ and $\mn$-configurations so the above table is valid for all cases listed in (\ref{Q111:case}). 
\section{A compendium of 7d Sasaki-Einstein manifolds}\label{App:SE}
%%%%%

In this appendix, we collect some known facts regarding the geometry and volumes of various Sasaki-Einstein manifolds. All 6d K\"{a}hler-Einstein metrics with line element $ds^2_6$ have been normalized such that 
\begin{equation}
R_{\mu\nu}^{(6)} = 8\,g_{\mu\nu}^{(6)} \, ,
\end{equation}
in what follows. See \cite{Castellani:1983yg} for an early reference on Freund-Rubin type AdS$_4$ solutions in 11d supergravity with internal Sasaki-Einstein manifolds.\\

\noindent \textbf{ADHM}: The $S^7$ metric is locally given in the canonical form 
\begin{equation}
\label{metric:SE} 
ds^2_{\text{SE}_7} = (d\psi + \sigma)^2 + ds^2_6 \, , 
\end{equation}
with~\cite{Bergman:2009zh}
%\footnote{In Appendix B of \cite{Bergman:2009zh}, take $\varphi^\text{there}=\psi^\text{here}$ and $\psi^\text{there}=\varphi^\text{here}$.}
%
\begin{equation}
	\begin{split}
		\sigma&=\fft12(\cos^2\xi-\sin^2\xi)d\varphi+\fft12\cos^2\xi\cos\theta_1d\phi_1+\fft12\sin^2\xi\cos\theta_2d\phi_2,\\
		ds^2_6&=d\xi^2+\cos^2\xi\sin^2\xi\left(d\varphi+\fft12\cos\theta_1d\phi_2-\fft12\cos\theta_2d\phi_2\right)^2\\
		&\quad+\fft14\cos^2\xi\left(d\theta_1^2+\sin^2\theta_1d\phi_1^2\right)+\fft14\sin^2\xi\left(d\theta_2^2+\sin^2\theta_2d\phi_2^2\right) .
	\end{split}
\end{equation}

The coordinate ranges for these angular coordinates on $S^7$ are
\begin{equation}
	0\leq2\psi\pm\varphi<4\pi\,,\quad 0\leq\xi\leq\fft\pi2\,,\quad 0\leq\theta_{1,2}\leq\pi\,,\quad 0\leq\phi_{1,2}<2\pi\,.
\end{equation}
To describe the orbifold $S^7/\mathbb Z_{N_f}$, consider the following reparametrization:
\begin{equation}
\begin{split}
	X_1&=e^{\mathrm{i}(\psi+\fft{\varphi+\phi_1}{2})}\cos\xi\cos\fft{\theta_1}{2}\,,\\
	X_2&=e^{\mathrm{i}(\psi+\fft{\varphi-\phi_1}{2})}\cos\xi\sin\fft{\theta_1}{2}\,,\\
	X_3&=e^{\mathrm{i}(\psi-\fft{\varphi-\phi_2}{2})}\sin\xi\cos\fft{\theta_2}{2}\,,\\
	X_4&=e^{\mathrm{i}(\psi-\fft{\varphi+\phi_2}{2})}\sin\xi\sin\fft{\theta_2}{2}\,,
\end{split}
\end{equation}
where the $X_i$ satisfy $\sum_{i=1}^4|X_i|^2=1$. In these embedding coordinates, the $\mathbb Z_{N_f}$ orbifold giving rise to the ADHM theory is given by~\cite{Mezei:2013gqa}
%\footnote{In the convention of \cite{Grassi:2014vwa} the $\mathbb Z_{N_f}$ orbifolding action corresponds to $(X_3,X_4)\sim (e^{\fft{2\pi i}{N_f}}X_3,e^{-\fft{2\pi i}{N_f}}X_4)$ but here we follow the convention of \cite{Mezei:2013gqa}.}
%
\begin{equation}
	(X_3,X_4)\sim e^{\fft{2\pi \mathrm{i}}{N_f}}(X_3,X_4) \, ,
\end{equation}
which is equivalent to the following identification
\begin{equation}
	2\psi-\varphi\sim2\psi-\varphi+\fft{4\pi}{N_f} \, , \label{S7:ZNf:range}
\end{equation}
in the canonical coordinates. The volume of $S^7/\mathbb Z_{N_f}$ can then be evaluated in a straightforward way as
\begin{equation}
	\text{vol}[S^7/\mathbb Z_{N_f}]=\fft{\pi^4}{3N_f} \, .
\end{equation}
%

%Note that (\ref{S7:ZNf:range}) is distinguished from the $\mathbb Z_{k}$ orbifolding of $S^7$ for the U$(N)_k\times$U$(N)_{-k}$ ABJM theory (see Appendix B of \cite{Bergman:2009zh} for example)
%%
%\begin{equation}
%	(X_1,X_2,X_3,X_4)\sim e^{\fft{2\pi i}{k}}(X_1,X_2,X_3,X_4)\quad\Leftrightarrow\quad \psi\sim\psi+\fft{2\pi}{k}.
%\end{equation}
%%

\medskip

\noindent$\bm{N^{0,1,0}}$: The $N^{0,1,0}$ metric is locally given in the canonical form (\ref{metric:SE}) with~\cite{Page:1984ac,Gauntlett:2005jb}
%\footnote{In Appendix A of \cite{Gauntlett:2005jb}, take $-\fft12\psi^\text{there}=\psi^\text{here}$.}
%
\begin{equation}
	\begin{split}
		\sigma&=-\fft12\cos\mu\sin\theta(\cos\phi\sigma_1-\sin\phi\sigma_2)+\fft14\cos\theta(1+\cos^2\mu)\sigma_3-\fft12\cos\theta d\phi\,,\\
		ds^2_6&=\fft14\left(d\theta-\cos\mu(\sin\phi\sigma_1+\cos\phi\sigma_2)\right)^2\\
		&\quad+\fft14\sin^2\theta\left(d\phi-\cos\mu\cot\theta(\cos\phi\sigma_1-\sin\phi\sigma_2)-\fft12(1+\cos^2\mu)\sigma_3\right)^2\\
		&\quad+\fft12\left(d\mu^2+\fft14\sin^2\mu(\sigma_1^2+\sigma_2^2)+\fft14\sin^2\mu\cos^2\mu\sigma_3^2\right),
	\end{split}\label{N010}
\end{equation}
where $\sigma_i$ are the right-invariant one forms on SU(2),
\begin{equation}
\begin{split}
	\sigma_1&=\sin\beta\,d\alpha-\cos\beta\sin\alpha\,d\gamma\,,\\
	\sigma_2&=\cos\beta\,d\alpha+\sin\beta\sin\alpha\,d\gamma\,,\\
	\sigma_3&=d\beta+\cos\alpha\,d\gamma\,.
\end{split}
\end{equation}
The coordinate ranges for a globally well-defined metric on $N^{0,1,0}$ are given as 
\begin{equation}
\begin{alignedat}{3}
	0&\leq\psi<\pi\,,&\qquad0&\leq\mu\leq\fft\pi2\,,&\qquad0&\leq\theta<\pi\,,\qquad 0\leq\phi<2\pi\,,\\
	0&\leq\alpha<\pi\,,&\qquad 0&\leq\beta<2\pi\,,&\qquad 0&\leq\gamma<4\pi\,.
\end{alignedat}
\end{equation}
%
%It is not straightforward to specify a further coordinate identification required to describe the orbifold $N^{0,1,0}/\mathbb Z_k$ in the canonical $N^{0,1,0}$ metric (\ref{metric:SE}) with (\ref{N010}). 

The $\mathbb Z_k$ orbifold action can be specified by understanding the canonical $N^{0,1,0}$ metric (\ref{N010}) as the 11d Taub-NUT spacetime obtained by uplifting the 10d geometry describing $k$ D6 branes extended along the AdS$_4\times\mathbb{RP}^3$ part of the AdS$_4\times\mathbb{CP}^3$ solution~\cite{Hikida:2009tp}. The volume of $N^{0,1,0}/\mathbb Z_{k}$ can be evaluated to give
\begin{equation}
	\text{vol}[N^{0,1,0}/\mathbb Z_{k}]=\fft{\pi^4}{8k} \, ,
\end{equation}
using the fact that the $\mathbb{Z}_k$ orbifold reduces the original $N^{0,1,0}$ volume by a factor of $k$.

\medskip

\noindent$\bm{V^{5,2}}$: The $V^{5,2}$ metric is locally given in the canonical form (\ref{metric:SE}) with~\cite{Bergman:2001qi}
%\footnote{In Appendix B of \cite{Bergman:2001qi}, take $\fft34\psi^\text{there}=\psi^\text{here}$ to satisfy the normalization (\ref{SE:normal}).}
%
\begin{equation}
	\begin{split}
		\sigma&=\fft38\cos\alpha(d\beta-\cos\theta_1d\phi_1-\cos\theta_2d\phi_2)\,,\\[1mm]
		ds^2_6&=\fft38d\alpha^2+\fft{3}{32}\sin^2\alpha\left(d\beta-\cos\theta_1d\phi_1-\cos\theta_2d\phi_2\right)^2\\
		&\quad+\fft{3}{32}(1+\cos^2\alpha)(d\theta_1^2+\sin^2\theta_1d\phi_1^2+d\theta_2^2+\sin^2\theta_2d\phi_2^2)\\
		&\quad-\fft{3}{16}\sin^2\alpha\cos\beta\left(d\theta_1d\theta_2-\sin\theta_1\sin\theta_2d\phi_1d\phi_2\right)\\
		&\quad+\fft{3}{16}\sin^2\alpha\sin\beta(\sin\theta_2d\theta_1d\phi_2+\sin\theta_1d\theta_2d\phi_1).
	\end{split}
\end{equation}
The coordinate ranges for a globally well-defined metric on $V^{5,2}$ are
\begin{equation}
	0\leq\psi<\fft32\pi\,,\quad 0\leq\alpha\leq\fft\pi2\,,\quad 0\leq\beta<4\pi\,,\quad 0\leq\theta_{1,2}\leq\pi\,,\quad 0\leq\phi_{1,2}<2\pi\,.
\end{equation}
To describe the orbifold $V^{5,2}/\mathbb Z_{N_f}$ we make the identification~\cite{Martelli:2009ga}
\begin{equation}
	\phi_2\sim\phi_2+\fft{2\pi}{N_f} \, .
\end{equation}
%
%Note that the periodic identification is along the $\phi_2$ coordinate to describe the $\mathbb Z_{N_f}{\subseteq}$U(1)$_b$ orbifolding of $V^{5,2}$, which breaks the isometry from SO(5)$\times$U(1)$_R$ to SU(2)$_r\times$U(1)$_b\times$U(1)$_R$, not along the $\psi$ coordinate for the $\mathbb Z_{N_f}{\subseteq}$U(1)$_R$ orbifolding as explained in section 3 of \cite{Martelli:2009ga}. 
The volume of $V^{5,2}/\mathbb Z_{N_f}$ can then be evaluated in a straightforward way and reads
\begin{equation}
	\text{vol}[V^{5,2}/\mathbb Z_{N_f}]=\fft{27\pi^4}{128N_f}\,.
\end{equation}

\medskip

\noindent$\bm{Q^{1,1,1}}$: The $Q^{1,1,1}$ metric is locally given in the canonical form (\ref{metric:SE}) with~\cite{Duff:1986hr,Franco:2009sp}
%\footnote{In (9.2.18) of \cite{Duff:1986hr}, take $\tau^\text{there}=4\psi^\text{here}$, $c^\text{there}=\fft14$, and $\Lambda_i^\text{there}=8$ to satisfy the normalization (\ref{SE:normal}). From \cite{Franco:2009sp} we take $\psi^\text{there}=4\psi^\text{here}$.}
%
\begin{equation}
\begin{split}
	\sigma&=\fft14\sum_{i=1}^3\cos\theta_i\,d\phi_i\,,\\
	ds^2_6&=\fft18\sum_{i=1}^3\left(d\theta_i^2+\sin^2\theta_i\,d\phi_i^2\right).
\end{split}
\end{equation}
The coordinate ranges for a globally well-defined metric are 
\begin{equation}
	0\leq\psi<\pi\,,\quad 0\leq\theta_{1,2,3}\leq\pi\,,\quad 0\leq\phi_{1,2,3}<2\pi\,.
\end{equation}
To describe the orbifold $Q^{1,1,1}/\mathbb Z_{N_f}$ we make the identification~\cite{Franco:2009sp}
\begin{equation}
	(\phi_2,\phi_3)\sim(\phi_2,\phi_3)+\bigg(\fft{2\pi}{N_f},\fft{2\pi}{N_f}\bigg) \, .
\end{equation}
The volume of $Q^{1,1,1}/\mathbb Z_{N_f}$ then reads
\begin{equation}
	\text{vol}[Q^{1,1,1}/\mathbb Z_{N_f}]=\fft{\pi^4}{8N_f} \, .
\end{equation}
%

%%%%%%%%%%%%%%%%%%%%%%%%%%%%%%%%%%%%%%%%%%

\bibliography{Index-others}

\providecommand{\href}[2]{#2}\begingroup\raggedright\begin{thebibliography}{10}

\bibitem{Pestun:2016zxk}
V.~Pestun et~al., {\it {Localization techniques in quantum field theories}},
  {\em J. Phys. A} {\bf 50} (2017), no.~44 440301,
  [\href{http://arxiv.org/abs/1608.02952}{{\tt arXiv:1608.02952}}].

\bibitem{Zaffaroni:2019dhb}
A.~Zaffaroni, {\it {AdS black holes, holography and localization}},  {\em
  Living Rev. Rel.} {\bf 23} (2020), no.~1 2,
  [\href{http://arxiv.org/abs/1902.07176}{{\tt arXiv:1902.07176}}].

\bibitem{Aharony:2008ug}
O.~Aharony, O.~Bergman, D.~L. Jafferis, and J.~Maldacena, {\it {N=6
  superconformal Chern-Simons-matter theories, M2-branes and their gravity
  duals}},  {\em JHEP} {\bf 10} (2008) 091,
  [\href{http://arxiv.org/abs/0806.1218}{{\tt arXiv:0806.1218}}].

\bibitem{Kapustin:2009kz}
A.~Kapustin, B.~Willett, and I.~Yaakov, {\it {Exact Results for Wilson Loops in
  Superconformal Chern-Simons Theories with Matter}},  {\em JHEP} {\bf 03}
  (2010) 089, [\href{http://arxiv.org/abs/0909.4559}{{\tt arXiv:0909.4559}}].

\bibitem{Fuji:2011km}
H.~Fuji, S.~Hirano, and S.~Moriyama, {\it {Summing Up All Genus Free Energy of
  ABJM Matrix Model}},  {\em JHEP} {\bf 08} (2011) 001,
  [\href{http://arxiv.org/abs/1106.4631}{{\tt arXiv:1106.4631}}].

\bibitem{Marino:2011eh}
M.~Marino and P.~Putrov, {\it {ABJM theory as a Fermi gas}},  {\em J. Stat.
  Mech.} {\bf 1203} (2012) P03001, [\href{http://arxiv.org/abs/1110.4066}{{\tt
  arXiv:1110.4066}}].

\bibitem{Drukker:2010nc}
N.~Drukker, M.~Marino, and P.~Putrov, {\it {From weak to strong coupling in
  ABJM theory}},  {\em Commun. Math. Phys.} {\bf 306} (2011) 511--563,
  [\href{http://arxiv.org/abs/1007.3837}{{\tt arXiv:1007.3837}}].

\bibitem{Herzog:2010hf}
C.~P. Herzog, I.~R. Klebanov, S.~S. Pufu, and T.~Tesileanu, {\it {Multi-Matrix
  Models and Tri-Sasaki Einstein Spaces}},  {\em Phys. Rev. D} {\bf 83} (2011)
  046001, [\href{http://arxiv.org/abs/1011.5487}{{\tt arXiv:1011.5487}}].

\bibitem{Bobev:2020egg}
N.~Bobev, A.~M. Charles, K.~Hristov, and V.~Reys, {\it {The Unreasonable
  Effectiveness of Higher-Derivative Supergravity in AdS$_4$ Holography}},
  {\em Phys. Rev. Lett.} {\bf 125} (2020), no.~13 131601,
  [\href{http://arxiv.org/abs/2006.09390}{{\tt arXiv:2006.09390}}].

\bibitem{Bobev:2021oku}
N.~Bobev, A.~M. Charles, K.~Hristov, and V.~Reys, {\it {Higher-derivative
  supergravity, AdS$_{4}$ holography, and black holes}},  {\em JHEP} {\bf 08}
  (2021) 173, [\href{http://arxiv.org/abs/2106.04581}{{\tt arXiv:2106.04581}}].

\bibitem{Bhattacharyya:2012ye}
S.~Bhattacharyya, A.~Grassi, M.~Marino, and A.~Sen, {\it {A One-Loop Test of
  Quantum Supergravity}},  {\em Class. Quant. Grav.} {\bf 31} (2014) 015012,
  [\href{http://arxiv.org/abs/1210.6057}{{\tt arXiv:1210.6057}}].

\bibitem{Dabholkar:2014wpa}
A.~Dabholkar, N.~Drukker, and J.~Gomes, {\it {Localization in supergravity and
  quantum $AdS_4/CFT_3$ holography}},  {\em JHEP} {\bf 10} (2014) 090,
  [\href{http://arxiv.org/abs/1406.0505}{{\tt arXiv:1406.0505}}].

\bibitem{Hristov:2022lcw}
K.~Hristov, {\it {ABJM at finite N via 4d supergravity}},  {\em JHEP} {\bf 10}
  (2022) 190, [\href{http://arxiv.org/abs/2204.02992}{{\tt arXiv:2204.02992}}].

\bibitem{Nosaka:2015iiw}
T.~Nosaka, {\it {Instanton effects in ABJM theory with general R-charge
  assignments}},  {\em JHEP} {\bf 03} (2016) 059,
  [\href{http://arxiv.org/abs/1512.02862}{{\tt arXiv:1512.02862}}].

\bibitem{Hatsuda:2016uqa}
Y.~Hatsuda, {\it {ABJM on ellipsoid and topological strings}},  {\em JHEP} {\bf
  07} (2016) 026, [\href{http://arxiv.org/abs/1601.02728}{{\tt
  arXiv:1601.02728}}].

\bibitem{Mezei:2013gqa}
M.~Mezei and S.~S. Pufu, {\it {Three-sphere free energy for classical gauge
  groups}},  {\em JHEP} {\bf 02} (2014) 037,
  [\href{http://arxiv.org/abs/1312.0920}{{\tt arXiv:1312.0920}}].

\bibitem{Bobev:2022jte}
N.~Bobev, J.~Hong, and V.~Reys, {\it {Large N Partition Functions, Holography,
  and Black Holes}},  {\em Phys. Rev. Lett.} {\bf 129} (2022), no.~4 041602,
  [\href{http://arxiv.org/abs/2203.14981}{{\tt arXiv:2203.14981}}].

\bibitem{Bobev:2022eus}
N.~Bobev, J.~Hong, and V.~Reys, {\it {Large N partition functions of the ABJM
  theory}},  {\em JHEP} {\bf 02} (2023) 020,
  [\href{http://arxiv.org/abs/2210.09318}{{\tt arXiv:2210.09318}}].

\bibitem{Bobev:2022wem}
N.~Bobev, S.~Choi, J.~Hong, and V.~Reys, {\it {Large N superconformal indices
  for 3d holographic SCFTs}},  {\em JHEP} {\bf 02} (2023) 027,
  [\href{http://arxiv.org/abs/2210.15326}{{\tt arXiv:2210.15326}}].

\bibitem{Liu:2017vll}
J.~T. Liu, L.~A. Pando~Zayas, V.~Rathee, and W.~Zhao, {\it {Toward Microstate
  Counting Beyond Large N in Localization and the Dual One-loop Quantum
  Supergravity}},  {\em JHEP} {\bf 01} (2018) 026,
  [\href{http://arxiv.org/abs/1707.04197}{{\tt arXiv:1707.04197}}].

\bibitem{Benini:2015noa}
F.~Benini and A.~Zaffaroni, {\it {A topologically twisted index for
  three-dimensional supersymmetric theories}},  {\em JHEP} {\bf 07} (2015) 127,
  [\href{http://arxiv.org/abs/1504.03698}{{\tt arXiv:1504.03698}}].

\bibitem{Benini:2015eyy}
F.~Benini, K.~Hristov, and A.~Zaffaroni, {\it {Black hole microstates in
  AdS$_{4}$ from supersymmetric localization}},  {\em JHEP} {\bf 05} (2016)
  054, [\href{http://arxiv.org/abs/1511.04085}{{\tt arXiv:1511.04085}}].

\bibitem{Benini:2016hjo}
F.~Benini and A.~Zaffaroni, {\it {Supersymmetric partition functions on Riemann
  surfaces}},  {\em Proc. Symp. Pure Math.} {\bf 96} (2017) 13--46,
  [\href{http://arxiv.org/abs/1605.06120}{{\tt arXiv:1605.06120}}].

\bibitem{Closset:2016arn}
C.~Closset and H.~Kim, {\it {Comments on twisted indices in 3d supersymmetric
  gauge theories}},  {\em JHEP} {\bf 08} (2016) 059,
  [\href{http://arxiv.org/abs/1605.06531}{{\tt arXiv:1605.06531}}].

\bibitem{Liu:2017vbl}
J.~T. Liu, L.~A. Pando~Zayas, V.~Rathee, and W.~Zhao, {\it {One-Loop Test of
  Quantum Black Holes in anti\textendash{}de Sitter Space}},  {\em Phys. Rev.
  Lett.} {\bf 120} (2018), no.~22 221602,
  [\href{http://arxiv.org/abs/1711.01076}{{\tt arXiv:1711.01076}}].

\bibitem{Bobev:2020pjk}
N.~Bobev, A.~M. Charles, and V.~S. Min, {\it {Euclidean black saddles and
  AdS$_{4}$ black holes}},  {\em JHEP} {\bf 10} (2020) 073,
  [\href{http://arxiv.org/abs/2006.01148}{{\tt arXiv:2006.01148}}].

\bibitem{Hosseini:2016tor}
S.~M. Hosseini and A.~Zaffaroni, {\it {Large $N$ matrix models for 3d ${\cal
  N}=2$ theories: twisted index, free energy and black holes}},  {\em JHEP}
  {\bf 08} (2016) 064, [\href{http://arxiv.org/abs/1604.03122}{{\tt
  arXiv:1604.03122}}].

\bibitem{Hosseini:2016ume}
S.~M. Hosseini and N.~Mekareeya, {\it {Large $N$ topologically twisted index:
  necklace quivers, dualities, and Sasaki-Einstein spaces}},  {\em JHEP} {\bf
  08} (2016) 089, [\href{http://arxiv.org/abs/1604.03397}{{\tt
  arXiv:1604.03397}}].

\bibitem{PandoZayas:2020iqr}
L.~A. Pando~Zayas and Y.~Xin, {\it {Universal logarithmic behavior in
  microstate counting and the dual one-loop entropy of $AdS_4$ black holes}},
  {\em Phys. Rev. D} {\bf 103} (2021), no.~2 026003,
  [\href{http://arxiv.org/abs/2008.03239}{{\tt arXiv:2008.03239}}].

\bibitem{Chester:2021gdw}
S.~M. Chester, R.~R. Kalloor, and A.~Sharon, {\it {Squashing, Mass, and
  Holography for 3d Sphere Free Energy}},  {\em JHEP} {\bf 04} (2021) 244,
  [\href{http://arxiv.org/abs/2102.05643}{{\tt arXiv:2102.05643}}].

\bibitem{Minahan:2021pfv}
J.~Minahan, U.~Naseer, and C.~Thull, {\it {Squashing and supersymmetry
  enhancement in three dimensions}},  {\em SciPost Phys.} {\bf 12} (2022),
  no.~1 025, [\href{http://arxiv.org/abs/2107.07151}{{\tt arXiv:2107.07151}}].

\bibitem{Jafferis:2011zi}
D.~L. Jafferis, I.~R. Klebanov, S.~S. Pufu, and B.~R. Safdi, {\it {Towards the
  F-Theorem: N=2 Field Theories on the Three-Sphere}},  {\em JHEP} {\bf 06}
  (2011) 102, [\href{http://arxiv.org/abs/1103.1181}{{\tt arXiv:1103.1181}}].

\bibitem{Bobev:2018uxk}
N.~Bobev, V.~S. Min, and K.~Pilch, {\it {Mass-deformed ABJM and black holes in
  AdS$_{4}$}},  {\em JHEP} {\bf 03} (2018) 050,
  [\href{http://arxiv.org/abs/1801.03135}{{\tt arXiv:1801.03135}}].

\bibitem{Bobev:2018wbt}
N.~Bobev, V.~S. Min, K.~Pilch, and F.~Rosso, {\it {Mass Deformations of the
  ABJM Theory: The Holographic Free Energy}},  {\em JHEP} {\bf 03} (2019) 130,
  [\href{http://arxiv.org/abs/1812.01026}{{\tt arXiv:1812.01026}}].

\bibitem{Corrado:2001nv}
R.~Corrado, K.~Pilch, and N.~P. Warner, {\it {An N=2 supersymmetric membrane
  flow}},  {\em Nucl. Phys. B} {\bf 629} (2002) 74--96,
  [\href{http://arxiv.org/abs/hep-th/0107220}{{\tt hep-th/0107220}}].

\bibitem{Azzurli:2017kxo}
F.~Azzurli, N.~Bobev, P.~M. Crichigno, V.~S. Min, and A.~Zaffaroni, {\it {A
  universal counting of black hole microstates in AdS$_{4}$}},  {\em JHEP} {\bf
  02} (2018) 054, [\href{http://arxiv.org/abs/1707.04257}{{\tt
  arXiv:1707.04257}}].

\bibitem{Jafferis:2010un}
D.~L. Jafferis, {\it {The Exact Superconformal R-Symmetry Extremizes Z}},  {\em
  JHEP} {\bf 05} (2012) 159, [\href{http://arxiv.org/abs/1012.3210}{{\tt
  arXiv:1012.3210}}].

\bibitem{Bobev:2017uzs}
N.~Bobev and P.~M. Crichigno, {\it {Universal RG Flows Across Dimensions and
  Holography}},  {\em JHEP} {\bf 12} (2017) 065,
  [\href{http://arxiv.org/abs/1708.05052}{{\tt arXiv:1708.05052}}].

\bibitem{Kapustin:2010xq}
A.~Kapustin, B.~Willett, and I.~Yaakov, {\it {Nonperturbative Tests of
  Three-Dimensional Dualities}},  {\em JHEP} {\bf 10} (2010) 013,
  [\href{http://arxiv.org/abs/1003.5694}{{\tt arXiv:1003.5694}}].

\bibitem{Benini:2009qs}
F.~Benini, C.~Closset, and S.~Cremonesi, {\it {Chiral flavors and M2-branes at
  toric CY4 singularities}},  {\em JHEP} {\bf 02} (2010) 036,
  [\href{http://arxiv.org/abs/0911.4127}{{\tt arXiv:0911.4127}}].

\bibitem{Bashkirov:2010kz}
D.~Bashkirov and A.~Kapustin, {\it {Supersymmetry enhancement by monopole
  operators}},  {\em JHEP} {\bf 05} (2011) 015,
  [\href{http://arxiv.org/abs/1007.4861}{{\tt arXiv:1007.4861}}].

\bibitem{Grassi:2014vwa}
A.~Grassi and M.~Marino, {\it {M-theoretic matrix models}},  {\em JHEP} {\bf
  02} (2015) 115, [\href{http://arxiv.org/abs/1403.4276}{{\tt
  arXiv:1403.4276}}].

\bibitem{Atiyah:1978ri}
M.~F. Atiyah, N.~J. Hitchin, V.~G. Drinfeld, and Y.~I. Manin, {\it
  {Construction of Instantons}},  {\em Phys. Lett. A} {\bf 65} (1978) 185--187.

\bibitem{Drukker:2011zy}
N.~Drukker, M.~Marino, and P.~Putrov, {\it {Nonperturbative aspects of ABJM
  theory}},  {\em JHEP} {\bf 11} (2011) 141,
  [\href{http://arxiv.org/abs/1103.4844}{{\tt arXiv:1103.4844}}].

\bibitem{Hatsuda:2012dt}
Y.~Hatsuda, S.~Moriyama, and K.~Okuyama, {\it {Instanton Effects in ABJM Theory
  from Fermi Gas Approach}},  {\em JHEP} {\bf 01} (2013) 158,
  [\href{http://arxiv.org/abs/1211.1251}{{\tt arXiv:1211.1251}}].

\bibitem{Gaiotto:2009tk}
D.~Gaiotto and D.~L. Jafferis, {\it {Notes on adding D6 branes wrapping RP**3
  in AdS(4) x CP**3}},  {\em JHEP} {\bf 11} (2012) 015,
  [\href{http://arxiv.org/abs/0903.2175}{{\tt arXiv:0903.2175}}].

\bibitem{Hohenegger:2009as}
S.~Hohenegger and I.~Kirsch, {\it {A Note on the holography of Chern-Simons
  matter theories with flavour}},  {\em JHEP} {\bf 04} (2009) 129,
  [\href{http://arxiv.org/abs/0903.1730}{{\tt arXiv:0903.1730}}].

\bibitem{Hikida:2009tp}
Y.~Hikida, W.~Li, and T.~Takayanagi, {\it {ABJM with Flavors and FQHE}},  {\em
  JHEP} {\bf 07} (2009) 065, [\href{http://arxiv.org/abs/0903.2194}{{\tt
  arXiv:0903.2194}}].

\bibitem{Cheon:2011th}
S.~Cheon, D.~Gang, S.~Kim, and J.~Park, {\it {Refined test of AdS4/CFT3
  correspondence for N=2,3 theories}},  {\em JHEP} {\bf 05} (2011) 027,
  [\href{http://arxiv.org/abs/1102.4273}{{\tt arXiv:1102.4273}}].

\bibitem{Santamaria:2010dm}
R.~C. Santamaria, M.~Marino, and P.~Putrov, {\it {Unquenched flavor and
  tropical geometry in strongly coupled Chern-Simons-matter theories}},  {\em
  JHEP} {\bf 10} (2011) 139, [\href{http://arxiv.org/abs/1011.6281}{{\tt
  arXiv:1011.6281}}].

\bibitem{Martelli:2009ga}
D.~Martelli and J.~Sparks, {\it {AdS$_4$/CFT$_3$ duals from M2-branes at
  hypersurface singularities and their deformations}},  {\em JHEP} {\bf 12}
  (2009) 017, [\href{http://arxiv.org/abs/0909.2036}{{\tt arXiv:0909.2036}}].

\bibitem{Jafferis:2009th}
D.~L. Jafferis, {\it {Quantum corrections to $\mathcal{N} = 2$ Chern-Simons
  theories with flavor and their AdS$_{4}$ duals}},  {\em JHEP} {\bf 08} (2013)
  046, [\href{http://arxiv.org/abs/0911.4324}{{\tt arXiv:0911.4324}}].

\bibitem{Franco:2008um}
S.~Franco, A.~Hanany, J.~Park, and D.~Rodriguez-Gomez, {\it {Towards M2-brane
  Theories for Generic Toric Singularities}},  {\em JHEP} {\bf 12} (2008) 110,
  [\href{http://arxiv.org/abs/0809.3237}{{\tt arXiv:0809.3237}}].

\bibitem{Franco:2009sp}
S.~Franco, I.~R. Klebanov, and D.~Rodriguez-Gomez, {\it {M2-branes on Orbifolds
  of the Cone over Q**1,1,1}},  {\em JHEP} {\bf 08} (2009) 033,
  [\href{http://arxiv.org/abs/0903.3231}{{\tt arXiv:0903.3231}}].

\bibitem{Cremonesi:2010ae}
S.~Cremonesi, {\it {Type IIB construction of flavoured ABJ(M) and fractional M2
  branes}},  {\em JHEP} {\bf 01} (2011) 076,
  [\href{http://arxiv.org/abs/1007.4562}{{\tt arXiv:1007.4562}}].

\bibitem{Hatsuda:2014vsa}
Y.~Hatsuda and K.~Okuyama, {\it {Probing non-perturbative effects in
  M-theory}},  {\em JHEP} {\bf 10} (2014) 158,
  [\href{http://arxiv.org/abs/1407.3786}{{\tt arXiv:1407.3786}}].

\bibitem{Chester:2020jay}
S.~M. Chester, R.~R. Kalloor, and A.~Sharon, {\it {3d $ \mathcal{N} $ = 4 OPE
  coefficients from Fermi gas}},  {\em JHEP} {\bf 07} (2020) 041,
  [\href{http://arxiv.org/abs/2004.13603}{{\tt arXiv:2004.13603}}].

\bibitem{Willett:2016adv}
B.~Willett, {\it {Localization on three-dimensional manifolds}},  {\em J. Phys.
  A} {\bf 50} (2017), no.~44 443006,
  [\href{http://arxiv.org/abs/1608.02958}{{\tt arXiv:1608.02958}}].

\bibitem{Closset:2012ru}
C.~Closset, T.~T. Dumitrescu, G.~Festuccia, and Z.~Komargodski, {\it
  {Supersymmetric Field Theories on Three-Manifolds}},  {\em JHEP} {\bf 05}
  (2013) 017, [\href{http://arxiv.org/abs/1212.3388}{{\tt arXiv:1212.3388}}].

\bibitem{Chester:2014fya}
S.~M. Chester, J.~Lee, S.~S. Pufu, and R.~Yacoby, {\it {The $ \mathcal{N}=8 $
  superconformal bootstrap in three dimensions}},  {\em JHEP} {\bf 09} (2014)
  143, [\href{http://arxiv.org/abs/1406.4814}{{\tt arXiv:1406.4814}}].

\bibitem{Benna:2008zy}
M.~Benna, I.~Klebanov, T.~Klose, and M.~Smedback, {\it {Superconformal
  Chern-Simons Theories and AdS(4)/CFT(3) Correspondence}},  {\em JHEP} {\bf
  09} (2008) 072, [\href{http://arxiv.org/abs/0806.1519}{{\tt
  arXiv:0806.1519}}].

\bibitem{Warner:1983vz}
N.~P. Warner, {\it {Some New Extrema of the Scalar Potential of Gauged $N=8$
  Supergravity}},  {\em Phys. Lett. B} {\bf 128} (1983) 169--173.

\bibitem{Bobev:2019zmz}
N.~Bobev and P.~M. Crichigno, {\it {Universal spinning black holes and theories
  of class $ \mathcal{R} $}},  {\em JHEP} {\bf 12} (2019) 054,
  [\href{http://arxiv.org/abs/1909.05873}{{\tt arXiv:1909.05873}}].

\bibitem{Gauntlett:2007ma}
J.~P. Gauntlett and O.~Varela, {\it {Consistent Kaluza-Klein reductions for
  general supersymmetric AdS solutions}},  {\em Phys. Rev. D} {\bf 76} (2007)
  126007, [\href{http://arxiv.org/abs/0707.2315}{{\tt arXiv:0707.2315}}].

\bibitem{Romans:1991nq}
L.~J. Romans, {\it {Supersymmetric, cold and lukewarm black holes in
  cosmological Einstein-Maxwell theory}},  {\em Nucl. Phys. B} {\bf 383} (1992)
  395--415, [\href{http://arxiv.org/abs/hep-th/9203018}{{\tt hep-th/9203018}}].

\bibitem{BenettiGenolini:2019jdz}
P.~Benetti~Genolini, J.~M. Perez Ipi\~na, and J.~Sparks, {\it {Localization of
  the action in AdS/CFT}},  {\em JHEP} {\bf 10} (2019) 252,
  [\href{http://arxiv.org/abs/1906.11249}{{\tt arXiv:1906.11249}}].

\bibitem{Benini:2016rke}
F.~Benini, K.~Hristov, and A.~Zaffaroni, {\it {Exact microstate counting for
  dyonic black holes in AdS4}},  {\em Phys. Lett. B} {\bf 771} (2017) 462--466,
  [\href{http://arxiv.org/abs/1608.07294}{{\tt arXiv:1608.07294}}].

\bibitem{Cassani:2011fu}
D.~Cassani and P.~Koerber, {\it {Tri-Sasakian consistent reduction}},  {\em
  JHEP} {\bf 01} (2012) 086, [\href{http://arxiv.org/abs/1110.5327}{{\tt
  arXiv:1110.5327}}].

\bibitem{Cassani:2012pj}
D.~Cassani, P.~Koerber, and O.~Varela, {\it {All homogeneous N=2 M-theory
  truncations with supersymmetric AdS4 vacua}},  {\em JHEP} {\bf 11} (2012)
  173, [\href{http://arxiv.org/abs/1208.1262}{{\tt arXiv:1208.1262}}].

\bibitem{Ferrero:2020twa}
P.~Ferrero, J.~P. Gauntlett, J.~M.~P. Ipi\~na, D.~Martelli, and J.~Sparks, {\it
  {Accelerating black holes and spinning spindles}},  {\em Phys. Rev. D} {\bf
  104} (2021), no.~4 046007, [\href{http://arxiv.org/abs/2012.08530}{{\tt
  arXiv:2012.08530}}].

\bibitem{Caldarelli:1998hg}
M.~M. Caldarelli and D.~Klemm, {\it {Supersymmetry of Anti-de Sitter black
  holes}},  {\em Nucl. Phys. B} {\bf 545} (1999) 434--460,
  [\href{http://arxiv.org/abs/hep-th/9808097}{{\tt hep-th/9808097}}].

\bibitem{Cassani:2021dwa}
D.~Cassani, J.~P. Gauntlett, D.~Martelli, and J.~Sparks, {\it {Thermodynamics
  of accelerating and supersymmetric AdS4 black holes}},  {\em Phys. Rev. D}
  {\bf 104} (2021), no.~8 086005, [\href{http://arxiv.org/abs/2106.05571}{{\tt
  arXiv:2106.05571}}].

\bibitem{SCI-WIP}
N.~Bobev, S.~Choi, J.~Hong, and V.~Reys, {\it {work in progress}},
  \href{http://arxiv.org/abs/2023}{{\tt 2023}}.

\bibitem{Inglese:2023wky}
M.~Inglese, D.~Martelli, and A.~Pittelli, {\it {The Spindle Index from
  Localization}},  \href{http://arxiv.org/abs/2303.14199}{{\tt
  arXiv:2303.14199}}.

\bibitem{Lezcano:2021qbj}
A.~G. Lezcano, J.~Hong, J.~T. Liu, and L.~A. Pando~Zayas, {\it {The
  Bethe-Ansatz approach to the $ \mathcal{N} $ = 4 superconformal index at
  finite rank}},  {\em JHEP} {\bf 06} (2021) 126,
  [\href{http://arxiv.org/abs/2101.12233}{{\tt arXiv:2101.12233}}].

\bibitem{Benini:2021ano}
F.~Benini and G.~Rizi, {\it {Superconformal index of low-rank gauge theories
  via the Bethe Ansatz}},  {\em JHEP} {\bf 05} (2021) 061,
  [\href{http://arxiv.org/abs/2102.03638}{{\tt arXiv:2102.03638}}].

\bibitem{Gang:2019jut}
D.~Gang and M.~Yamazaki, {\it {Expanding 3d $ \mathcal{N} $ = 2 theories around
  the round sphere}},  {\em JHEP} {\bf 02} (2020) 102,
  [\href{http://arxiv.org/abs/1912.09617}{{\tt arXiv:1912.09617}}].

\bibitem{Airy-WIP}
N.~Bobev, P.-J. De~Smet, J.~Hong, V.~Reys, and X.~Zhang, {\it {work in
  progress}},  \href{http://arxiv.org/abs/2023}{{\tt 2023}}.

\bibitem{Hosseini:2019iad}
S.~M. Hosseini, K.~Hristov, and A.~Zaffaroni, {\it {Gluing gravitational blocks
  for AdS black holes}},  {\em JHEP} {\bf 12} (2019) 168,
  [\href{http://arxiv.org/abs/1909.10550}{{\tt arXiv:1909.10550}}].

\bibitem{Choi:2019dfu}
S.~Choi and C.~Hwang, {\it {Universal 3d Cardy Block and Black Hole Entropy}},
  {\em JHEP} {\bf 03} (2020) 068, [\href{http://arxiv.org/abs/1911.01448}{{\tt
  arXiv:1911.01448}}].

\bibitem{Hristov:2021qsw}
K.~Hristov, {\it {4d $ \mathcal{N} $ = 2 supergravity observables from
  Nekrasov-like partition functions}},  {\em JHEP} {\bf 02} (2022) 079,
  [\href{http://arxiv.org/abs/2111.06903}{{\tt arXiv:2111.06903}}].

\bibitem{Hristov:2022plc}
K.~Hristov, {\it {Maximally symmetric nuts in 4d \ensuremath{\mathscr{N}} = 2
  higher derivative supergravity}},  {\em JHEP} {\bf 02} (2023) 110,
  [\href{http://arxiv.org/abs/2212.10590}{{\tt arXiv:2212.10590}}].

\bibitem{Bergman:2009zh}
O.~Bergman and S.~Hirano, {\it {Anomalous radius shift in AdS(4)/CFT(3)}},
  {\em JHEP} {\bf 07} (2009) 016, [\href{http://arxiv.org/abs/0902.1743}{{\tt
  arXiv:0902.1743}}].

\bibitem{Hristov:2018lod}
K.~Hristov, I.~Lodato, and V.~Reys, {\it {On the quantum entropy function in 4d
  gauged supergravity}},  {\em JHEP} {\bf 07} (2018) 072,
  [\href{http://arxiv.org/abs/1803.05920}{{\tt arXiv:1803.05920}}].

\bibitem{Hristov:2019xku}
K.~Hristov, I.~Lodato, and V.~Reys, {\it {One-loop determinants for black holes
  in 4d gauged supergravity}},  {\em JHEP} {\bf 11} (2019) 105,
  [\href{http://arxiv.org/abs/1908.05696}{{\tt arXiv:1908.05696}}].

\bibitem{BenettiGenolini:2023rkq}
P.~Benetti~Genolini, A.~Cabo-Bizet, and S.~Murthy, {\it {Supersymmetric phases
  of AdS$_4$/CFT$_3$}},  \href{http://arxiv.org/abs/2301.00763}{{\tt
  arXiv:2301.00763}}.

\bibitem{Castellani:1983yg}
L.~Castellani, L.~J. Romans, and N.~P. Warner, {\it {A Classification of
  Compactifying Solutions for $d=11$ Supergravity}},  {\em Nucl. Phys. B} {\bf
  241} (1984) 429--462.

\bibitem{Page:1984ac}
D.~N. Page and C.~N. Pope, {\it {New Squashed Solutions of $D=11$
  Supergravity}},  {\em Phys. Lett. B} {\bf 147} (1984) 55--60.

\bibitem{Gauntlett:2005jb}
J.~P. Gauntlett, S.~Lee, T.~Mateos, and D.~Waldram, {\it {Marginal deformations
  of field theories with AdS(4) duals}},  {\em JHEP} {\bf 08} (2005) 030,
  [\href{http://arxiv.org/abs/hep-th/0505207}{{\tt hep-th/0505207}}].

\bibitem{Bergman:2001qi}
A.~Bergman and C.~P. Herzog, {\it {The Volume of some nonspherical horizons and
  the AdS / CFT correspondence}},  {\em JHEP} {\bf 01} (2002) 030,
  [\href{http://arxiv.org/abs/hep-th/0108020}{{\tt hep-th/0108020}}].

\bibitem{Duff:1986hr}
M.~J. Duff, B.~E.~W. Nilsson, and C.~N. Pope, {\it {Kaluza-Klein
  Supergravity}},  {\em Phys. Rept.} {\bf 130} (1986) 1--142.

\end{thebibliography}\endgroup
\bibliographystyle{JHEP}

\end{document}